\documentclass{article}[12pt]
\usepackage{authblk}
\usepackage{tabto} 
\usepackage{relsize} 
\usepackage{graphicx} 
\usepackage{tikz} 
\usepackage{float} 
\usetikzlibrary{calc,arrows}
\usepackage{amsmath,amssymb} 
\newcommand{\tarc}{\mbox{\large$\frown$}}
\newcommand{\arc}[2][-3ex]{{#2}{\kern #1{\raisebox{1.5ex}{\tarc}}}}
\usepackage{amsfonts} 
\usepackage[export]{adjustbox} 
\usepackage{algpseudocode}
\usepackage{algorithm}
\usetikzlibrary{calc,arrows}
\usepackage{todonotes}
\usepackage{comment}
\usepackage{wrapfig}
\usepackage{hyperref} 
\hypersetup{
    colorlinks = true,
    linkcolor={blue!80!red},
    citecolor={green!60!blue},
    urlcolor={red!70!blue}
}
\usepackage{siunitx} 
\usepackage{appendix}
\usepackage{multirow} 
\usepackage[
backend=biber,
style=alphabetic,
sorting=ynt
]{biblatex}
\usepackage{geometry}
\usepackage{array}
\usepackage{longtable}

\def\Xint#1{\mathchoice
   {\XXint\displaystyle\textstyle{#1}}%
   {\XXint\textstyle\scriptstyle{#1}}%
   {\XXint\scriptstyle\scriptscriptstyle{#1}}%
   {\XXint\scriptscriptstyle\scriptscriptstyle{#1}}%
   \!\int}
\def\XXint#1#2#3{{\setbox0=\hbox{$#1{#2#3}{\int}$}
     \vcenter{\hbox{$#2#3$}}\kern-.5\wd0}}

\def\dashint{\Xint-}

\title{Falling plates with leading-edge vortex shedding}
\author[1]{Yu Jun Loo \thanks{looyujun@umich.edu}}
\author[1]{Silas Alben}
\affil[1]{Department of Mathematics, University of Michigan, Ann Arbor, MI 48109, USA}
\date{}
\begin{document}

\maketitle

\begin{abstract}
We develop a new numerical method for thin plates falling in inviscid fluid that allows for leading-edge vortex shedding. The inclusion of leading-edge shedding restores physical dynamics to vortex-sheet models of falling bodies, and for the first time large-amplitude fluttering and tumbling are observed in inviscid simulations. Leading-edge shedding is achieved by introducing a novel quadrature rule and smoothing procedure for the Birkhoff-Rott equations. The smoothing error is controlled by a novel fencing procedure. We find a transition point between fluttering and tumbling that is consistent with previous viscous simulations and experiments, and other falling motions such as looping, autorotation are also observed as the plate density increases. The dipole street wakes behind the fluttering plates resemble those in experiments. We consider plates bent into V shapes and study the effects of density and bending angle on the qualitative falling dynamics. At small densities, increasing the bending angle stabilizes the falling motion into fluttering, while at large densities, decreasing the bending angle stabilizes the falling motion into autorotation. In the autorotation regime, the magnitude of angular velocity increases as time cubed before it reaches a terminal angular velocity, and in the fluttering regime, the fluttering frequency scales as the $-1/2$ power of $R_1$, the plate density.

\end{abstract}

\section{Introduction}
Among the simplest examples of fluid-structure interaction is a body falling freely through fluid, driven only by gravity. This system shows many of the complex dynamics that occur in fluid-structure interactions in the natural environment. The physical mechanisms involved in passive falling dynamics at high Reynolds numbers are relevant to the settling of dandelion seeds \cite{cummins2018separated} and the passive gliding of swifts \cite{thomas_passive}, for example. The underlying passive dynamics of these systems has inspired energy-efficient designs for microscale gliders \cite{unpoweredflight}.
\newline

 In the experiments of \cite{Field1997} on falling disks, the motions were classified into four categories. These included steady falling (constant orientation), periodic fluttering (side-to-side motion), tumbling (end-over-end rotation), and chaotic motions (alternating between tumbling and fluttering). For macroscopic falling objects, the Reynolds number $\gtrapprox 10^2$ \cite{cummins2018separated, ANDERSEN_PESAVENTO_WANG_2005, tam2010tumbling}. At these relatively high Reynolds numbers relevant to this study, no steady falling occurs \cite{Field1997}. Instead, the falling motions consist primarily of side-to-side fluttering and end-over-end tumbling. At these same Reynolds numbers, the high numerical resolution required for Navier-Stokes simulations of these problems can make investigation of the long-time dynamics across parameter spaces prohibitively expensive. Recently, several authors have developed less expensive inviscid vortex-shedding models as approximations to these high-Reynolds-number systems, for both rigid \cite{jones2005falling, michelin2009unsteady, sohn} and flexible falling bodies \cite{albenFall}.
\newline

\cite{jones2005falling} presented some of the first inviscid simulations of a falling plate, using shed vortex sheets similar to the present study. \cite{michelin2009unsteady} used a related inviscid model, the Brown-Michael method, with the vortex sheets replaced by point vortices whose strengths and positions evolve in time. These studies were confined to the period of initial growth of small perturbations from broadside-on falling ($t\lessapprox15$) where the plate weakly oscillates, and halted before steady-state large-amplitude fluttering or tumbling could occur. This was due primarily to the computational difficulties associated with computing leading-edge vortex shedding \cite{jones2005falling, kansoRot}.
\newline

To address these computational challenges, \cite{albenFall} developed an approach for flexible falling plates where vorticity release from body edges halts intermittently when the fluid velocity at the edge is directed onto the body. At those instants the flow approximates that with an attached leading edge vortex. Experiments have found that a body's ability to passively bend sometimes allows the leading-edge vortex to remain attached rather than being shed \cite{kim2011flexibility}. Halting leading-edge shedding intermittently allowed simulations to be extended beyond the point where leading-edge shedding would normally occur and was also used in \cite{kansoRot} and \cite{sohn} for hovering and falling rigid bodies respectively. 
\newline

However, in most real situations, vorticity is continually shed from the leading edge, which dramatically changes the pressure loading on the body. \cite{PanXiaoLE} demonstrated this effect, finding that including leading-edge shedding greatly improved the accuracy of thrust forces in an inviscid boundary element model of flow past a flapping foil. In the aforementioned inviscid models without leading-edge shedding, many of the dynamics that are commonly observed in both experiments and direct Navier-Stokes simulations of falling bodies, such as tumbling and side-to-side fluttering, fail to occur. Instead, the models settle into long periods of gliding and diving where the body moves nearly tangentially to its leading edge. \cite{albenFall} found that flexibility can destabilize the body from this state. In this work, we will demonstrate a robust method for continual leading-edge shedding that results in more realistic dynamics.  A related body of work has developed the concept of a leading-edge suction parameter (LESP), to determine when leading-edge shedding occurs in discrete vortex models similar to the one used in this study \cite{ramesh2012theoretical, narsipur}. With this approach, shedding is still halted intermittently (but less frequently than in \cite{albenFall}), and when shedding occurs the vortex sheet is discontinuous across the leading edge \cite{ramesh2012theoretical} \cite[182]{eldredge_inviscid_flows}.
\newline

In this study, we propose a series of improvements to the numerical scheme described in \cite{albenFlex}. These improvements stabilize the interaction between the falling body and the vortex sheet shed from the leading edge,  allowing for continual shedding from the leading edge. To demonstrate the method, we use it to study the long-time behavior of falling plates, both flat and curved, in an inviscid fluid. With leading-edge shedding present, both fluttering and periodic tumbling are observed for the first time in inviscid simulations.
\newline
 
The first systematic study of falling plates was given by \cite{smith1971autorotating}, which classified the plate dynamics by the dimensionless moment of inertia $I^*$. In the present study, we will classify the motions across $R_1 = \frac{3 \pi}{4}I^*$ \cite{sohn}. For small values of $I^*$, the plates flutter side to side. As $I^*$ increases, the plate dynamics transition to a tumbling motion where the plate falls end-over-end. The transition region between fluttering and tumbling was found to be $0.2 < I^* < 0.3$, for thickness to length ratios in the range $[1/4 , 2/5]$. This transition region was further verified by both the direct Navier-Stokes simulations of \cite{wang_numerical} who determined it as $0.21< I^* < 0.31$ for thickness to length ratios $[1/20, 1/4]$, as well as the inviscid simulations of \cite{sohn} who determined the transition point as $I^* \approx 0.27$ for a plate of zero thickness. According to the bifurcation diagram in \cite{Xiang2018}, the transition between fluttering and tumbling occurs at $I^* \approx 0.2$ for thickness to length ratios in the range $[1/200,3/80]$. The experiments of \cite{belmonte1998flutter} classified falling plates by the square of the dimensionless mass density $\text{Fr}= R_1^2$ and showed for the range of thickness to length ratios $[1/320,1/5]$, that the transition from fluttering to tumbling occurs at $\text{Fr} = 0.67 \pm 0.05$, or $I^* \approx 0.4$,  slightly higher than the aforementioned studies suggest. These experimental results were reproduced numerically in \cite{Rana2020} which determined the transition region to be $0.24< I^* < 0.4$ for the thickness to length ratios $[1/14,1/5]$, and attributes the difference in transition regions from \cite{wang_numerical} to the sensitivity of the system to initial conditions. On the other hand, \cite{sohn} suggests that the differences may be due to the experimental set up. The simulations in \cite{wang_numerical} suggest that decreasing the thickness to length ratio lowers the transition point, shifting the transition region downward. In this study, we consider plates of zero thickness. 
 \newline 
 
 Many other recent works have studied rounded bodies falling in a fluid. Direct Navier-Stokes simulations of falling thin ellipses were performed in \cite{pesavento2004falling, ANDERSEN_PESAVENTO_WANG_2005, andersen2005unsteady, wu_ellipse}. In \cite{wu_ellipse}, the ellipses were classified based on their dimensionless moment of inertia $I^*$, and their aspect ratio. The effects of both $I^*$ and the Reynolds number $\text{Re}$ on the resulting dynamics were studied. Near the transition between fluttering and tumbling,  $0.12 < I^* < 0.35$,  chaotic motions appeared where the ellipse alternated between fluttering and tumbling at irregular intervals. For falling circular disks, decreasing $I^*$ (or $\text{Fr}$) introduces 3D dynamics. The side-to-side fluttering motion transitions smoothly to an elliptical then circular spiraling motion \cite{zhong2011experimental}. \cite{heisinger2014coins} studied the phase space of falling coins and obtained statistics for the coins' landing positions. Despite working in 3D, the major falling modes could still be classified as fluttering, tumbling, or chaotic falling. 
 \newline 

Multiple falling plates were studied in \cite{KushwahaMultiPlates}. After an initial period of interaction, all plates eventually settle into either fluttering or tumbling motions. The separated wake induced by porosity was observed to have stabilizing effects on a Dandelion seed in \cite{cummins2018separated}. Effects of porosity and flexibility on stabilizing the dynamics of falling thin objects were studied by \cite{ledda2019flow} and \cite{tam2010tumbling} respectively. In \cite{ledda2019flow}, porosity was shown to stabilize the passive fluttering motion of a falling dandelion seed into steady falling. A related stabilization is caused by holes in a falling coin, studied by \cite{vincent2016holes}.
\newline

To validate these inviscid approaches, several studies have compared inviscid models with viscous simulations and experiments. Remarkable agreement between dye-visualized shed vortex sheets in experiments \cite{didden_experiments} and inviscid computations were shown by \cite{nitsche1994numerical} for a flow through a circular tube. Good agreement between experiments and inviscid simulations for the forces and flows around a flapping foil were shown by \cite{PanXiaoLE}, as long as leading-edge shedding was accounted for. In \cite{sheng2012simulating}, without leading-edge vortex shedding, there was less agreement between the forces computed with inviscid and direct Navier-Stokes methods, which could also be due to the absence of skin friction. In the present study we incorporate leading-edge shedding as well as skin friction effects. 
\newline

\section{Model}
\vspace{-0.5cm}
\begin{figure}[H]
    \centering
    \includegraphics[width = 1\textwidth, trim = {4cm 11cm 4cm 2.2cm}, clip]{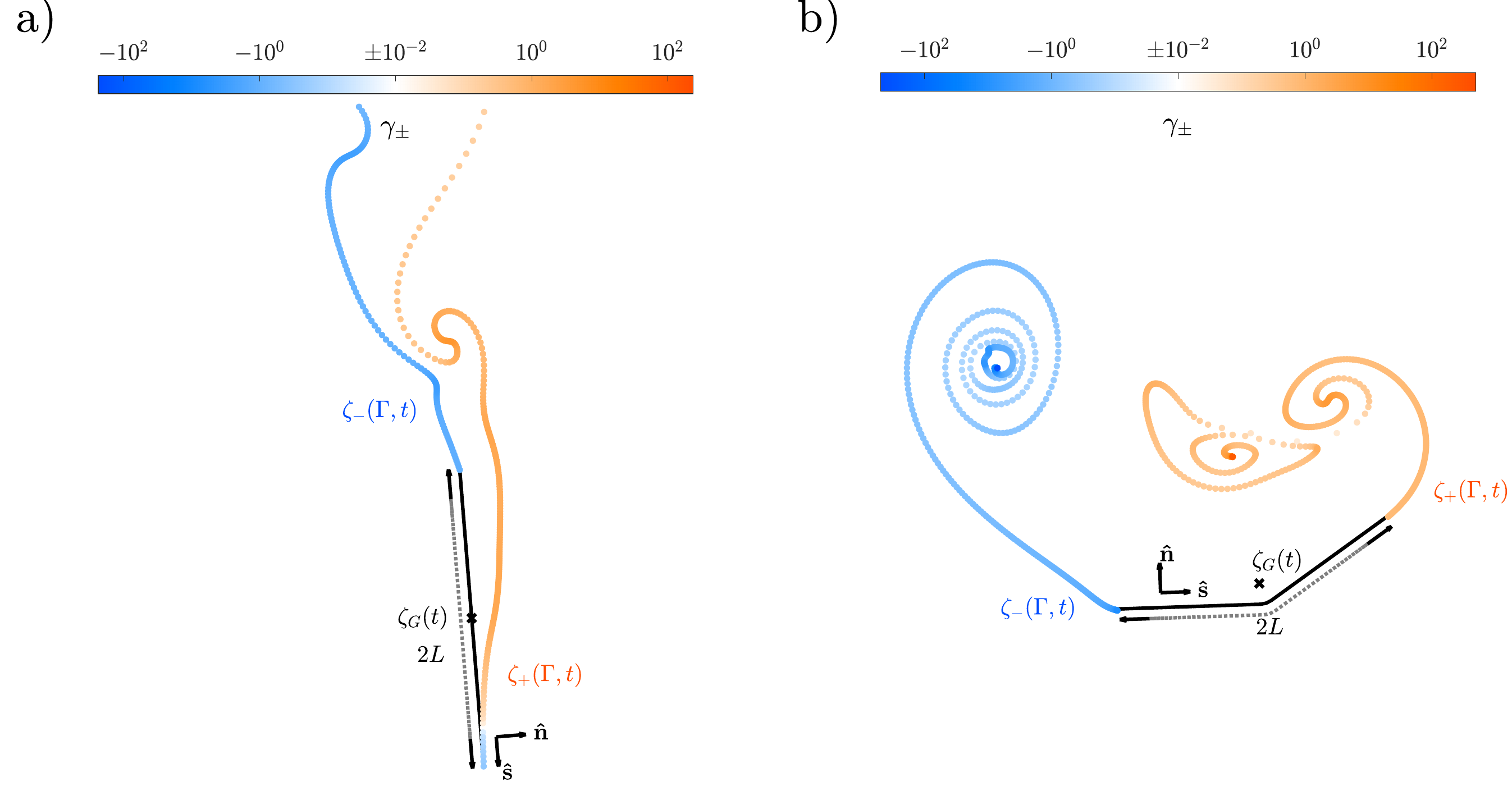}
    \caption{(a) A flat and (b) V-shaped plate falling under gravity, with center of mass $\zeta_G$ and length $2L$. It has unit tangent and normal vectors $\hat{\textbf{s}} $ and $\hat{\textbf{n}} $ respectively. Off the edges, vortex sheets are being shed with their strength $\gamma_\pm$ colored with a logarithmic scale. They are parameterized by $\zeta_\pm(\Gamma,t)$ where $\Gamma$ is the circulation.}
    \label{Falling V-Flyer Model}
\end{figure}
Many previous works have studied falling flat plates \cite{jones2005falling, belmonte1998flutter, sohn, pomerenk2024aerodynamicequilibriaflightstability}. Here we study this case as well as the more general class of rigid V-shaped plates, with an example shown in figure \ref{Falling V-Flyer Model}. We have also considered other plate shapes but focus on 
V-shaped plates because they yield a wider range of dynamics.
The plates are represented by a 1D curve in the 2D $x$-$y$ fluid plane. Both the body and flow extend uniformly with width $W$ in the $z$ (out-of plane) direction. 
The body has length $2L$ in the $x$-$y$ plane and thickness $h \ll L$. The thickness of the body is approximated as 0 relative to the characteristic length scale $L$. The $x$-$y$ flow velocity is written as the complex function $u(z) + iv(z)$, where $z = x + iy$ and $i = \sqrt{-1}$. The body position is written as a complex-valued curve 
\begin{align}
\zeta(s,t) = \zeta_G(t) + \zeta_0(s)e^{i\beta(t)} \label{Body}
\end{align}
with $\zeta_G(t) = x_G(t) + iy_G(t)$ the center of mass,  $\zeta_0(s)$ the mean-zero (smooth) shape of the body, fixed in time, and $\beta(t)$ the body's orientation angle, with initial value $\beta(0) = 0$. The arc length parameter $s \in [-L, L]$ along the body. The unit tangent vector along the body is $\hat{\textbf{s}} = \partial_s\zeta(s,t)$ and the unit normal is $\hat{\textbf{n}} = i\hat{\textbf{s}}$. We use $+$ to denote the the side of the body towards which $\hat{\textbf{n}}$ points and $-$ for the other side. We also use $+$ for the $s = L$ end of the body and $-$ for the $s = -L$ end.
\newline

Note that the logarithmic scales in figure  \ref{Falling V-Flyer Model} (and several subsequent figures) are actually two logarithmic scales joined together, to show both signs of vortex-sheet strength $\gamma$. Points with with small $|\gamma|$ ($\leq 10^{-2}$) are colored white. 

\subsection{Rigid-body equations}
\subsubsection{Force balance in an inviscid fluid}
A local force balance on a small portion of the body lying in the interval $[s- \frac{\Delta s}{2}, s + \frac{\Delta s}{2}]$ gives
\begin{equation}
    \rho_b h W \Delta s \partial_{tt}\zeta(s,t) = -W \Delta s [p]^+_-(s)\hat{\textbf{n}}(s)  - i \rho_b h W \Delta s g + f_{int}. \label{FBal1}
\end{equation}
The left side is the body inertia, with $\rho_b$ the body mass per unit volume. The first term on the right side is the pressure force, with $[p]^+_- = p^+ - p^-$ the jump in pressure across the body. The second term on the right side is the gravitational force, with $g$ the gravitational acceleration. The third term on the right side, $f_{int}$, is the sum of the internal forces exerted on this portion of the body by the neighboring portions, in order to maintain the rigid shape of the body.  
Summing over the entire body on both sides of the equation and taking the limit $\Delta s \to 0$, we obtain the equation for the center-of-mass motion:
\begin{equation}
    \rho_b h W 2L \partial_{tt}\zeta_G(t) = -W \int_{-L}^{L}[p]^+_-(s)\hat{\textbf{n}}(s) \, ds  - i \rho_b h W 2L g \label{DimensionalForceBalance}.
\end{equation}
The net contribution from the internal forces vanishes because they appear in equal and opposite pairs, by Newton's third law.

\subsubsection{Nondimensionalization}
To nondimensionalize the equations, we take the half-length of the body, $L$, as the characteristic length scale. We define the characteristic velocity $U$ as an approximation of the body's terminal speed, its time-averaged speed in the steady-state falling motion reached after the initial transient dynamics. At steady state the left side of 
\eqref{DimensionalForceBalance} has time-average zero and the pressure and gravity terms on the right side balance. The high-Reynolds-number scaling of $[p]^+_-$ is $\rho_f U^2$, e.g.~by considering the Bernoulli equation. Thus $\rho_f U^2 \sim \rho_b h g$ so we take $U = \sqrt{\frac{\rho_b h g}{\rho_f}}$ as the characteristic velocity scale.
Choosing $\rho_f U^2$ for the characteristic pressure scale, we can now nondimensionalize the variables as follows:
\begin{equation}
    t = \widetilde{t}\ \frac{L}{U},\ \ \  \zeta_G = \widetilde{\zeta_G}\ L,\ \ \  [p]^+_- = \widetilde{[p]^+_-}\ \rho_f U^2,\ \ \  s = \widetilde{s}\ L.
\end{equation}
Here, the variables with tildes, $\widetilde{t},\  \widetilde{\zeta},\  \widetilde{[p]^+_-},\  \widetilde{s}$, denote their dimensionless counterparts. After substituting these variables into \eqref{DimensionalForceBalance} and suppressing the tildes, we obtain the dimensionless force balance equation, essentially equivalent to the versions in \cite{albenFall} and \cite{sohn}:
\begin{equation}
    R_1\partial_{tt}\zeta_G(t) = -\frac{1}{2}\int_{-1}^{1}[p]^+_-(s)\hat{\textbf{n}}(s)\, ds - i.
    \label{ForceBalance}
\end{equation}
Here, $R_1 = \frac{\rho_b h }{\rho_f L}$ is the dimensionless mass density of the body, and measures the importance of body inertia relative to fluid inertia. This quantity is equal to $\text{Fr}$, the dimensionless Froude number as defined in \cite{sohn} \cite{jones2005falling}, and equal to $\text{Fr}^2$ as defined in \cite{belmonte1998flutter} and \cite{Rana2020}. 

\subsubsection{Torque balance in an inviscid fluid} 
The angular momentum about the origin of the small portion of the body lying in the interval $[s- \frac{\Delta s}{2}, s + \frac{\Delta s}{2}]$ is
\begin{align}
\mathbf{r}\times (m\mathbf{v}) = \zeta(s,t) \times \rho_b Wh\Delta s \partial_t \zeta(s,t) \label{AngMom}
\end{align}
where $\times$ denotes the cross product, assuming the operands are mapped from complex scalars to 3-vectors, e.g.~from $z = a + ib$ to $\mathbf{z} = (a,b,0)^T$. Then the cross product of two complex scalars, $\mathbf{z}_1\times \mathbf{z}_2$, is given by $-\Re(iz_1\overline{z_2})$. We set the time-rate-of-change of the angular momentum \eqref{AngMom} equal to the torque about the origin exerted by the forces in \eqref{FBal1}:
\begin{align}
 \zeta(s,t) \times \rho_b Wh\Delta s \partial_{tt} \zeta(s,t) =  \zeta(s,t) \times (-W \Delta s [p]^+_-(s)\hat{\textbf{n}}(s)  - i \rho_b h W \Delta s g + f_{int}) \label{TorqueBal1}
\end{align}
We insert $\zeta(s,t)$ from \eqref{Body} and sum \eqref{TorqueBal1} over the body, taking the limit $\Delta s \to 0$. The resulting integrals are simplified using \eqref{DimensionalForceBalance} and that the facts that $\zeta_0(s)$ has $s$-mean zero and the internal forces exert no net torque on the body. We obtain
\begin{equation}
    \rho_b Wh\int_{-L}^{L}|\zeta_0(s)|^2\, ds\ \partial_{tt}\beta(t) = W\int_{-L}^L\Re{\bigg(i\zeta_0(s)e^{i\beta(t)}\overline{\big([p]^+_-(s)\hat{\textbf{n}}(s)\big)}\bigg)}\, ds.
    \label{DimensionalTorqueBalance}
\end{equation}
Finally, by nondimensionalizing again as above we obtain 

\begin{equation}
    I_G\partial_{tt}\beta(t) = \int_{-1}^1\Re{\bigg(i\zeta_0(s)e^{i\beta(t)}\overline{\big([p]^+_-(s)\hat{\textbf{n}}(s)\big)}\bigg)}\, ds.
    \label{TorqueBalance}
\end{equation}
Here, $I_G = R_1\int_{-1}^{1}|\zeta(s,t) - \zeta_G(s)|^2\, ds =  R_1\int_{-1}^{1}|\zeta_0(s)|^2\, ds$ is the (dimensionless) moment of inertia of the body about its center of mass. Henceforth, all equations are dimensionless.




 
\subsection{Fluid equations}
In the presence of small viscosity, the fluid velocity matches the body velocity on its surface, but rapidly transitions to other values as one moves away from the body through thin boundary layers along the body surface. In the limit of zero viscosity, a boundary layer shrinks to a zero-thickness vortex sheet along the body surface, and the velocity tangent to the body surface jumps discontinuously across the vortex sheet. In the limit of zero body thickness, the vortex sheets on the two sides of the body coincide and are a single vortex sheet called the ``bound" vortex sheet \cite{eldredge_inviscid_flows}.
\newline

With small viscosity, vorticity diffuses slightly from the boundary layers along the body but mainly enters the fluid at ``separation points" where the boundary layer is advected into the flow, particularly at the sharp edges of the body. In our zero-viscosity model the situation is similar---there is no diffusion of vorticity from the vortex sheets, only advection from the separation points, which we assume to be located only at the two edges of the body.  Consequently, two ``free" vortex sheets emanate from the body, one from each edge. The bound vortex sheet along the body and the free vortex sheets form one continuous vortex sheet (discretized in figure  \ref{Falling V-Flyer Model}), which we compute at each time step.
\newline

The body and flow are initially at rest with zero vorticity. As  $t$ increases from 0, gravity causes the body to accelerate with a downward component. As the body moves there is a corresponding fluid motion that can be represented in terms of the strengths and positions of the bound and free vortex sheets. As time increases, the lengths of the free vortex sheets increase at each edge, starting from zero. 
The bound vortex sheet position is the same as the body, $\zeta(s,t), -1 \leq s \leq 1$. The free vortex sheets emanating from the $s = \pm 1$ edges are denoted $\zeta_{+}(s,t),\ s \in [1,s_{+} + 1]$ and $\zeta_{-}(s,t),\ s \in [-s_{-} -1,-1]$. Here, $s_{\pm}$ are the lengths of the positive and negative vortex sheets respectively, and the vortex sheet strengths $\gamma_{\pm}(s,t)$ are defined analogously.
\newline

The strength of the entire (bound and free) vortex sheet is denoted $\gamma(s,t)$, with $s \in[-1 -s_-, 1 + s_+]$. We denote the restriction of $\gamma(s,t)$ to the negative, positive, and bound vortex sheets as $\gamma_-(s,t)$, $\gamma_+(s,t)$, and $\gamma_b(s,t)$ respectively.  
\newline

Integrating the vortex strengths $\gamma_b$, and $\gamma_\pm$ against the Biot-Savart kernel, we obtain an equation for the conjugate fluid velocity  $w(z) = u(z) - iv(z)$,
 \begin{equation}
    u(z) - i v(z) = \frac{1}{2\pi i}\int_{-1}^{1}{\frac{\gamma_b(s,t)}{z - \zeta(s,t)}\, ds}  +\frac{1}{2\pi i}\int_{-s_{-} -1}^{-1}{\frac{\gamma_{-}(s,t)}{z - \zeta_{-}(s,t)}\, ds} + \frac{1}{2\pi i}\int_{1}^{s_{+} + 1}{\frac{\gamma_{+}(s,t)}{z - \zeta_{+}(s,t)}\, ds}.
\end{equation}

Let $\Gamma(s,t)$ denote the arc length integral of $\gamma$ along the positive and negative vortex sheets, defined piecewise:
\begin{align}
\Gamma(s,t) = \begin{cases}
    \int_{-s_--1}^s \gamma(s',t)\,ds', \quad s \in [-s_--1, -1] \\
   \int_{1}^{s}\gamma_+(s',t)\,ds', \quad s \in [1, 1+s_+]
\end{cases} 
\end{align}

and $\Gamma_\pm(t)$ denote the total circulations in the positive and negative vortex sheets, i.e.
\begin{align}
\Gamma_+(t) \equiv \int_{1}^{1+s_{+}}\gamma_{+}(s,t)\, ds \quad ; \quad \Gamma_-(t) \equiv \int_{-s_{-}-1}^{-1}\gamma_{-}(s,t)\, ds.
\end{align}

Then, by reparameterizing the positions of the free vortex sheets in terms of $\Gamma$, we may rewrite the equation for the conjugate velocity as
\begin{equation}
     u(z) - i v(z) = \frac{1}{2\pi i}\int_{-1}^{1}{\frac{\gamma_b(s,t)}{z - \zeta(s,t)}\, ds} + \frac{1}{2\pi i}\int_{0}^{\Gamma_-(t)}{\frac{d\Gamma}{z - \zeta_{-}(\Gamma,t)}} + \frac{1}{2\pi i}\int_{0}^{\Gamma_+(t)}{\frac{d\Gamma}{z - \zeta_{+}(\Gamma,t)}}. \label{ConjFluidVel}
\end{equation}
Note that $\Gamma_+(t)$ is the negative of the quantity with this name in \cite{Jones2003TheSF,albenFlex} and
$\Gamma(s,t)$ differs from the quantity with the same name in those works by an additive constant ($\Gamma_+(t)$) on the positive vortex sheet.

\subsubsection{Birkhoff-Rott equation}
By demanding that the pressure be a continuous field across the free vortex sheets, the free vortex sheets must evolve with velocities equal to the average of the fluid velocities across their surfaces \cite[84]{eldredge_inviscid_flows}. A consequence of the Sokhotski–Plemelj formulae is that the principal value of a singular integral is the average of its limiting values on either side. A brief discussion of this formula is included in appendix \ref{log singularity appendix}. Hence, by evaluating \eqref{ConjFluidVel} on the free sheets in the principle-value sense, we obtain the following Birkhoff-Rott equations for the (conjugate) fluid velocity of the free sheets.
\begin{multline}
     \partial_{t}\bar\zeta_{\pm}(\Gamma,t) = \frac{1}{2\pi i}\int_{-1}^{1}{\frac{\gamma_b(s,t)}{\zeta_{\pm}(\Gamma,t) - \zeta(s,t)}\, ds} + \frac{1}{2\pi i}\int_{0}^{\Gamma_-(t)}{\frac{d\Gamma'}{\zeta_{\pm}(\Gamma,t) - \zeta_{-}(\Gamma',t)}} \\ + 
     \frac{1}{2\pi i}\int_{0}^{\Gamma_+(t)}{\frac{d\Gamma'}{\zeta_{\pm}(\Gamma,t) - \zeta_{+}(\Gamma',t)}}. \label{Birkhoff-Rott}
\end{multline}
For the computational method, we will modify the Birkhoff-Rott equations as described in section \ref{Velocity Smoothing}.

\subsubsection{Blob regularization} \label{Blob Regularization}
The Birkhoff-Rott equations for the evolution of the free vortex sheets are ill-posed, and develop a curvature singularity at a finite critical time \cite{moore1979spontaneous, Krasny_1986, Shelley_1992}. Krasny showed that the Birkhoff-Rott equations can be evolved numerically past the critical time by introducing a smoothing parameter $\delta$ to ``desingularize" the equations \cite{krasny1986desingularization}:

\begin{multline}
     u(z) - iv(z) = \frac{1}{2\pi i}\int_{-1}^{1}{\frac{\gamma_b(s,t)}{z- \zeta(s,t)}\, ds} + \frac{1}{2\pi i}\int_{0}^{\Gamma_-(t)}{\frac{\overline{z - \zeta_{-}(\Gamma',t)}}{|z - \zeta_{-}(\Gamma',t)|^2 + \delta^2}d\Gamma'} \\ +
     \frac{1}{2\pi i}\int_{0}^{\Gamma_+(t)}{\frac{\overline{z - \zeta_{+}(\Gamma',t)}}{|z - \zeta_{+}(\Gamma',t)|^2 + \delta^2}d\Gamma'}. \label{ConjVelSmoothed}
\end{multline}
The presence of $\delta$ smooths the kernel into a vortex blob, with core size determined by $\delta$ \cite{krasny1990computing}. Here, and in all that follows, we take $\delta = 0.2$ as is done in \cite{nitsche1994numerical}, which shows good agreement with experiment. This choice was also made in several other studies \cite{sohn,alben2013efficient,kansoHov,albenFall}. As discussed in \cite{krasny1986desingularization, lingXuStartingVortex}, decreasing $\delta$ increases the number of spiral turns in the vortex cores; A similar effect occurs when increasing the Reynolds number in viscous simulations. Indeed, $\delta$ can be thought of as a form of numerical viscosity that stabilizes the otherwise ill-posed evolution \cite{Krasny_1986} of vortex layers in the limit of infinite Re. In some studies, $\delta$ is associated with a specific physical viscosity \cite{lingXuStartingVortex}, and hence a specific Reynolds number, but we do not take this point of view in this study. In appendix \ref{refinement in delta section} we show examples of the effect of varying $\delta$ on the dynamics of the flat plate. When $\delta$ ranges from 0.1 to 0.3, the effect of $\delta$ on the qualitative features of the steady state dynamics is often small, but may be large, particularly at $R_1$ near transitions in dynamical states, where the simulations are more sensitive to the parameters.

\subsubsection{No-penetration condition \& Kelvin's circulation theorem}
Let the body velocity $\partial_t \zeta(s,t)$ have tangential and normal components $\tau$ and $\nu$ respectively: 
\begin{eqnarray}
    \partial_t\zeta(s,t) = \tau(s,t)\hat{\textbf{s}}(s,t) + \nu(s,t)\hat{\textbf{n}}(s,t).
\end{eqnarray} 
Taking the limit that $z$ approaches the body in \eqref{ConjVelSmoothed},
analogous to \eqref{Birkhoff-Rott} we have
\begin{multline}
    \partial_t \bar \zeta(s,t) = \mu(s,t)\overline{\hat{\textbf{s}}(s,t)} + \xi(s,t)\overline{\hat{\textbf{n}}(s,t)} = \frac{1}{2\pi i}\dashint_{-1}^{1}{\frac{\gamma_b(s',t)}{\zeta(s,t) - \zeta(s',t)}\, ds'} \\ +  \frac{1}{2\pi i}\int_{0}^{\Gamma_-(t)}{\frac{\overline{\zeta(s,t) - \zeta_{-}(\Gamma,t)}}{|\zeta(s,t) - \zeta_{-}(\Gamma,t)|^2 + \delta^2}d\Gamma}  + \frac{1}{2\pi i}\int_{0}^{\Gamma_+(t)}{\frac{\overline{\zeta(s,t) - \zeta_{+}(\Gamma,t)}}{|\zeta(s,t) - \zeta_{+}(\Gamma,t)|^2 + \delta^2}d\Gamma}.\label{muksi}
\end{multline}
Here, $\mu$ and $\xi$ are the averages of the tangential and normal velocities on the two sides of the body. The no-penetration condition states that the fluid and body velocities have the same components normal to the body at the the body surface:
\begin{equation}
    \nu(s,t) = \xi(s,t).
\end{equation}
That is, 
\begin{multline}
    \Re\Big(\hat{\textbf{n}}(s,t)\partial_t\bar{\zeta}(s,t) \Big) = 
        \Re\Bigg(\hat{\textbf{n}}(s,t)\bigg(\frac{1}{2\pi i}\dashint_{-1}^{1}{\frac{\gamma_b(s',t)}{\zeta(s,t) - \zeta(s',t)}\, ds'} + \frac{1}{2\pi i}\int_{0}^{\Gamma_-(t)}{\frac{\overline{\zeta(s,t) - \zeta_{-}(\Gamma,t)}}{|\zeta(s,t) - \zeta_{-}(\Gamma,t)|^2 + \delta^2}d\Gamma} \\ + \frac{1}{2\pi i}\int_{0}^{\Gamma_+(t)}{\frac{\overline{\zeta(s,t) - \zeta_{+}(\Gamma,t)}}{|\zeta(s,t) - \zeta_{+}(\Gamma,t)|^2 + \delta^2}d\Gamma}\bigg)\Bigg),
    \label{NoPenetration}
\end{multline}
 (Note that in complex notation, the dot product of two vectors $\textbf{z}_1\cdot\textbf{z}_2$ is given by $\Re(\overline{z_1}z_2)$.) Given known positive and negative circulations, the no-penetration condition \eqref{NoPenetration} uniquely determines $\gamma_b$ provided we add the additional constraint that the total circulation in the fluid vanishes \cite{Golberg1990IntroductionTT} \cite[285]{eldredge_inviscid_flows}. Indeed, since the body is released from rest, the initial circulation present in fluid is zero. Combined with Kelvin's circulation theorem, which asserts that the total circulation is independent of time \cite[36]{eldredge_inviscid_flows}, we must have
\begin{equation} \label{Kelvin's Circulation Theorem}
    \Gamma_+(t) + \Gamma_-(t) + \Gamma_b(t) = 0.
\end{equation}
Here, $\Gamma_b(t) = \int_{-1}^1\gamma_b(s,t)\, ds$ is the total circulation on the body. Alongside the no-penetration condition \eqref{NoPenetration}, this is an additional constraint that must be satisfied by the bound vortex strength $\gamma_b$.

\subsection{Pressure-jump equation} \label{Pressure Jump Section}
By considering the Bernoulli equation in the reference frame of the moving body, and taking its difference when evaluated on both sides of the body, the following pressure-jump equation can be obtained \cite[85]{eldredge_inviscid_flows} \cite{point_vortex_attraction}.
\begin{equation}
    [p]^+_-(s,t) = \partial_t\big(\Gamma_-(t) + \Gamma_b(s,t)\big) + (\mu(s,t) - \tau(s,t))\gamma_b(s,t) \label{Bernoulli}
\end{equation}

Here, $\Gamma_b(s,t) = \int_{-1}^s\gamma_b(s,t)\, ds$, and $\mu(s,t)$ is the $\hat{\textbf{s}}$ component of (\ref{muksi}).

The equation for the pressure jump \eqref{Bernoulli} is identical to that in \cite{kansoHov} and \cite{sohn}. However, we note that there is another, equally natural choice for the pressure-jump equation:
\begin{equation}
    \begin{cases}
        \partial_s [p]^+_-(s,t) = \partial_t\gamma_b(s,t) + \partial_s\big((\mu(s,t) - \tau(s,t))\gamma_b(s,t)\big), \label{1stOrderBernoulli} \\
        [p]^+_-(-1,t) = 0
    \end{cases}
\end{equation}
This equation was used with success in \cite{albenFall} and \cite{Mavroyiakoumou_Alben_2020}, and can be obtained for instance, by differentiating \eqref{Bernoulli} with respect to arc length. The edge condition follows from the Birkhoff-Rott equations which demand that the pressure jump across the free sheet vanishes \cite{eldredge_inviscid_flows}. Hence, by the continuity of the pressure jump from the body to free sheets, $[p]^+_-(\pm1,t) = 0$. The two equations \eqref{Bernoulli} and \eqref{1stOrderBernoulli} are therefore equivalent provided that the constraint
\begin{equation}
    \partial_t\Gamma_-(t)+ (\mu(-1,t) - \tau(-1,t))\gamma_b(-1,t) = 0
\end{equation}
is satisfied. However, the situation is complicated by the fact that in a numerical implementation, the quantities  $\partial_t\Gamma_-(t) + (\mu(\pm 1,t) - \tau(\pm 1,t))\gamma_b(\pm 1,t)$ may cease to be zero, leading to different approximate formulations of the pressure equation. Indeed, although both pressure equations are equivalent in a mathematical sense, they become distinct with the blob regularization of the singular integrals in \eqref{NoPenetration}, as discussed next. 

\subsection{Kutta condition} \label{Kutta Condition Section}
The system of equations \eqref{ForceBalance}, \eqref{TorqueBalance}, 
\eqref{Birkhoff-Rott}, and \eqref{NoPenetration}--\eqref{Bernoulli} contains two functions that are as yet undetermined: $\Gamma_-(t)$ and $\Gamma_+(t)$, the cumulative circulations released at each edge. They are determined at each time $t$ by the Kutta condition,  the physical constraint that the flow velocity be bounded at the sharp edges of the body. In the seminal work \cite{Jones2003TheSF}, Jones shows that the bounded velocity constraint has multiple formulations that are mathematically equivalent:

\begin{equation}
    \begin{cases}
    &\partial_t\Gamma_-(t) + \big(\mu(- 1,t) - \tau(-1,t)\big)\gamma_b(-1,t) = 0, \\
    &\partial_t\Gamma_+(t) - \big(\mu(  1,t) - \tau(1,t)\big)\gamma_b(1,t) = 0
    \end{cases} \label{Pressure Vanishes Kutta Condition}
\end{equation}

\begin{equation}
    \begin{cases}
    &\gamma_b(-1,t) = \gamma_-(-1,t), \\
    &\gamma_b(1,t) = \gamma_+(1,t)
    \end{cases} \label{Continuous gamma Kutta Condition}
\end{equation}

\begin{equation}
    \begin{cases}
    &\sigma(-1,t) = 0, \\
    &\sigma(1,t) = 0,
    \end{cases}
    \ \ \ \gamma_b(s,t) = \frac{\sigma(s,t)}{\sqrt{1 - s^2}} \label{Suction Vanishes Kutta Condition}
\end{equation}

Here, $\sigma(\pm1,t)$ is the ``suction parameter" at the $\pm$ edge \cite[178]{eldredge_inviscid_flows}. It can be shown that the generic solution to \eqref{NoPenetration} $\gamma_b$ has inverse square root singularities and is thus of the form $\gamma_b(s,t) = \frac{\sigma(s,t)}{\sqrt{1 - s^2}}$ for some continuous $\sigma(s,t)$ \cite{Muskhelishvili}. \eqref{Suction Vanishes Kutta Condition} is therefore equivalent to removing the singularities and leaving $\gamma_b$  finite at the edges. In \cite{Jones2003TheSF}, it is shown using explicit formulae for $\gamma_b$ that enforcing the constraints \eqref{Pressure Vanishes Kutta Condition} is equivalent to removing these singularities. Hence both Kutta conditions \eqref{Pressure Vanishes Kutta Condition} and \eqref{Suction Vanishes Kutta Condition} are equivalent. Furthermore, as shown in section \ref{Pressure Jump Section}, \eqref{Pressure Vanishes Kutta Condition} is really the assertion that the pressure jump is continuous from the body to the free vortex sheet. \eqref{Continuous gamma Kutta Condition} asserts that the vortex sheet strength is continuous from the body to the free vortex sheet. Since a version of \eqref{Bernoulli} holds for the entire vortex sheet (both free and bound parts), in general the pressure jump $[p]^+_-$ is continuous if and only if the vortex strength $\gamma$ is continuous. Hence, mathematically speaking, equations \eqref{Pressure Vanishes Kutta Condition}--\eqref{Suction Vanishes Kutta Condition} are all equivalent, and can be viewed as ``the" Kutta condition.
\newline

However, with a nonzero blob regularization $\delta$ of the free sheets but not the bound sheet in \eqref{NoPenetration}, \eqref{Pressure Vanishes Kutta Condition}--\eqref{Suction Vanishes Kutta Condition} become three distinct conditions.
In the standard vortex shedding method used in previous studies, including those from the present authors, exactly one point is released onto the free sheets at each time step. The value of the circulation at this newly released point is chosen so that the computed solution satisfies one of \eqref{Pressure Vanishes Kutta Condition}--\eqref{Suction Vanishes Kutta Condition}. An alternative method that we have implemented is satisfying more than one of \eqref{Pressure Vanishes Kutta Condition}--\eqref{Suction Vanishes Kutta Condition} simultaneously at each time step, by releasing the corresponding number of vortex sheet points from each edge. Such a method is only possible because these equations are nonequivalent. However, the careful investigation of such methods is beyond the scope of this paper. 
\newline

Based on experimentation, we have selected \eqref{Suction Vanishes Kutta Condition} as the Kutta condition for the present study because it seems to be the most numerically stable.
\newline

There is another Kutta condition, distinct from \eqref{Pressure Vanishes Kutta Condition}--\eqref{Suction Vanishes Kutta Condition}, that has also been used (e.g.~\cite{Jones2003TheSF,sohn}):
\begin{equation}
    \begin{cases}
    &\gamma_b(-1,t) = 0 , \\
    &\gamma_b(1,t) = 0.
    \end{cases} \label{gamma Vanishes Kutta Condition}
\end{equation}
Unlike \eqref{Pressure Vanishes Kutta Condition}--\eqref{Suction Vanishes Kutta Condition}, this Kutta condition applies the principle of bounded velocity not to the original, singular equations, but to the blob-regularized equations. This Kutta condition is strictly stronger than \eqref{Suction Vanishes Kutta Condition}, since if the bound circulation $\gamma_b$ is to attain a finite value at the edges of the body, the suction must vanish there as well. It suppresses not only the intrinsic inverse square root singularities that arise in the general solution of $\gamma_b$, but also logarithmic singularities that arise by blob regularizing only the free, not bound, sheets. To explain the logarithmic singularities, let us fix $t$ for the moment and suppose for simplicity that our plate is flat and parameterized by $\zeta(s,t) = s$, $-1 < s < 1$. Then, as noted in \cite{Jones2003TheSF}, and \cite[42]{Muskhelishvili}, we may write the plate's contribution to the (conjugate) fluid velocity as 
\begin{align*}
    \frac{1}{2\pi i}\int_{-1}^{1} \frac{\gamma_b(s,t)}{z - \zeta(s,t)}\, ds &= \frac{1}{2\pi i}\int_{-1}^{1} \frac{\gamma_b(s,t) - \gamma_b(\pm 1,t)}{z -s}\, ds + \frac{1}{2\pi i}\gamma_b(\pm 1,t)\int_{-1}^1\frac{ds}{z-s}\\
    &= \frac{\gamma_b(\pm 1,t)}{2\pi i }\log(\frac{z + 1}{z - 1}) + \Phi_\pm(z)\\
\end{align*}
Here, $\Phi_\pm(z) =  \frac{1}{2\pi i}\int_{-1}^{1} \frac{\gamma_b(s,t) - \gamma_b(\pm 1,t)}{z -s}\, ds$ is a bounded function near the $\pm$ edge. This is because the numerator of the integrand vanishes at the $\pm$ edge, canceling the singularity there. A justification of this subtle point in a more general case is included in appendix \ref{log singularity appendix}. Thus when \eqref{ConjVelSmoothed} is used to compute the fluid velocity with one of \eqref{Pressure Vanishes Kutta Condition}--\eqref{Suction Vanishes Kutta Condition}, logarithmic singularities occur at the edges of the body where $\gamma_b$ is nonzero. This is because although the singular integrals in \eqref{ConjFluidVel} 

\begin{align}\frac{1}{2\pi i}\int_{-s_{-} -1}^{-1}{\frac{\gamma_{-}(s,t)}{z - \zeta_{-}(s,t)}\, ds} &= \frac{1}{2\pi i}\int_{0}^{\Gamma_-(t)}{\frac{d\Gamma'}{z - \zeta_{-}(\Gamma',t)}}, \nonumber \\
\frac{1}{2\pi i}\int_{1}^{1 + s_+}{\frac{\gamma_{+}(s,t)}{z - \zeta_{+}(s,t)}\, ds} &= \frac{1}{2\pi i}\int_{0}^{\Gamma_+(t)}{\frac{d\Gamma'}{z - \zeta_{+}(\Gamma',t)}}
\end{align}
likewise have logarithmic singularities at the edges of the body that cancel those of the plate, their blob regularized counterparts in \eqref{ConjVelSmoothed},
\begin{equation} \label{smoothed sheet contribution}
\frac{1}{2\pi i}\int_{0}^{\Gamma_-(t)}{\frac{\overline{z - \zeta_{-}(\Gamma,t)}}{|z - \zeta_{-}(\Gamma,t)|^2 + \delta^2}d\Gamma}, \ \
    \frac{1}{2\pi i}\int_{0}^{\Gamma_+(t)}{\frac{\overline{z - \zeta_{+}(\Gamma,t)}}{|z - \zeta_{+}(\Gamma,t)|^2 + \delta^2}d\Gamma}
\end{equation}

do not. Indeed, as their kernels are smooth, the resulting contribution of \eqref{smoothed sheet contribution} is smooth as well. Hence, unless the Kutta condition \eqref{gamma Vanishes Kutta Condition} is used, the total fluid velocity has a logarithmic singularity at the plate edges.
\newline

In general, however, when computing the singular integral with a quadrature rule with weights $w_j$, $$\frac{1}{2\pi i}\int_{-1}^{1}{\frac{\gamma_b(s,t)}{z - s} \, ds \approx \sum_j w_j \frac{\gamma_j}{z - s_j}}$$ one introduces first-order (simple pole) singularities as $z$ approaches the grid points $s_j$ of the body. These first-order singularities are much stronger than the logarithmic singularities at the edges of the body. Hence these logarithmic singularities can be, and are often ignored. In section \ref{LogDesingularizationOfBody} we will propose a quadrature method that does not introduce any first-order singularities. In section \ref{Velocity Smoothing} we will describe how we also remove the logarithmic singularities. These two methods help stabilize the flow near the edges and allow for more robust and accurate computation of leading-edge shedding in the simulation. 

\section{Numerical method}
\label{numerical method section}
\begin{longtable}{|>{\centering\arraybackslash}m{3cm}|>{\centering\arraybackslash}m{8cm}|>{\centering\arraybackslash}m{4cm}|}
\hline
\textbf{Name of equation} & \textbf{Equation} & \textbf{Discretization methods} \\
\hline
Force balance equation &
\begin{equation} \label{force balance table}
    R_1\partial_{tt}\zeta_G(t) = -\frac{1}{2}\int_{-1}^{1}[p]^+_-(s)\hat{\textbf{n}}(s)\, ds - i
\end{equation}
& Second-order backwards differentiation, trapezoidal rule \\
\hline
Torque-balance equation &
\begin{equation} \label{torque balance table}
    I_G\partial_{tt}\beta(t) = \int_{-1}^1\Re{\bigg(i\zeta_0(s)e^{i\beta(t)}\overline{\big([p]^+_-(s)\hat{\textbf{n}}(s)\big)}\bigg)}\, ds
\end{equation}& 
Second-order backwards differentiation, trapezoidal rule
\\
\hline

No-penetration condition \& Kelvin's circulation theorem &
\begin{multline} \label{no penetration table}
    \Re\Big(\hat{\textbf{n}}(s,t)\partial_t\bar\zeta(s,t) \Big) = \\ 
        \Re\Bigg(\hat{\textbf{n}}(s,t)\bigg(\frac{1}{2\pi i}\dashint_{-1}^{1}{\frac{\gamma_b(s',t)}{\zeta(s,t) - \zeta(s',t)}\, ds'}\\
        + \frac{1}{2\pi i}\int_{0}^{\Gamma_-(t)}{\frac{\overline{\zeta(s,t) - \zeta_{-}(\Gamma,t)}}{|\zeta(s,t) - \zeta_{-}(\Gamma,t)|^2 + \delta^2}d\Gamma} \\
        + \frac{1}{2\pi i}\int_{0}^{\Gamma_+(t)}{\frac{\overline{\zeta(s,t) - \zeta_{+}(\Gamma,t)}}{|\zeta(s,t) - \zeta_{+}(\Gamma,t)|^2 + \delta^2}d\Gamma}\bigg)\Bigg),
\end{multline}

\begin{equation} \label{kelvin table}
    \Gamma_+(t) + \Gamma_-(t) + \Gamma_b(t) = 0
\end{equation} 
&
See section \ref{solving no penetration}
\\
\hline
Suction Kutta condition &
\begin{equation} \label{kutta condition table}
    \sigma(1,t) = \sigma(-1,t) = 0
\end{equation}
&
See section \ref{Suction Section}
\\
\hline
Pressure-jump equation & 
\begin{multline} \label{pressure jump equation table}
    [p]^+_-(s,t) = \partial_t\big(\Gamma_-(t) + \Gamma_b(s,t)\big) \\ + (\mu(s,t) - \tau(s,t))\gamma_b(s,t),
\end{multline}
\begin{multline} \label{muEqn}
    \mu(s,t) = \Re\Bigg(\hat{\textbf{s}}(s,t)\bigg(\frac{1}{2\pi i}\dashint_{-1}^{1}{\frac{\gamma_b(s',t)}{\zeta(s,t) - \zeta(s',t)}\, ds'} \\ + \frac{1}{2\pi i}\int_{0}^{\Gamma_-(t)}{\frac{\overline{\zeta(s,t) - \zeta_{-}(\Gamma,t)}}{|\zeta(s,t) - \zeta_{-}(\Gamma,t)|^2 + \delta^2}d\Gamma} \\ + \frac{1}{2\pi i}\int_{0}^{\Gamma_+(t)}{\frac{\overline{\zeta(s,t) - \zeta_{+}(\Gamma,t)}}{|\zeta(s,t) - \zeta_{+}(\Gamma,t)|^2 + \delta^2}d\Gamma}\bigg)\Bigg)
\end{multline}

&

See section \ref{numerical birkhoff rott}
\\
\hline
Birkhoff-Rott equations &
\begin{multline} \label{birkhoff rott table}
     \partial_{t}\bar\zeta_{\pm}(\Gamma,t)= \frac{1}{2\pi i}\int_{-1}^{1} \frac{\gamma_b(s,t)}{\zeta_\pm(\Gamma,t) - \zeta(s,t)}\, ds\\
    + \frac{1}{2\pi i}\int_{0}^{\Gamma_-(t)}{\frac{\overline{\zeta_{\pm}(\Gamma,t) - \zeta_{-}(\Gamma',t)}}{|\zeta_{\pm}(\Gamma,t) - \zeta_{-}(\Gamma',t)|^2 + \delta^2}d\Gamma'}\\  +
     \frac{1}{2\pi i}\int_{0}^{\Gamma_+(t)}{\frac{\overline{\zeta_{\pm}(\Gamma,t) - \zeta_{+}(\Gamma',t)}}{|\zeta_{\pm}(\Gamma,t) - \zeta_{+}(\Gamma',t)|^2 + \delta^2}d\Gamma'}
\end{multline}
& 
See section \ref{numerical birkhoff rott}\\
\hline
\end{longtable}
\vspace{-0.5cm}
\hspace{0.02cm}
\begin{table}[H]
    \centering
    \begin{tabular}{|c|c|}
    \hline
    \textbf{Name of Variable} & \textbf{Unknown Variable}\\
    \hline
    Center-of-mass position & $\zeta_G(t)$ \\
    Orientation angle & $\beta(t)$ \\
    Bound vortex sheet strength & $\gamma(s,t)$ \\
    Circulation of $\pm$ sheet & $\Gamma_\pm(t)$ \\
    Pressure jump & $[p]^+_-(s,t)$ \\ 
    Position of the $\pm$ sheet & $\zeta_\pm(\Gamma,t)$ \\
    
    \hline
    \end{tabular}
    \label{unknown variables}
\end{table}

These 6 equations form a closed system that allows us to determine the 6 unknown functions which completely determine the state of the system. The main innovations in the algorithm presented lie in the treatment of the Birkhoff-Rott equations, and to a smaller extent, the pressure equation. A complete discussion of the modifications made to these equations as well as their discretizations can be found in section \ref{numerical birkhoff rott}. A detailed outline of the steps of the algorithm can be found in section \ref{outline of vortex shedding algorithm}. The algorithm can be loosely described as follows. It is an implicit-explicit method that treats the body variables (body position, bound vortex sheet strength, etc.) implicitly, and the free vortex sheets explicitly. The body is initialized at rest with zero pressure jump and vortex sheet strength. At each time step we perform the following procedure. The free sheets are evolved explicitly forward by one time step $dt$ using the Birkhoff-Rott equations \eqref{birkhoff rott table}. The body is then evolved implicitly forward by $dt$ using the force- \eqref{force balance table} and torque-balance \eqref{torque balance table} equations. Simultaneously, a new mesh point on the $\pm$ free sheet is placed on the corresponding $\pm$ edge that connects to it. This mesh point has circulation determined by the suction Kutta condition \eqref{kutta condition table}. In solving for the forces that dictate the body's motion, the pressure-jump equation \eqref{pressure jump equation table} is solved for the pressure, and the no-penetration condition \eqref{no penetration table} is solved for the bound vortex sheet strength. All these quantities are solved for implicitly along with the body's new position using Broyden's method. The procedure is then repeated. 
\newline

\subsection{No-penetration condition \& Kelvin's circulation theorem} \label{solving no penetration}
We handle the time derivatives with second-order backward differentiation. The body's arc-length parameter $s$ is discretized with two interlaced grids, the Chebyshev nodes of the first and second kinds respectively:
\begin{align*}
   s^{1\text{st}}_j = -\cos(\pi  (2j +  1) /(2n)), \text{ } j = 0,\ldots,n -1; \quad   s^{2\text{nd}}_j = -\cos(\pi  j /n), \text{ } j = 0,\ldots,n.
\end{align*}
We seek to compute the bound vortex sheet strength $\gamma_b(s,t)$ at the 
$\{s^{2\text{nd}}_j\}$ by enforcing the no-penetration condition at the 
$\{s^{1\text{st}}_j\}$, similar to the collocation method for the airfoil equation in
\cite{Golberg1990IntroductionTT}.  
Hence, we discretize the singular integral in \eqref{no penetration table} as 
\begin{equation}\label{gamma equation 1-n}
    \frac{1}{2\pi i}\int_{-1}^{1}{\frac{\gamma_b(s',t)}{\zeta^{1\text{st}}_{j} - \zeta(s',t)}\, ds'}  \approx \sum_{k = 0}^{n} w_k \frac{\gamma_j^{2 \text{nd}}}{\zeta^{1\text{st}}_{j} - \zeta_k^{2\text{nd}}}
\end{equation}
and discretize Kelvin's circulation theorem \eqref{kelvin table} using
\begin{equation}\label{gamma equation n}
    \Gamma_b(t) = \int_{-1}^1\gamma_b(s,t)\, ds \approx \sum_{k = 0}^{n} w_k \gamma^{2\text{nd}}_j.
\end{equation}
Here, $w_k$ are the weights of the trapezoidal rule on the grid points of the second kind. The remaining integrals in \eqref{no penetration table} need special treatment. Their discretization is discussed in section \ref{LogDesingularizationOfBody} alongside the similar terms in the Birkhoff-Rott equation \eqref{birkhoff rott table}. By substituting \eqref{gamma equation 1-n} into \eqref{no penetration table}, and \eqref{gamma equation n} into \eqref{kelvin table} we obtain a linear system of $n + 1$ equations in the $n + 1$ unknowns $\gamma_j^{2\text{nd}}$. Solving these yields the $n + 1$ values $\gamma_j^{2\text{nd}}$.

\subsection{Kutta condition} \label{Suction Section}
To discretize \eqref{kutta condition table} we need to obtain a discretization of the suction function $\sigma(s,t)$, which we denote $\sigma_j^\text{2nd}$. The discretized Kutta condition can then be written as
\begin{align}\label{kutta equation discretized}
    \sigma_0^\text{2nd} &= 0, \nonumber\\
    \sigma_n^\text{2nd} &= 0
\end{align}

If we were to apply the formula $\sigma(s,t) = \sqrt{1 - s^2}\gamma(s,t)$ to discretize $\sigma$, then we would have $\sigma_j^\text{2nd} = \sqrt{1 - (s^{\text{2nd}}_j)^2}\ \gamma^{\text{2nd}}_j$. Regardless of the values of $\gamma_j^\text{2nd}$, this would force $\sigma_j^\text{2nd} = \sqrt{1 - 1}\ \gamma_j^\text{2nd} = 0$ for $j = 0, n$. Equation \eqref{kutta equation discretized} will then hold trivially, making it redundant. $\sigma$ is thus discretized as follows. First, we use Chebyshev interpolation to interpolate $\gamma$ from the points of the second kind to the first kind, hence obtaining the $n$ values $\gamma^{1\text{st}}_j$. We may thus form $\sigma^{1\text{st}}_j = \gamma^{1\text{st}}_j \sqrt{1 - s_j^{1\text{st}}}$. Now, we may interpolate back to the points of the second kind using Chebyshev interpolation again to obtain the $n+1$ points $\sigma^{2\text{nd}}_j$. Henceforth we suppress the superscript ``$2\text{nd}$" for readability so that $s_j = s_j^{2\text{nd}}$, $\zeta_j = \zeta_j^{2\text{nd}}$, and $\gamma_j = \gamma_j^{2\text{nd}}$.

\subsection{Birkhoff-Rott and pressure-jump equations}
\label{numerical birkhoff rott}
We now discuss the the numerical treatment of both the Birkhoff-Rott equations and the pressure-jump equation. Third-order Adams-Bashforth is used to evolve the former, while second-order backward differentiation is used for the latter. The principal issue we address is the discretization of the Cauchy integral 
\begin{equation}
\frac{1}{2\pi i}\int_{-1}^{1}{\frac{\gamma_b(s',t)}{z - \zeta(s',t)}\, ds'} \label{body_contribution_cauchy_integral}
\end{equation}
with $ z = \zeta(s,t)$ or $\zeta_\pm(\Gamma,t)$. We begin by developing a simple quadrature rule to compute such integrals.
\subsubsection{Log quadrature} \label{LogDesingularizationOfBody}
Recall the Birkhoff-Rott equations \eqref{birkhoff rott table} for the velocity of the free sheets. In computing the velocity induced by the body, one evaluates the Cauchy integral \eqref{body_contribution_cauchy_integral} with $z$ on the free sheet. Should $z$ lie on the body as well, this Cauchy integral is of principal-value type, and there are simple quadrature methods that compute the integral accurately \cite{boikov2001numerical}. When $z$ is far from the body, the integrand is smooth, and standard quadrature methods such as the trapezoidal rule yield accurate results. However, in simulating the long-time behavior of a falling body, the free vortex sheets often lie very close to the body but not on it. In this near-singular integral situation, the standard quadrature methods are inaccurate, and the quadrature methods for principal valued integrals do not apply. A new approach is needed. Recently, Nitsche proposed to evaluate 1D near-singular integrals based on the addition of error-correcting terms \cite{NitscheNearSingularIntegral}. It was employed in \cite{sohn} to compute the vortex wake of a falling plate. Here, we propose an alternative method to handle these near-singular integrals that achieves similar results, but in the opinion of the authors, is easier to implement. This method can be applied to flexible bodies as well.
\newline

As noted in section \ref{Kutta Condition Section}, discretizing this singular integral using the trapezoidal rule:
$$\frac{1}{2\pi i}\int_{-1}^{1}{\frac{\gamma_b(s,t)}{z - \zeta(s,t)} ds \approx \sum_j w_j \frac{\gamma_j}{z - \zeta_j}},$$
introduces first-order singularities on each of the grid points of the body. Here $w_j$ are the quadrature weights. While the computed velocity at $z$ blows up as it approaches the body, the true velocity remains bounded. Near the body, this mismatch between the computed and true velocities gives rise to significant errors. The principle behind the trapezoidal rule is that the integrand can be approximated by a piecewise linear function, which can then be integrated exactly. In the method described below, instead of approximating the singular integrand as a piecewise-linear function, we approximate a smooth term inside the integrand as a 
piecewise-linear function, and then compute the integral analytically, thereby obtaining an accurate approximation.
\newline

To begin, recall that we discretize the body's arc-length parameter $s$ using a Chebyshev grid points of the second kind with $n + 1$ points. Let them be $s_j = s_j^{2\text{nd}} =  -\cos(\frac{j\pi}{n})$, $j = 0,1,2,\ldots, n$. Let $\Gamma_j = \Gamma_b(s_j,t)$ and $\zeta^{2\text{nd}}_j = \zeta_j = \zeta(s_j,t)$, and recall that the circulation on the body is given by $\Gamma_b(s,t)$ where
\begin{equation*}
    \Gamma_b(s,t) = \int_{-1}^{s}\gamma_b(s',t)\, ds'.
\end{equation*}
Then, we may write 
\begin{align*}
    \int_{-1}^{1} \frac{\gamma_b(s,t)}{z - \zeta(s,t)}\, ds  &= \int_{\Gamma_0}^{\Gamma_n}\frac{d\Gamma}{z - \zeta(\Gamma,t)} \\
                                                          &= \sum_{j =0}^{n-1}\int_{\Gamma_j}^{\Gamma_{j+1}}\frac{d\Gamma}{z - \zeta(\Gamma,t)}.
\end{align*}
Now, by analogy with the trapezoidal rule, we make the assumption that $\zeta(\Gamma,t)$ is a linear function of $\Gamma$ on each subinterval $[\Gamma_j,\Gamma_{j+1}]$ and integrate exactly. On each subinterval $[\Gamma_j,\Gamma_{j+1}]$, we have $$\zeta(\Gamma,t) = a_j\Gamma + b_j$$ where $$a_j = \frac{\zeta_{j+1} - \zeta_{j}}{\Gamma_{j+1} - \Gamma_{j}}$$
and $$b_j = \frac{\zeta_j\Gamma_{j+1} - \zeta_{j+1}\Gamma_{j}}{\Gamma_{j+1} - \Gamma_{j}}.$$

It immediately follows that
\begin{align*}
    \int_{\Gamma_j}^{\Gamma_{j+1}}\frac{d\Gamma}{z - \zeta(\Gamma,t)} &\approx -\frac{1}{a_j}\int_{\Gamma_j}^{\Gamma_{j+1}}\frac{d\Gamma}{\Gamma - \frac{z - b_j}{a_j}} \\
    &=  -\frac{1}{a_j}\big(\log(a_j\Gamma_{j+1} + b_j - z) - \log(a_j\Gamma_{j}+ b_j - z )\big) \\
    &=  -\frac{1}{a_j}\big(\log(\zeta_{j+1} - z) - \log(\zeta_{j} - z)\big)
\end{align*}

Therefore, 

\begin{equation}
    \frac{1}{2\pi i}\int_{-1}^{1} \frac{\gamma_b(s,t)}{z - \zeta(s,t)}\, ds \approx \frac{-1}{2\pi i}\sum_{j=0}^{n} \frac{1}{a_j} \big(\log(\zeta_{j+1} - z) - \log(\zeta_{j} - z)\big)\label{log quadrature for singular body}
\end{equation}

In summary, to compute near-singular integrals without sampling the integrand, we first subdivide the body into smaller subintervals. On each subinterval, we assume the body position $\zeta$ is a piecewise linear function of circulation, and evaluate the integral exactly. Then, we sum the contributions of each segment to obtain the total contribution of the body to the flow. By doing so, we weaken the singularities in the computed flow from $O(\frac{1}{z-\zeta_j})$ to  $O(\log(z-\zeta_j))$, which allows for a more accurate computation of the velocity of the free vortex sheets. A similar method was used to compute frictional forces accurately in \cite{alben2013optimizing}. Similar formulae can be obtained even in the presence of blob regularization. Indeed, for consistency, we evaluate the contributions of the free sheets to the fluid velocity using a similar formula

\begin{multline}
    \frac{1}{2\pi i}\mathlarger{\int_0^{\Gamma_\pm(t)}}\frac{\overline{z - \zeta_\pm(\Gamma',t)}}{|z - \zeta_\pm(\Gamma',t)|^2 + \delta^2}d\Gamma' \approx \mathlarger{\sum_{j=0}^{k_\pm}}\frac{-1}{a_j 2\pi i}\mathlarger{\Bigg(} \frac{1}{2}\log{\bigg(\frac{(\Gamma_{\pm,j+1} -c_j^R)^2 + (c_j^I)^2 + \delta^2}{(\Gamma_{\pm,j} -c_j^R)^2 + (c_j^I)^2 + \delta^2}\bigg)} \\ + \frac{ic_j^I}{(c_j^I)^2 + \delta^2}\bigg( \arctan\big(\frac{\Gamma_{\pm,j+1}-c_j^R}{\sqrt{(c_j^I)^2 +\delta^2}}\big) - \arctan\big(\frac{\Gamma_{\pm,j}-c_j^R}{\sqrt{(c_j^I)^2 +\delta^2}}\big) \bigg)\mathlarger{\Bigg).} \label{log quadrature for smooth sheets}
\end{multline}
Here, $c_j = \frac{z -b_j}{a_j}$, and $a_j$, $b_j$ are given as before. $c_j^R = \Re{c_j}$ and $c_j^I = \Im{c_j}$ so that $c_j = c_j^R + ic_j^I$. See appendix \ref{log quadrature with regularization} for more details. Note that the formula \eqref{log quadrature for singular body} is the limit of \eqref{log quadrature for smooth sheets} (but for the bound instead of free vortex sheets) as $\delta \rightarrow 0$.
\newline

The method described thus far can be interpreted in the following way. It treats the plate as a $\textit{hybrid}$ between a $0$-dimensional and a $1$-dimensional vortex structure. Indeed, when using equation \eqref{gamma equation 1-n}, one implicitly solves for $\gamma_b$ by treating the plate as a collection of $n + 1$ $0$-dimensional point vortices, each with circulation $\Gamma_j = w_j \gamma^{2\text{nd}}_j, \text{ } j = 0,1,\ldots,n$. Later, however, when using $\gamma_b$ to compute the velocity induced by the bound vortex sheet, it is treated instead as a $1$-dimensional vortex structure to reduce the $1/(z-\zeta_j)$ singularities, to $\log(z-\zeta_j)$ singularities via integration in equation \eqref{log quadrature for singular body}. Note, however, that both equation \eqref{gamma equation 1-n} and equation \eqref{log quadrature for singular body} are both formally second order discretizations of the same integral (after a change of variables). In appendix \ref{numerical parameters and grid refinement studies} we show that we achieve expected second-order convergence in space with this hybrid approach.

\subsubsection{Velocity-smoothing the pressure-jump and Birkhoff-Rott equations}\label{Velocity Smoothing}
As discussed in both section \ref{Kutta Condition Section}, and appendix \ref{log singularity appendix}, the Cauchy integral
\begin{equation}
    \frac{1}{2\pi i}\int_{-1}^{1} \frac{\gamma_b(s,t)}{z - \zeta(s,t)}\, ds 
\end{equation}
can introduce logarithmic singularities into the flow at the edges of the body. These logarithmic singularities are analogous to the logarithmic singularities on the grid points of the body introduced by the log quadrature rule \eqref{log quadrature for singular body}. Methods have been developed to eliminate these 
logarithmic singularities that appear on the body's edges without forcing $\gamma$ to vanish at the edges.  Two such methods appear in \cite{albenRegularize}, ``tapered smoothing" and ``velocity smoothing." In the tapered-smoothing method, $\delta$ is no longer uniform, but rather a function of the arc length parameter of the vortex sheet, $\delta(s)$.
As $s \to \pm 1$, the edges of the body, $\delta$ tapers to $0$, thereby removing the logarithmic singularities on the body's edges. Tapered smoothing has been applied successfully in several studies such as \cite{sohn} and \cite{albenFoils}.
\newline

For falling bodies, however, the dynamics of the system is dominated by leading-edge effects. The computed dynamics are thus sensitive to the motions of the vortex sheets near the leading edge. As observed in \cite{albenRegularize}, accurately computing the motions of the sheets regularized by $\delta$ requires the local mesh spacing on the vortex sheets to be proportional to $\delta$. Indeed, this is a general feature of vortex blob methods, which require the vortex cores to overlap in order for convergence properties to hold \cite{koumoutsakos1993direct}. Since our equations are nondimensionalized by the terminal velocity of the body, the speed of the body is $O(1)$. Hence as the points on the vortex sheet are released from the body edges at each time step and advect with the local flow speed, the mesh spacing near the body edges needs to be proportional to $dt \cdot O(1)  = O(dt)$ for accurate evolution of the sheets near the body edges. In the tapered-smoothing method, since $\delta$ tapers to $0$ near the edge, it becomes difficult to evolve the sheets near the edges accurately without a priori knowledge of the initial shape of the vortex sheet leaving the edge. (See \cite{albenRegularize} for one such example.) Therefore, a method that accurately computes the flow of the free sheets near the edge while simultaneously suppressing the logarithmic singularities there is desirable.
\newline

Here we propose a lightly modified version of the method of velocity smoothing that allows for exactly that. It can be described as follows. When computing the velocities of the free sheets close to the body in \eqref{birkhoff rott table}, the Cauchy integral representing the contribution of the bound vortex sheet to the fluid velocity is smoothed as well. That is, for points sufficiently close to the body, the integral 

\begin{equation}
    \frac{1}{2\pi i}\int_{-1}^{1} \frac{\gamma_b(s,t)}{\zeta_\pm(\Gamma,t) - \zeta(s,t)}\, ds 
    \label{unsmoothedBodyVel}
\end{equation}

is replaced with 
\begin{equation}
    \frac{1}{2\pi i}\int_{-1}^{1} \frac{\overline{\zeta_\pm(\Gamma,t) - \zeta(s,t)}}{|\zeta_\pm(\Gamma,t) - \zeta(s,t)|^2 +\delta^2}\gamma_b(s,t)\, ds
    \label{smoothedBodyVel}
\end{equation}

for such points, with $\delta$ uniform on the free sheets. Near the body, one favors \eqref{smoothedBodyVel} which suppresses the logarithmic singularities nearby. On the other hand, further from the body, the singularities are weak enough that one instead favors \eqref{unsmoothedBodyVel} which occurs in the exact no-penetration condition. Let $l(\zeta_\pm)$ denote the physical distance from $\zeta_\pm$ to the body (i.e.~$l(\zeta_\pm) = \min_{-1 \leq s\leq 1}(|\zeta_\pm - \zeta(s,t)|)$). Then consider a $C^\infty$ smooth transition function 
$$B(l) = \begin{cases}
          0 \text{,\  if }l = 0, \\
          \frac{\exp(-\delta/l)}{\exp(-\delta/l) + \exp(-\delta/(\delta -l))} \text{,\  if }0 < l < \delta, \\
          1  \text{,\  if $l \geq \delta$}.
         \end{cases}$$  The contribution of the body to the velocity of the free sheets at $\zeta_\pm$ is given by 

\begin{equation}
     \frac{B(l(\zeta_\pm))}{2\pi i}\int_{-1}^{1} \frac{\gamma_b(s,t)}{\zeta_\pm - \zeta(s,t)}\, ds + 
    \frac{ 1 - B(l(\zeta_\pm))}{2\pi i}\int_{-1}^{1} \frac{\overline{\zeta_\pm - \zeta(s,t)}}{|\zeta_\pm - \zeta(s,t)|^2 +\delta^2}\gamma_b(s,t)\, ds. \label{VelocitySmoothing}
\end{equation}

The blending function $B$ was chosen since it smoothly transitions from $0$ to $1$ over a distance $\delta$. Moreover, $$\frac{d^m}{dl^m}B(l) = 0 \text{ for every integer } m \geq 0 \text{ and } l = 0,\ \delta.$$ That is to say, the function $B$ approaches $0$ and $1$ with infinite order at $l = 0$ and $l = \delta$ respectively. Hence, as $\zeta_\pm$ approaches the edge, the velocity induced by the body on $\zeta_\pm$ blends from singular to smoothed with infinite order over a distance $\delta$. As observed in \cite{albenRegularize}, this procedure improves the smoothing error from $O(\delta^{1/2})$ to $O(\delta^{3/2})$ in simple model problems. 
\newline

Thus, in this study, we evolve the sheets with the velocity-smoothed Birkhoff-Rott equations given by
\begin{multline}
     \partial_{t}\bar\zeta_{\pm}(\Gamma,t)= \frac{B(l(\zeta_\pm))}{2\pi i}\int_{-1}^{1} \frac{\gamma_b(s,t)}{\zeta_\pm(\Gamma,t) - \zeta(s,t)}\, ds + 
    \frac{ 1 - B(l(\zeta_\pm))}{2\pi i}\int_{-1}^{1} \frac{\overline{\zeta_\pm(\Gamma,t) - \zeta(s,t)}}{|\zeta_\pm(\Gamma,t) - \zeta(s,t)|^2 +\delta^2}\gamma_b(s,t)\, ds \\
    + \frac{1}{2\pi i}\int_{0}^{\Gamma_-(t)}{\frac{\overline{\zeta_{\pm}(\Gamma,t) - \zeta_{-}(\Gamma',t)}}{|\zeta_{\pm}(\Gamma,t) - \zeta_{-}(\Gamma',t)|^2 + \delta^2}d\Gamma'}  +
     \frac{1}{2\pi i}\int_{0}^{\Gamma_+(t)}{\frac{\overline{\zeta_{\pm}(\Gamma,t) - \zeta_{+}(\Gamma',t)}}{|\zeta_{\pm}(\Gamma,t) - \zeta_{+}(\Gamma',t)|^2 + \delta^2}d\Gamma'}. \label{BirkhoffRottSmoothed}
\end{multline}
instead of (\ref{birkhoff rott table}). 
We handle the pressure-jump equation in a similar way. That is, when computing $\mu(s,t)$ in the pressure-jump equation \eqref{pressure jump equation table}, we smooth plate's contribution to the fluid velocity $\mu(s,t)$ as well. Hence we use the following equation for $\mu(s,t)$ instead of \eqref{muEqn}:

\begin{multline}
    \mu(s,t) = \Re\Bigg(\hat{\textbf{s}}(s,t)\bigg(\frac{1}{2\pi i}\int_{-1}^{1} \frac{\overline{\zeta_\pm(\Gamma,t) - \zeta(s,t)}}{|\zeta_\pm(\Gamma,t) - \zeta(s,t)|^2 +\delta^2}\gamma_b(s,t)\, ds + \frac{1}{2\pi i}\int_{0}^{\Gamma_-(t)}{\frac{\overline{\zeta(s,t) - \zeta_{-}(\Gamma,t)}}{|\zeta(s,t) - \zeta_{-}(\Gamma,t)|^2 + \delta^2}d\Gamma} \\ + \frac{1}{2\pi i}\int_{0}^{\Gamma_+(t)}{\frac{\overline{\zeta(s,t) - \zeta_{+}(\Gamma,t)}}{|\zeta(s,t) - \zeta_{+}(\Gamma,t)|^2 + \delta^2}d\Gamma}\bigg)\Bigg).\label{smooth bernoulli pressure jump}
\end{multline}

In equations \eqref{BirkhoffRottSmoothed} and \eqref{smooth bernoulli pressure jump}, the smoothed Cauchy integrals are discretized using the log-quadrature rule discussed in the preceding section, \ref{LogDesingularizationOfBody}.

\subsubsection{Controlling the no-penetration error with sub-step fencing} \label{Sub-StepFencing}

Following \cite{albenFlex}, each time step consists of two sub-steps. During the first sub-step, the free sheets are evolved (explicitly) using the Birkhoff-Rott equation \eqref{birkhoff rott table} and the log-quadrature rule discussed in section \ref{LogDesingularizationOfBody}. In the second sub-step, the body is evolved (implicitly) using both the force-balance \eqref{force balance table} and torque-balance \eqref{torque balance table} equations. During this sub-step, the vortex strength $\gamma_b$ and pressure jump $[p]^+_-$ are determined using the no-penetration equation \eqref{no penetration table} and pressure-jump equation \eqref{pressure jump equation table} respectively. As discussed in section \ref{solving no penetration}, during the implicit sub-step where the body moves, the no-penetration equation \eqref{no penetration table} is used to obtain $\gamma_b$ as a function of the quantities $\zeta(s,t)$, $\partial_t\zeta(s,t)$, $\zeta_\pm(\Gamma,t)$ and $\Gamma(s,t)$ by inverting a Cauchy integral.
\newline

This approach is insufficient to enforce the no-penetration condition strictly because the free vortex sheets and body may interpenetrate during both the implicit and explicit sub-steps. During the implicit sub-step, the free sheets are fixed, and the body moves through the fluid. Although the no-penetration condition is used to determine the bound vortex sheet strength $\gamma_b$, the plate may pass through the free vortex sheet on this sub-step due to temporal and spatial discretization errors. During the explicit sub-step, the free sheets may pass through the body for the same reason. An additional source of error on the explicit sub-step is the use of velocity smoothing, which introduces additional error into the no-penetration condition. Indeed, since $\gamma_b$ is obtained with zero smoothing on the plate, by advecting the vortex sheets with smoothing on the plate in the Birkhoff Rott equations \eqref{BirkhoffRottSmoothed}, one introduces an error of size $O(\delta)$ into the no-penetration condition near the body as the price for removing logarithmic singularities in the flow. Note that there is an analogous source of error during the implicit sub-step incurred by blob regularizing the free sheets.
\newline

Interpenetration of the body and free sheets can be controlled by the following procedure termed ``sub-step fencing." After the body moves in the implicit sub-step, or the sheets move in the explicit sub-step, we identify all points on the free vortex sheets that pass through the plate. We decompose their displacements into components normal and tangential to the body. We set the normal component to be the distance to the body minus a small value (10$^{-6}$). A crucial detail is that this sub-step fencing procedure needs to be applied during each iteration of nonlinear solver used to solve the discretized system of equations in the implicit sub-step. In effect, the velocity of each crossing point is projected tangentially along the body, ensuring that streamlines do not pass through the body, which obeys the no-penetration condition. 
\newline

Another approach to fencing, used by \cite{PanXiaoLE}, is to project the points only once at the end of each full time step, and place them a fixed distance away from the body. We find that the more-frequent fencing that we use---during each sub-step, including within the iterative solver---is necessary because if any point on the free sheet passes through the body at some sub-step, the discontinuity of the fluid velocity across the body introduces a large error in the velocities computed at the next sub-step. Furthermore, as the free sheets are evolved using the multi-step method third-order Adams-Bashforth, third-order accuracy is only possible if the velocity field varies smoothly over the trajectory of the point. For a point that passes through the body, however, this condition is violated, and we lose third-order accuracy.

\subsection{Outline of the coupled fluid-body solver}
\label{outline of vortex shedding algorithm}
We now outline the major steps of the algorithm, focusing on the novel aspects. A refinement study can be found for select cases in appendix \ref{refinement study}.

\begin{algorithm}[H]
\caption{Coupled fluid-body solver for a falling rigid body} \label{coupled fluid body solver}
\begin{algorithmic}[0]
\For{each time step $k = 1, 2, \ldots$}
    \State \textbf{Step 1:} Evolve vortex sheets explicitly to obtain $\zeta_\pm(\Gamma,t)$: 
    \State \quad $\bullet$ Use the velocity-smoothed Birkhoff-Rott equations \eqref{BirkhoffRottSmoothed} (section \ref{Velocity Smoothing}).
    \State \quad $\bullet$ Use the log-desingularized quadrature rules \eqref{log quadrature for singular body}, \eqref{log quadrature for smooth sheets} to discretize the integrals (section \ref{LogDesingularizationOfBody}).
    \\
    \State \textbf{Step 2:} Sub-step fence:
    \State \quad $\bullet$ Project all crossing vortex points normally to the body surface to prevent crossing.  (section \ref{Sub-StepFencing}).
    \\
    \State \textbf{Step 3:} Compute the 5 (real) unknown body variables at the current time step implicitly by solving \\
    \quad\quad\quad\quad\quad\ \   the following nonlinear system of 5 (real) discretized equations iteratively:
    \State \quad \textbf{Equations for the nonlinear solver:}
    \State \quad\quad $\bullet$ Force balance equation \eqref{force balance table} (2 real equations) 
    \State \quad\quad $\bullet$ Torque balance equation \eqref{torque balance table}
    \State \quad\quad $\bullet$ Kutta condition on each edge \eqref{kutta condition table} (2 real equations)
    \\
    \State \quad \textbf{Unknown variables for the nonlinear solver:}
    \State \quad\quad $\bullet$ Center-of-mass position $\zeta_G(t)$ (2 real variables)
    \State \quad\quad $\bullet$ Angle of rotation from initial rest state $\beta(t)$ 
    \State \quad\quad $\bullet$ Total circulation of sheets $\Gamma_\pm(t)$ (2 real variables)
    \\
    \State \quad \textbf{At each iteration of the nonlinear solver:}
    \State \quad\quad $\bullet$ Sub-step fence again, and project all crossing vortex points onto plate as it moves (section \ref{Sub-StepFencing}). 
    \State \quad\quad $\bullet$ Use the collocation method to obtain $\gamma_b(s,t)$  (section \ref{solving no penetration}).
    \State \quad\quad $\bullet$ Use the pressure-jump equations \eqref{pressure jump equation table} to compute the pressure jump $[p]^+_-$ from $\gamma_b(s,t)$.
    \State \quad\quad $\bullet$ Interpolate $\gamma_b(s,t)$ to obtain $\sigma(\pm1,t)$ for use in the Kutta condition  (section \ref{Suction Section}).
\EndFor
\end{algorithmic}
\end{algorithm}
    
\section{Dynamics of falling plates}
In this section, we study the dynamics of both flat plates and a family of V-shaped plates obtained by bending the flat plate symmetrically about its center. In the system of equations described at the beginning of section \ref{numerical method section}, the only free parameter is the density $R_1$ which we vary in this study. In section \ref{flat plate section} we study flat plates over a large range of densities $0 <R_1 \leq 1000$ and describe the resulting falling motions. In section \ref{vplate section} we study the general class of V-shaped plates over a smaller range of densities, $0 \leq R_1 \leq 5$ to accommodate the larger class of plate geometries. For both flat and V-shaped plates, we sample the space of possible steady-state falling motions by initializing the plate with orientation angles $0 \leq \beta(0) \leq 45^\circ$. The values of the density $R_1$ and initial angle $\beta(0)$ considered are listed later, in table \ref{flat plate parameters table} for the flat plates and table \ref{V-plate parameters table} for the V-shaped plates. In all that follows, we use $dt = 0.012$ for the time step size and $n = 100$ for the number of grid points on the plate. A more detailed summary of the numerical parameters used, including how the number of grid points on the free sheets is determined, is included in appendix \ref{numerical parameters and grid refinement studies}.

\subsection{The flat plate}
\label{flat plate section}
The flat plate is a special case of the V-shaped plates that can move purely tangentially by translating in the in-plane direction. In the inviscid model without viscous skin friction, such a motion can result in very small or even zero vortex sheet strengths and pressure forces on the plate, and thus unrealistically large plate velocities under gravitational forces. To obtain realistic dynamics for the flat plate, we add an additional term to the right-hand side of the force balance equation \eqref{force balance table} to account for skin friction as was done in \cite{albenFoils}. We now make a brief digression to describe how skin friction is included into our numerical model of the flat plate.

\subsubsection{Including skin friction}
We assume that the skin friction at a given point on the flat plate can be approximated by that of the Blasius boundary layer on a flat plate in a steady oncoming flow. The velocity in the Blasius skin friction formula is set to $V_{avg}$, the (tangential) difference between the plate and flow velocity, averaged along the plate and over both sides. This model is similar to that in \cite{albenFoils} (which approximated $V_{avg}$ by the negative plate velocity). The traction is thus equal to that which would be induced by a Blasius boundary layer attached to the flat plate. For the moment, let us use dimensional variables. Following the assumption, the surface traction induced by skin friction is approximately \cite{batchelor1967introduction}
 \begin{equation}
    \tau_{skin} = \frac{1}{3}\exp(i\beta)\rho_f\hat{V}_{avg} \sqrt{\nu\frac{|V_{avg}|^3}{L + \hat{V}_{avg}s}}. 
 \end{equation}
Here, $\nu$ is the kinematic viscosity, $\hat{V}_{avg} = \frac{V_{avg}}{|V_{avg}|}$, and $\beta$ is the angle of the plate (relative to the horizontal axis). As the body has two sides, total viscous drag is given by 

\begin{align}
    F_{visc} &= 2\exp(i\beta)\int_{-L}^{L} \frac{1}{3}\rho_f\hat{V}_{avg} \sqrt{\nu\frac{|V_{avg}|^3}{L + \hat{V}_{avg} s}} \, ds= \frac{4\sqrt{2L\nu}}{3}\exp(i\beta)\rho_f|V_{avg}|^{\frac{3}{2}}\hat{V}_{avg}.
\end{align}

We add this term to the right-hand side of (\ref{DimensionalForceBalance}) to obtain
\begin{equation}
    \rho_b h W 2L \partial_{tt}\zeta_G(s) = W \int_{-L}^{L}[p]^+_-(s)\hat{\textbf{n}}(s) \, ds  - i \rho_b h W 2L g +   \frac{4\sqrt{2L\nu}}{3}\exp(i\beta)\rho_f|V_{avg}|^{\frac{3}{2}}\hat{V}_{avg}.
\end{equation}

Nondimensionalising as before,
\begin{equation}\label{Plate Force Balance}
    R_1 \partial_{tt}\zeta_G(s) = -\frac{1}{2} \int_{-1}^{1}[p]^+_-(s)\hat{\textbf{n}}(s) \, ds  - i +   \frac{2\sqrt{2}}{3\sqrt{Re}}\exp(i\beta)|V_{avg}|^{\frac{3}{2}}\hat{V}_{avg}.
\end{equation}
Here, $Re = \frac{UL}{\nu}$ is the Reynolds number, which we fix as $Re = 10^3$ for the remainder of this study. As shown above, the drag induced by skin friction is $O(1/\sqrt{Re})$. Following the analysis in \cite{CurvedBoundaryLayer}, the necessary modification to \eqref{Plate Force Balance} to account for either a curved body, or a body traveling in a circular arc, requires the inclusion of additional terms of a higher order, $O(1/Re)$. Hence, in the inviscid limit, \eqref{Plate Force Balance} can be thought of as a low-order model for plates, either flat or slightly curved, and traveling in either straight lines or circular arcs. As will be shown in subsequent sections, the generic behavior of a falling flat plate is gliding along a circular arc with small curvature. Hence for a majority of the motions considered, the Blasius boundary layer approximation may be used as a low-order model. Although also valid for slightly curved bodies, in this study, we only apply skin friction to flat plates. For simplicity, we neglect the effect of nearby free vortex sheets on the skin friction.

\subsubsection{Physical and numerical parameter values}
The plates are released from rest at a fixed initial angle $\beta(0)$. Each free vortex sheet $\zeta_\pm$ is initialized as a single line segment with two endpoints, located at distances of $10^{-5}$ and $2\times10^{-5}$ away from the $\pm$ edges, respectively, in the direction tangent to the plate. Both free vortex sheets have zero initial circulations. We discretize the body's arc-length parameter $s$ using a Chebyshev grid with $n = 100$ points (section \ref{solving no penetration}), and use a time step $dt = 0.012$. More details can be found in appendix \ref{numerical parameters and grid refinement studies}. In table \ref{flat plate parameters table} we list the values of $R_1$ and initial angles used to obtain the results in the subsequent subsections. The generic steady state behavior is the same across initial angles. To sample the space of steady state trajectories thoroughly we have chosen 11 evenly spaced initial angles between $0^\circ$ and $45^\circ$ for each value of $R_1$.

\begin{table}[H]
    \centering
    \renewcommand{\arraystretch}{1.2}
    \begin{tabular}{|c|c|}
        \hline
         Initial angle & $0^\circ, 4.5^\circ, 9^\circ, 13.5^\circ, \ldots, 45^\circ$  \\
         \hline
         $R_1$ & 0,\ 0.01,\ \ldots,\ 0.1,\ 0.2,\ \ldots,\ 5,\ 10,\ 100,\ 500,\ 1000 \\ 
         \hline
    \end{tabular}
    \caption{Values of the physical parameters used for the falling flat plates}
    \label{flat plate parameters table}
\end{table}

In table \ref{flat plate steady state dynamics table} we describe the steady-state dynamical motions observed in different $R_1$ ranges. Each of the dynamical motions---``fluttering," ``mixed," ``tumbling," ``looping," and ``autorotating"---are described and defined as they arise in the subsequent subsections.

\begin{table}[H]
    \centering
\renewcommand{\arraystretch}{1.2}
\begin{tabular}{|>{\centering\arraybackslash}m{3cm}|>{\centering\arraybackslash}m{2cm}|>{\centering\arraybackslash}m{2cm}|>{\centering\arraybackslash}m{2cm}|>{\centering\arraybackslash}m{2cm}|>{\centering\arraybackslash}m{2cm}|}
    \hline
    $R_1$ & [0, 0.2) & [0.2, 0.7) & [0.7, 1.6) & [1.6, 2.8) & [2.8, 1000) \\[5pt]
    \hline
    Steady-state dynamics & Fluttering & Mixed & Tumbling & Looping & Autorotating \\
    \hline
\end{tabular}
    \caption{Ranges of $R_1$ and corresponding steady-state dynamics.}
    \label{flat plate steady state dynamics table}
\end{table}

\subsubsection{$R_1 < 0.2$: the pure fluttering regime}

\begin{figure}[H]
    \centering

\includegraphics[width=1\textwidth, trim=3cm 2cm 3cm 2cm, clip]{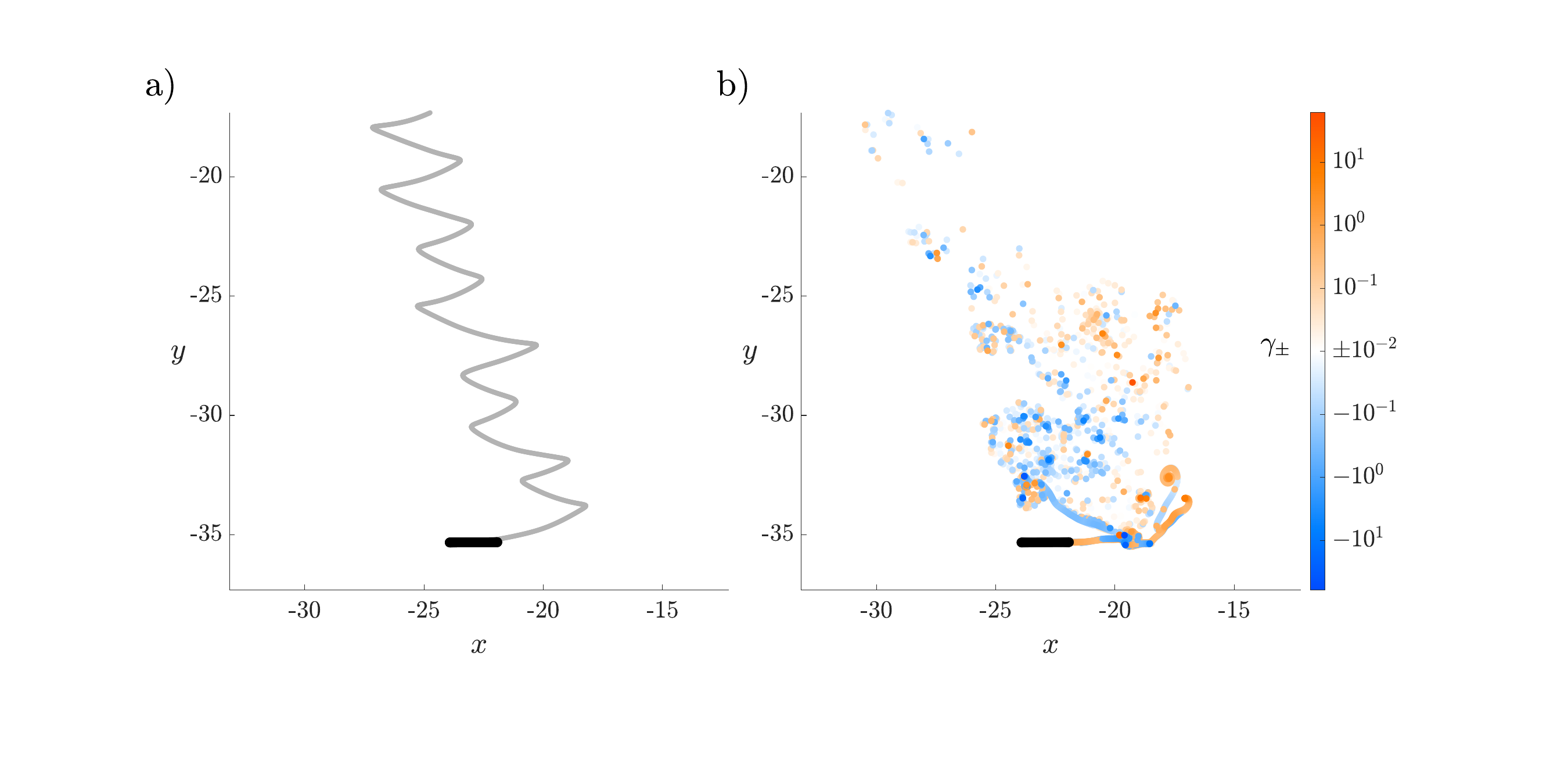}
    \caption{Side-by-side comparison between (a) the trajectory of the center of mass and (b) the chaotic vortex wake formed by a falling plate with $R_1 = 0.02$ undergoing small-amplitude fluttering. The plate was released with initial angle $\beta(0) = 25^\circ$.}
    \label{wake behind 0.02 figure}
\end{figure}

In this $R_1$ range the dominant steady-state motion is small-amplitude side-to-side fluttering as shown in figure \ref{wake behind 0.02 figure}. In panels (a) and (b) we show the trajectory of the center of mass and the vortex wake formed behind a plate undergoing small-amplitude fluttering at $R_1 = 0.02$. In this study, a body is said to be ``fluttering" if both its angular velocity and the horizontal velocity of its center of mass change sign as it falls, corresponding to a side-to-side oscillation while falling. The speed of the center of mass often vanishes, resulting in the sharp cusps in its trajectory in figure \ref{wake behind 0.02 figure}(a). Panel (b) shows that the vortex wake is highly irregular, its structure disturbed by repeated collisions with the fluttering plate.

\begin{figure}[H]
    \centering
    \includegraphics[width= 0.95\textwidth, trim=1.9cm 11cm 1.9cm 11cm, clip]{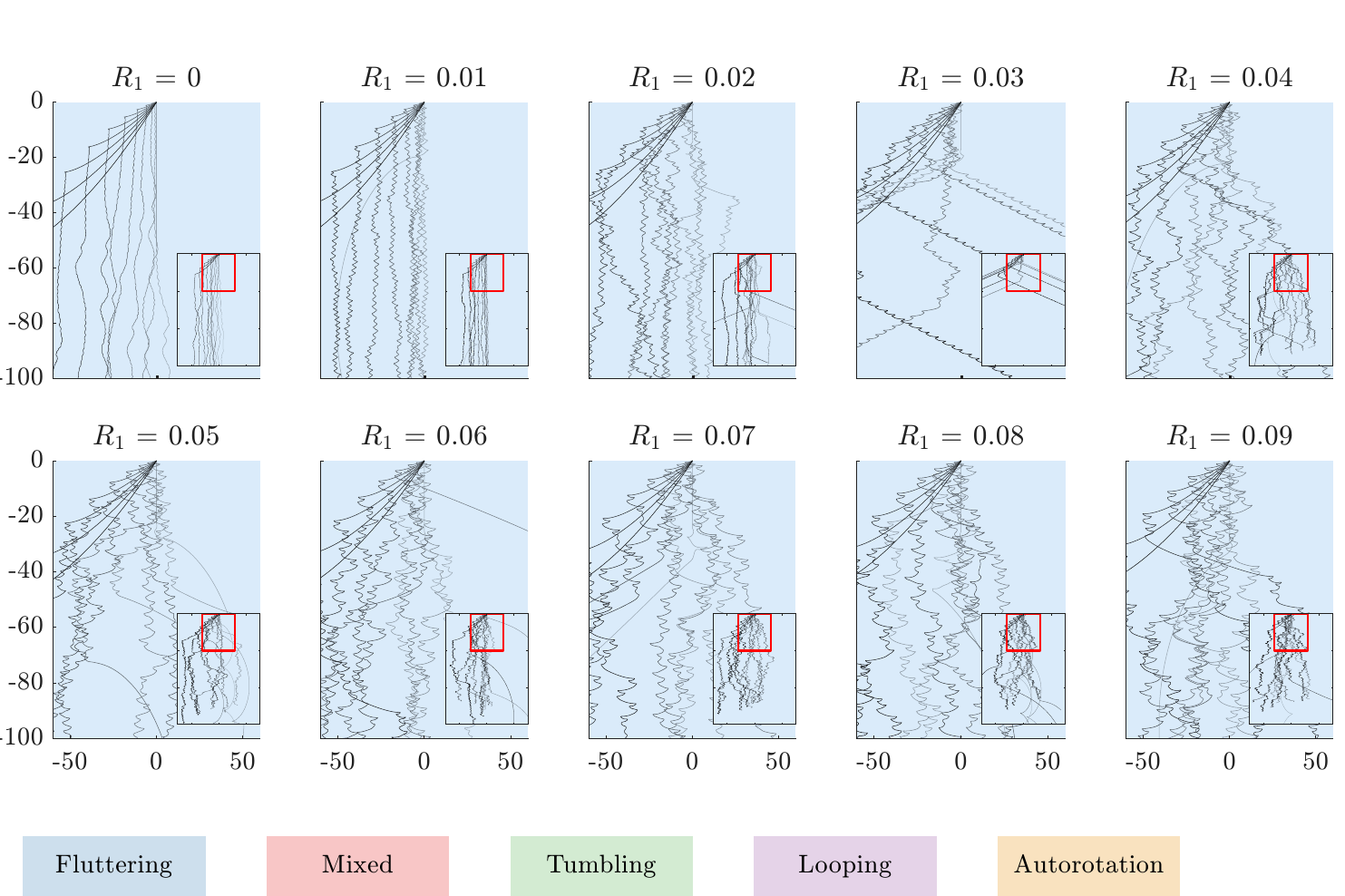}
    \caption{For $R_1 = 0, 0.01, \ldots, 0.09$, snippets of the center-of-mass trajectories of falling plates for eleven initial angles within $[0,\pi/4]$ during $0\leq t \lessapprox 500$. The darker trajectories correspond to larger initial angles. The plots are shaded according to the qualitative dynamics exhibited by the steady state. Within this $R_1$ range, only fluttering motions occur. Each panel contains an inset that shows the large-scale features of the same trajectories, with a red rectangle indicating the region shown in the main panel. }
    \label{plate small R1}
\end{figure}

In figure \ref{plate small R1}, in each panel labeled by $R_1$, we show the trajectories of the center of mass of a falling plate for various initial angles listed in table \ref{flat plate parameters table}. The trajectories are darker for larger initial angles. Each panel contains an inset showing a larger range. The red rectangle in the inset indicates the close-up portion of the trajectories shown in the main panel. Each panel is shaded according to the steady-state dynamics exhibited by the falling plates. As seen in the panels of figure \ref{plate small R1} (and later in figure \ref{plate medium R1} for $R_1 < 0.7$), increasing $R_1$ increases the side-to-side fluttering amplitude. As can be seen from figure \ref{plate small R1}, increasing the initial angle increases the duration of the initial period of gliding, but has no obvious effect on the steady-state dynamics. Occasionally, the side-to-side fluttering motion is punctuated by a much faster diving motion, shown the large arcs in figure \ref{plate small R1} when $R_1 \geq 0.05$. These occur when the plate collides with its own vortex wake, which can sometimes destabilize it into this faster falling motion.

\vspace{-0.5cm}

\begin{figure}[H]
    \centering    \includegraphics[width=1\textwidth, trim=4cm 4cm 4cm 3cm, clip]{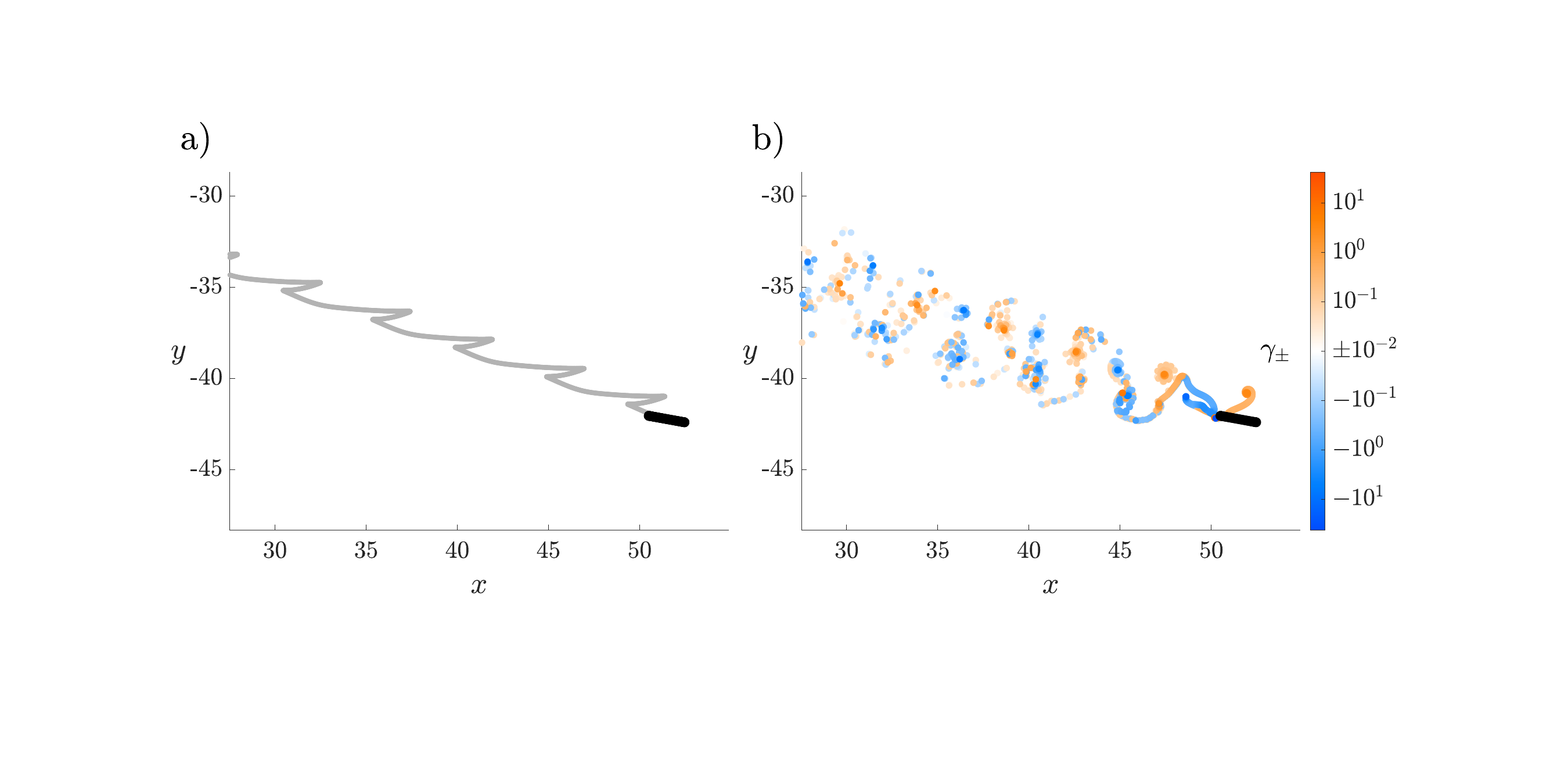}
    \caption{The trajectory of the center of mass (a) and the vortex wake (b) formed by a falling plate with $R_1 = 0.03$ undergoing progressive fluttering. The plate was released with initial angle $\beta(0) = 25^\circ$.}
    \label{wake behind 0.03 figure}
\end{figure}

\begin{wrapfigure}[21]{l}{0.3\textwidth} 
\includegraphics[width=0.325\textwidth, trim = 1cm  0 0 1.5cm]{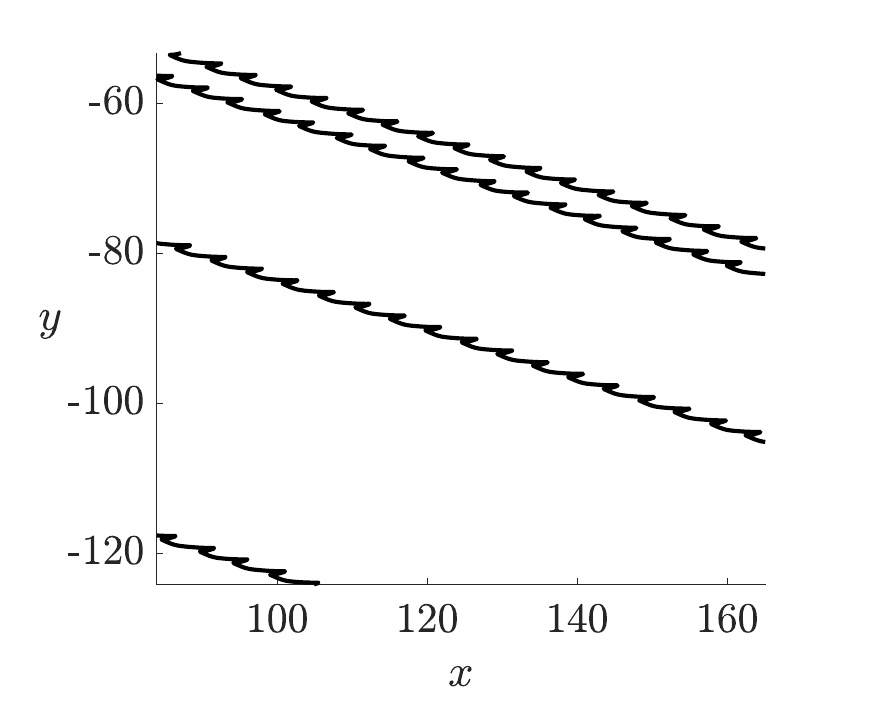}
    \caption{Plate trajectories when $R_1 = 0.03$ for initial angles $4.5^\circ,\ 22.5^\circ,\ 36^\circ, \ 45^\circ$. All trajectories simulated (including those not shown) converge to steady-state periodic fluttering at this value of $R_1$.}
    \label{progressive fluttering figure}
\end{wrapfigure}

As suggested by the vortex wake shown in figure \ref{wake behind 0.02 figure}(b), the steady-state side-to-side fluttering motions within this regime are chaotic, i.e.~sensitive to both initial conditions and numerical parameters. This is shown more clearly in appendix \ref{numerical parameters and grid refinement studies}. These fluttering motions can change sharply with $R_1$, and for some values of $R_1$ can switch dramatically from irregular and erratic, to periodic. For example, all trajectories are eventually periodic at $R_1 = 0.03$ but rarely at 0.02 and not at 0.04. The wake behind one such periodic trajectory is shown in figure \ref{wake behind 0.03 figure}, and periodic trajectories with $R_1 = 0.03$ and four initial angles are shown in figure \ref{progressive fluttering figure}.
\newline

When $R_1 = 0.03$ all falling plates oscillate about a diagonal path with a fixed nonzero angle from vertical (see figure \ref{progressive fluttering figure}). This is because the translational and rotational velocities are periodic, the latter with mean zero. Hence, the plate traverses a fixed distance in both the horizontal and vertical directions in each period. A close-up of these periodic trajectories is shown in figure \ref{progressive fluttering figure}. This staircase-like periodic fluttering motion observed here at $R_1 = 0.03$ was also found by \cite{ristroph_gliders} and \cite{pomerenk2024aerodynamicequilibriaflightstability} and termed  ``progressive fluttering." There, this falling mode was found at $I^* = 0.1$, or $R_1 \approx 0.23$. However, in their model, the center of equilibrium of the plate was shifted to one side which could explain the difference. Progressive fluttering is also observed for some initial angles when $R_1 = 0.02$ as can be seen from the straight diagonal lines in the inset of figure \ref{plate small R1} at this $R_1$.
\newline

\subsubsection{$0.2 \leq R_1 < 0.7$: the mixed-fluttering-and-tumbling regime}
\begin{figure}[H]
    \centering
\includegraphics[width=\textwidth, trim=4cm 5.5cm 4cm 5cm, clip]{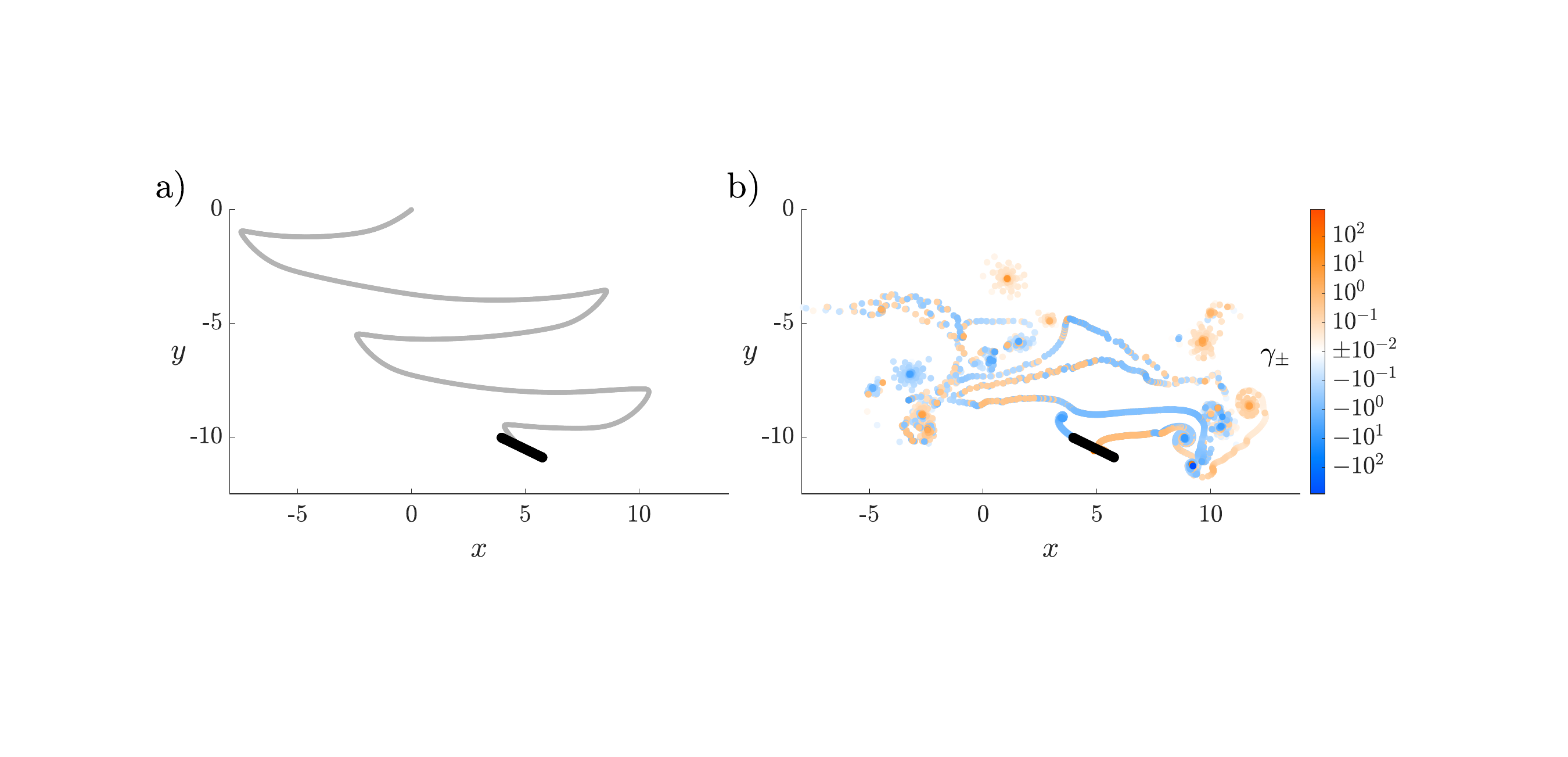}
    \caption{The trajectory of the center of mass (a) and the vortex wake (b) formed by a falling plate with $R_1 = 0.3$ undergoing large-amplitude fluttering. The plate was released with initial angle $\beta(0) = 25^\circ$.}
    \label{wake behind 0.3 figure}
\end{figure}

\begin{figure}[H]
    \centering
    \includegraphics[width=0.8\linewidth,trim= 0cm 0cm 0cm 0cm, clip]{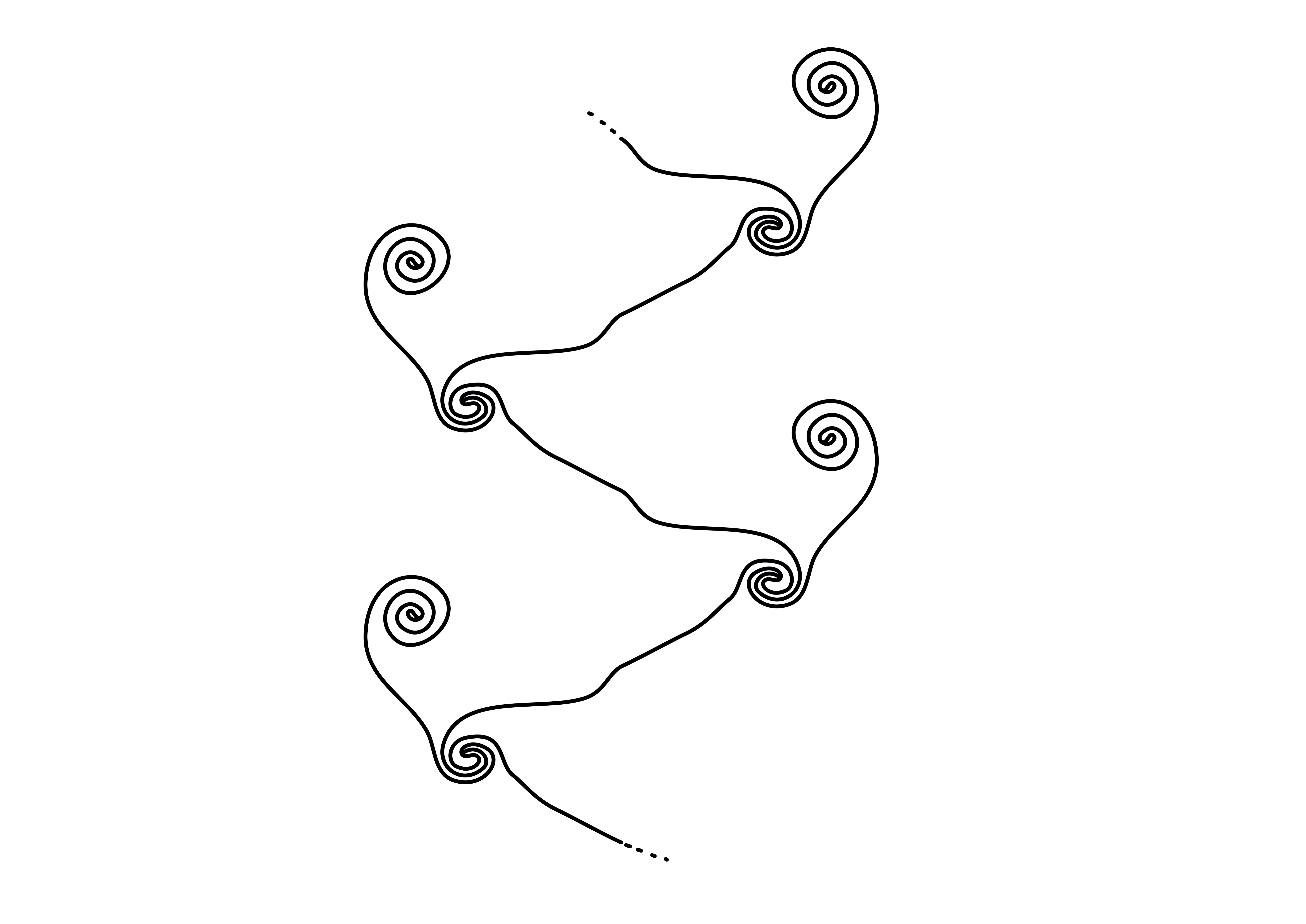}
    \caption{Schematic diagram showing the typical dipole vortex wake structure behind a plate undergoing large-amplitude fluttering.}
    \label{wake behind fluttering schematic figure}
\end{figure}

\begin{figure}[H]
    \centering
\includegraphics[width=0.9\textwidth, trim=0cm 1cm 0cm 1cm, clip]{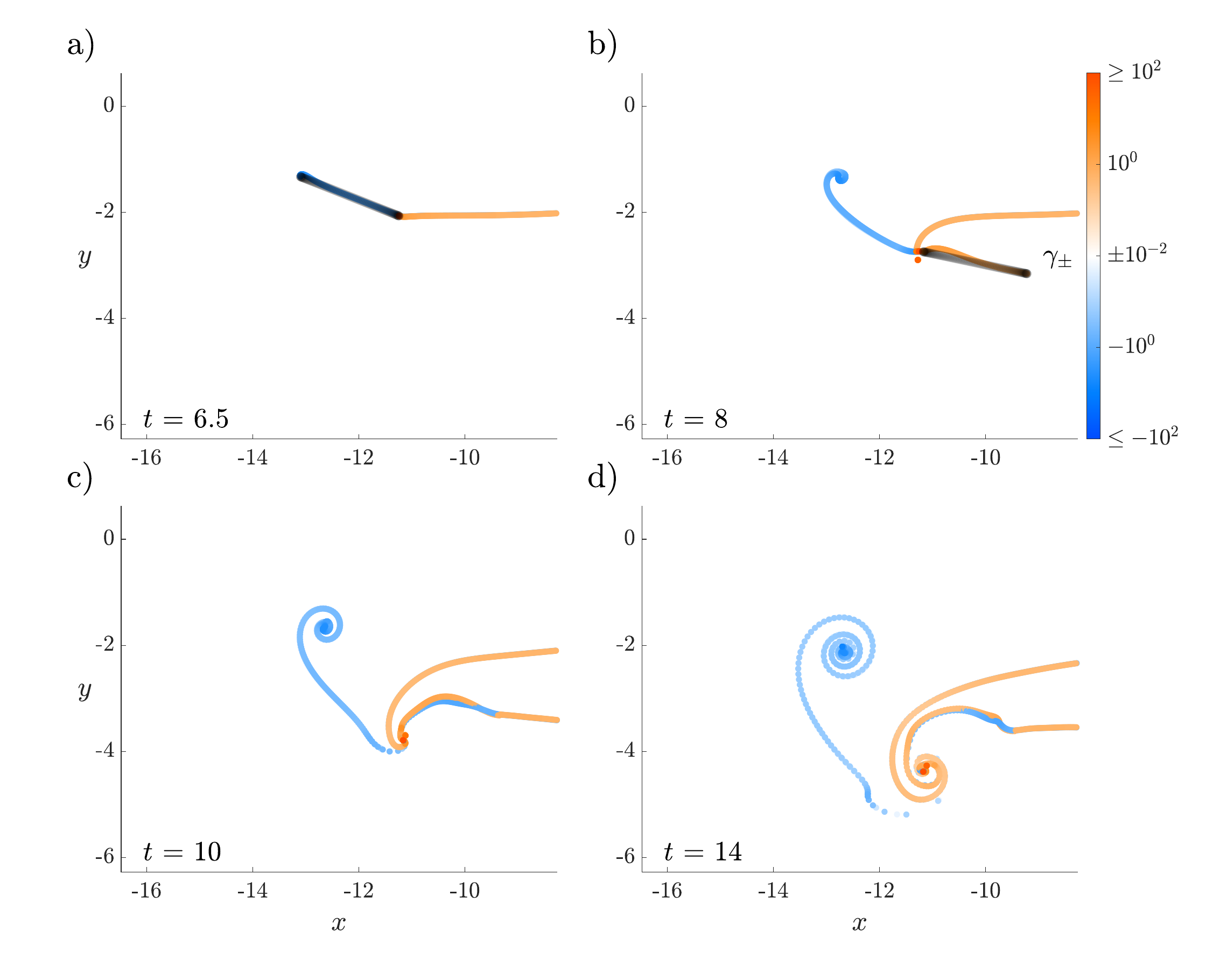}
    \caption{The formation and evolution of a dipole shed from a fluttering plate with $R_1 = 0.2$. Panels (a-d) show snapshots at times $t = 6.5,\ 8,\ 10,\ 14$, and the dipole that forms as the plate reverses direction, with a cusp in its trajectory. The plate was released with initial angle $\beta(0) = 25^\circ$.}
    \label{formation of a dipole figure}
\end{figure}

As $R_1$ increases past $0.1$, the fluttering amplitude begins to increase, resulting in a more regular vortex wake behind the fluttering plates. An example is shown in figure \ref{wake behind 0.3 figure}(b).  Figure \ref{wake behind fluttering schematic figure} shows a schematic diagram of this vortex wake.  It consists of an array of asymmetric dipoles formed by large and small vortex cores. The large vortex is shed from the leading edge as it decelerates, while the smaller vortex is shed from the same edge as the plate begins to accelerate. Figure \ref{formation of a dipole figure} shows an example of the formation of such a dipole, at four successive times in panels (a-d).

The formation of the dipole in figure \ref{formation of a dipole figure} can be explained intuitively as follows. When the plate stops moving leftward and starts moving rightward (panel (a)), the angular momentum of the plate rapidly changes sign \textit{twice}. In tandem with each of these large angular impulses is a corresponding release of vorticity into the fluid. The first sign change occurs just before it reverses direction, and its leading edge (left edge) becomes its trailing edge. Corresponding to this large angular impulse, a large vortex is formed at the left edge. This negative (blue) vortex is shown in panels (a-b) of figure \ref{formation of a dipole figure}. As the plate begins to move rightward, its angular velocity changes sign again. In tandem with this second angular impulse, another vortex is released from the right (leading) edge as it moves. This is shown in panels (b-c) of figure \ref{formation of a dipole figure}. Together, these two vortices from one of the dipoles shown in the schematic shown in figure \ref{wake behind fluttering schematic figure}, and can be seen in panels (c-d) of figure \ref{formation of a dipole figure}. This transient dipole wake structure resembles the photographs of the experiments of \cite{belmonte1998flutter} and can be seen in the supplementary movie 
``\href{https://drive.google.com/file/d/1d0ePShyGJktTPMsPEhrCLNT8uM9sacfY/view?usp=drive_link}{movie\_0.35\_flatplate.avi}".
\begin{figure}[H]
    \centering
    \includegraphics[width=0.95\textwidth, trim=1.9cm 11cm 1.9cm 11cm, clip]{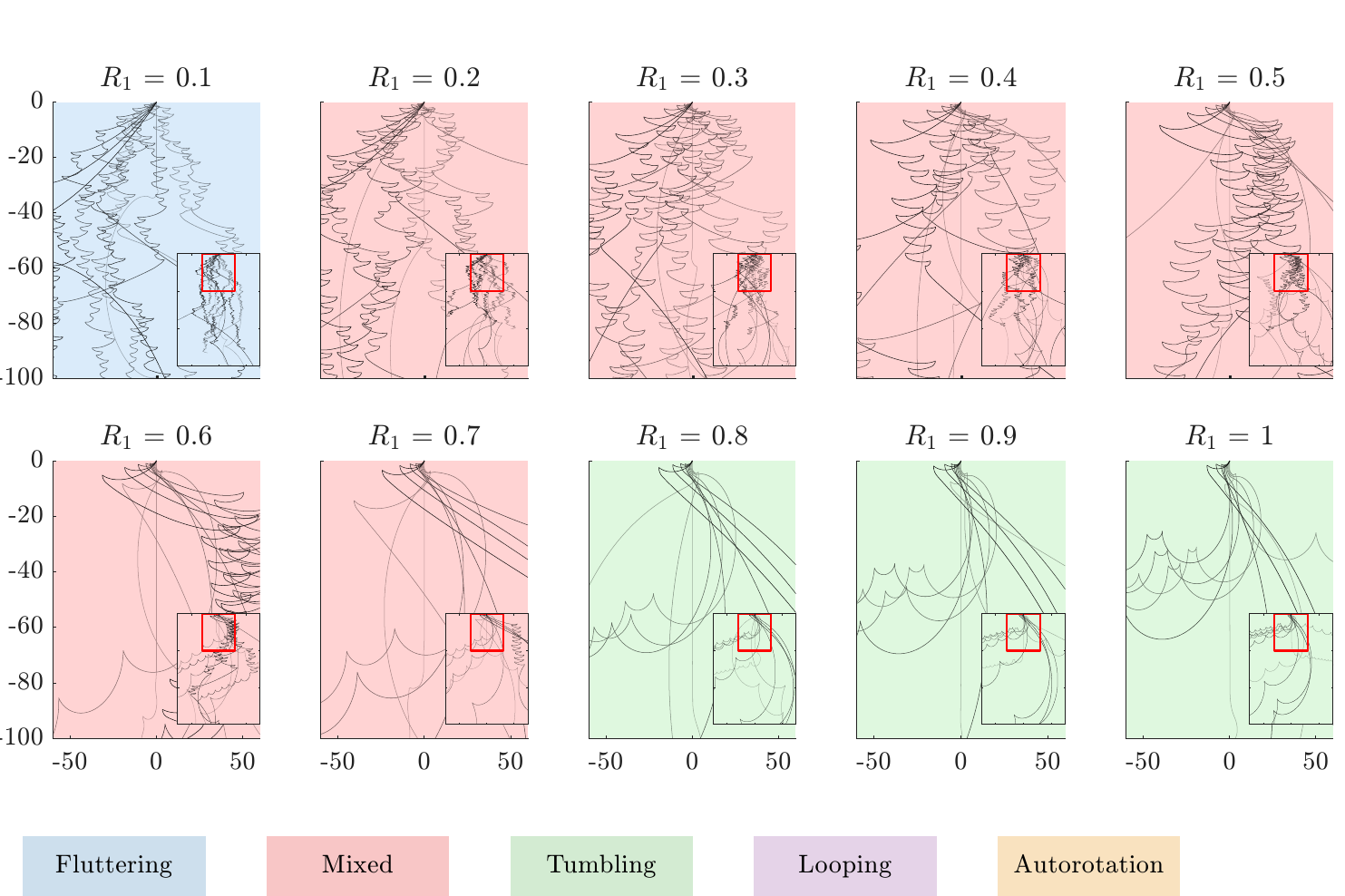}
    \caption{For $R_1 = 0.1, 0.2, \ldots, 1$, snippets of the center-of-mass trajectories of falling plates for eleven initial angles within $[0,\pi/4]$ during $0\leq t \lessapprox 500$. The darker trajectories correspond to larger initial angles. The plots are shaded according to the qualitative dynamics exhibited by the steady state. Each panel contains an inset that shows the large-scale features of the same trajectories, with a red rectangle indicating the region shown in the main panel. }
    \label{plate medium R1}
\end{figure}

Alongside this large-amplitude fluttering, beginning at $R_1 \approx 0.2$ another falling motion emerges: tumbling. This transition is shown in panels $R_1$ = 0.1 and
0.2 of figure \ref{plate medium R1}. The typical wake behind a tumbling trajectory is shown in figure \ref{wake behind 1 figure}, and can be seen more clearly in the supplementary movie ``\href{https://drive.google.com/file/d/1EHKvYWzmVjnOF_NBo3JkeaqUmCdvhucU/view?usp=drive_link}{movie\_1.2\_flatplate.avi}". In this study, a plate is said to be tumbling if, like fluttering, the plate center-of-mass velocity vanishes periodically, but unlike fluttering, its angular velocity never changes sign. As for fluttering, cusps occur in the trajectories for tumbling when the center-of-mass velocity vanishes (e.g.~figure \ref{wake behind 1 figure}(a)). As the density of the plate ($R_1$) increases, the fluid pressure force that restores the plate to the broadside-down position becomes insufficient to overcome the plate's inertia. The plate thus falls end-over-end, and the trajectory is a series of curved arcs joined by cusps. 
\newline

Figure \ref{plate medium R1} shows center-of-mass trajectories for plates with  $0.1 \leq R_1 \leq 1$. Unlike the fluttering motions, whose amplitudes increase as $R_1$ increases (figure \ref{plate medium R1}, $R_1 = 0.1, \ldots, 0.5$), the amplitudes of the tumbling motions generally decrease as $R_1$ increases (figure \ref{plate medium R1}, $R_1 = 0.6, \ldots, 1$). Here, the fluttering amplitude refers to the mean distance traveled during each plate oscillation cycle, while the tumbling amplitude refers to the the mean distance traveled during each plate rotation cycle.
While the fluttering motions are non-periodic and somewhat erratic, all the tumbling motions here are eventually periodic (as we can see from plots of the angular velocity versus time). There is an intuitive explanation for this difference. The dynamics of large clusters of vortices are inherently chaotic \cite{aref1983integrable}. Typically the fluttering plate is close to a large cluster of vortices, and the non-periodicity of the plate motion seems to reflect that of the surrounding vortices. By contrast, a tumbling plate often moves away from its vortex wake into a smoother surrounding flow, allowing for more regular plate dynamics. This can also be seen from the numerical refinement study in appendix \ref{numerical parameters and grid refinement studies} which suggests that among all the falling motions observed, only the tumbling motions are \textit{not} chaotic, in the sense that the other types of motion are highly sensitive to parameters. Within this transition regime, the plates can switch between fluttering and tumbling motions as can be seen in the inset of the $R_1 = 0.5$ panel in figure \ref{plate medium R1}.

\subsubsection{$0.7 \leq R_1 < 1.6$: the pure tumbling regime}
\begin{figure}[H]
    \centering
\includegraphics[width=0.9\textwidth, trim=5cm 1.5cm 5cm 0cm, clip]{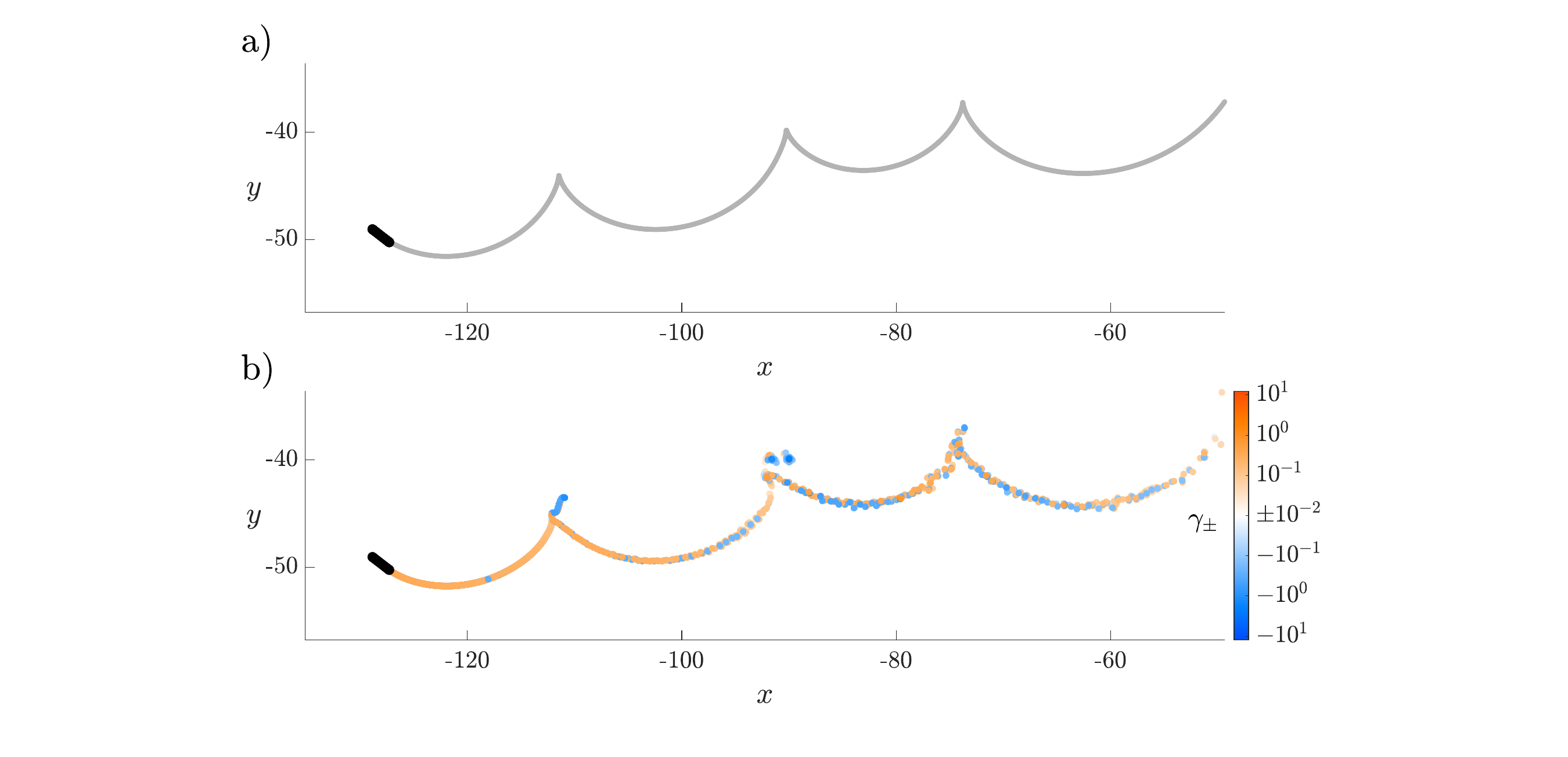}
    \caption{The trajectory of the center of mass (a) and the vortex wake (b) formed by a falling plate with $R_1 = 1$ undergoing tumbling. The plate was released with initial angle $\beta(0) = 25^\circ$.}
    \label{wake behind 1 figure}
\end{figure}

Between $R_1= 0.6$ and $0.7$, steady-state fluttering motions cease,  and only tumbling motions remain. The transition between fluttering and tumbling was found at $I^* \approx 0.3$ in \cite{smith1971autorotating} \cite{wang_numerical} \cite{sohn}, which corresponds to $R_1 \approx 0.7$ here, since $R_1 = \frac{3\pi}{4}I^*$ discussed in \cite{sohn}. It should be noted, however, that unlike these other studies, here we consider a range of initial angles in $[0, \pi/4]$. The fact that the angular velocity of the plate undergoing a tumbling motion never changes sign is what distinguishes it from fluttering. Indeed, this is  what is shown panel (b) of figure \ref{wake behind 1 figure} with the bright orange (positive) vortex sheet released from the trailing edge of the plate as it tumbles. Unlike fluttering, in which vorticity is predominantly shed from the leading edge, in the tumbling motion vorticity is predominantly shed from the trailing edge. This may be one reason that although brief large-amplitude motions similar to looping and tumbling were found in a similar model without leading-edge shedding in \cite{sohn}, no large-amplitude fluttering appeared.
\newline 

\begin{figure}[H]
    \centering
    \includegraphics[width=0.7\linewidth, trim = 0 0.25cm 0 0, clip]{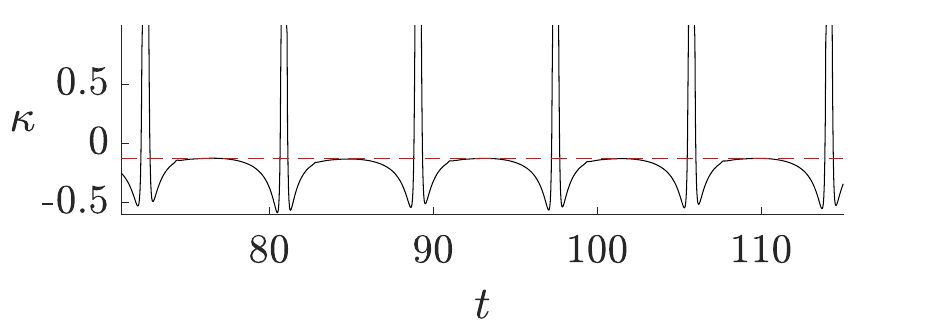}
    \caption{The curvature of the center-of-mass trajectory of a tumbling plate with $R_1 = 1.2$. The red dotted line at $\kappa = -0.125$ shows that the curvature within each tumbling period is relatively constant. The motion of this plate is shown in the supplementary movie ``\href{https://drive.google.com/file/d/1EHKvYWzmVjnOF_NBo3JkeaqUmCdvhucU/view?usp=drive_link}{movie\_1.2\_flatplate.avi}".}
    \label{constant curvature while tumbling figure}
\end{figure}

\begin{figure}[H]
    \centering
    \includegraphics[width=0.7\linewidth, trim=0cm 1cm 0 0cm, clip]{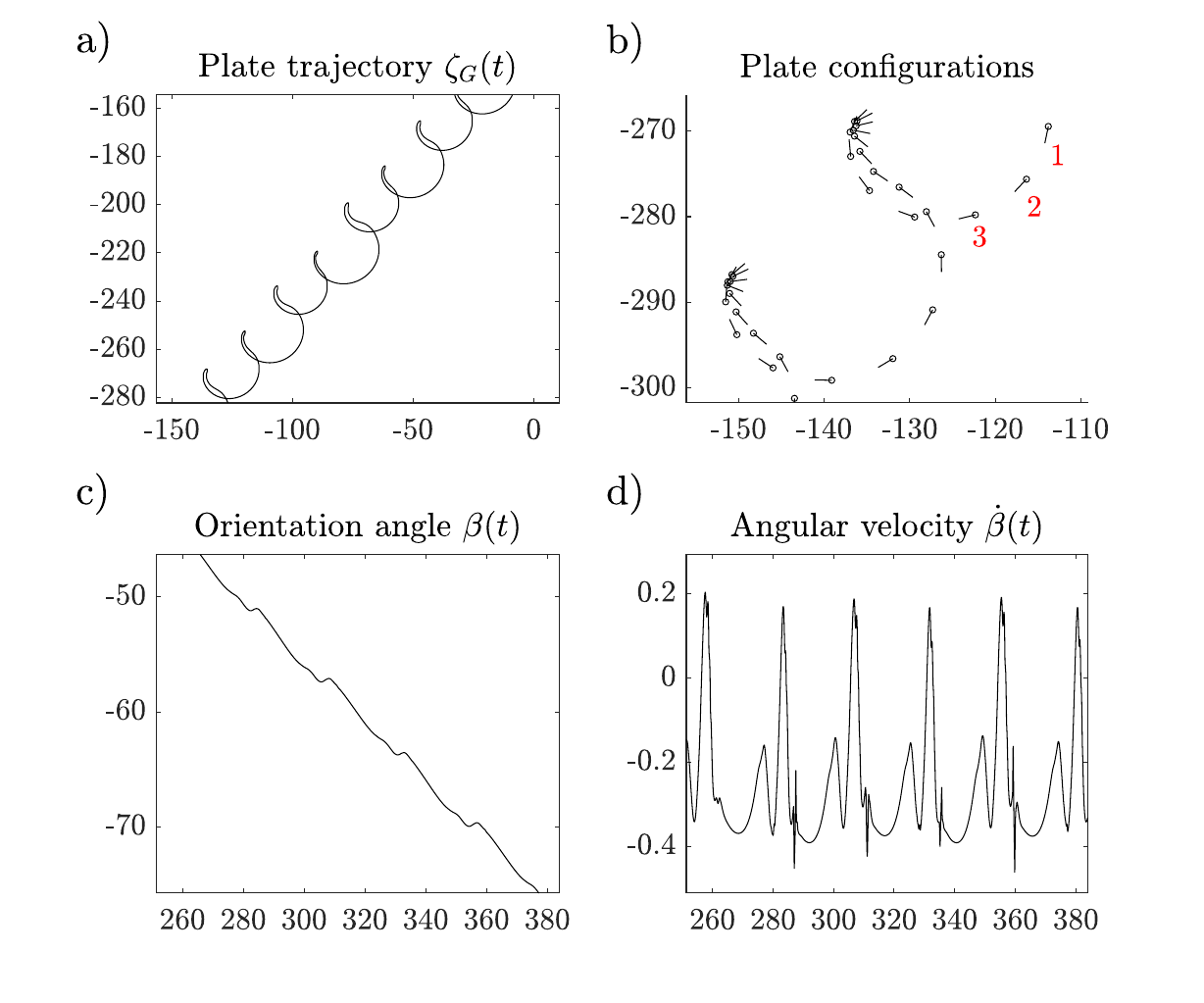}
    \caption{A special falling motion that occurs when $R_1 = 1.4$, for $5$ out of $11$ initial angles simulated.}
    \label{loop_and_tumble_figure}
\end{figure}

Figure \ref{wake behind 1 figure} shows that the maximum magnitude of vortex strength in the wake of a tumbling plate is only $\approx 10$. In contrast, figure \ref{wake behind 0.3 figure} shows that for large amplitude fluttering the maximum vortex strength $\approx 1000$. Intuitively, the release of vorticity corresponds to a change in angular momentum of the plate. Hence, as the plate releases relatively little vorticity while tumbling, we might expect that it should travel with relatively constant angular velocity. If the plate travels approximately tangentially to its trajectory and has constant angular velocity, then the curvature of its trajectory would be relatively constant (i.e.~it would resemble a circular arc). This is seen in figure \ref{wake behind 1 figure} and more clearly in figure \ref{constant curvature while tumbling figure} where the curvature $\kappa$ of the center-of-mass trajectory of a tumbling plate with $R_1 = 1.2$ is shown. The red dotted line at $\kappa = -0.125$ shows that for much of the tumbling period, the curvature is approximately constant. The large spikes in curvature mark the endpoints of each tumbling period, where a curvature singularity (i.e.~a cusp) forms. In
figure \ref{loop_and_tumble_figure}(a), the circular arcs formed by a tumbling plate's trajectory are also readily apparent.

As in the pure fluttering regime, tumbling motions may differ dramatically from one another. Near $R_1 = 1.4$, a special quasi-periodic tumbling motion occurs, shown in figure \ref{loop_and_tumble_figure}. For most initial angles, the plates tumble normally, forming the characteristic sharp cusps in their trajectories (figure \ref{wake behind 1 figure}). For apparently random initial angles, the plate overshoots as it falls, forming a loop as it tips over and slices through its wake and trajectory. Despite the close interaction between the plate and its wake, the system remains quasi-periodic. In panel (a) we show the trajectory of the center of mass and the loops it forms, and in panel (b) we show a close-up of the plate configurations as it performs a loop. We number the first three snapshots in red to indicate the direction of motion. In panels (c) and (d) respectively, we show the orientation angle and angular velocity, which is close to periodic. This hybrid falling motion appears near the transition between the tumbling and looping regimes ($R_1 \approx 1.6)$, and blends the two falling motions.

\subsubsection{$1.6 \leq R_1 < 2.8$: the looping regime}
\begin{figure}[H]
    \centering
    \includegraphics[width=\textwidth, trim=4cm 3.5cm 4cm 2.5cm, clip]{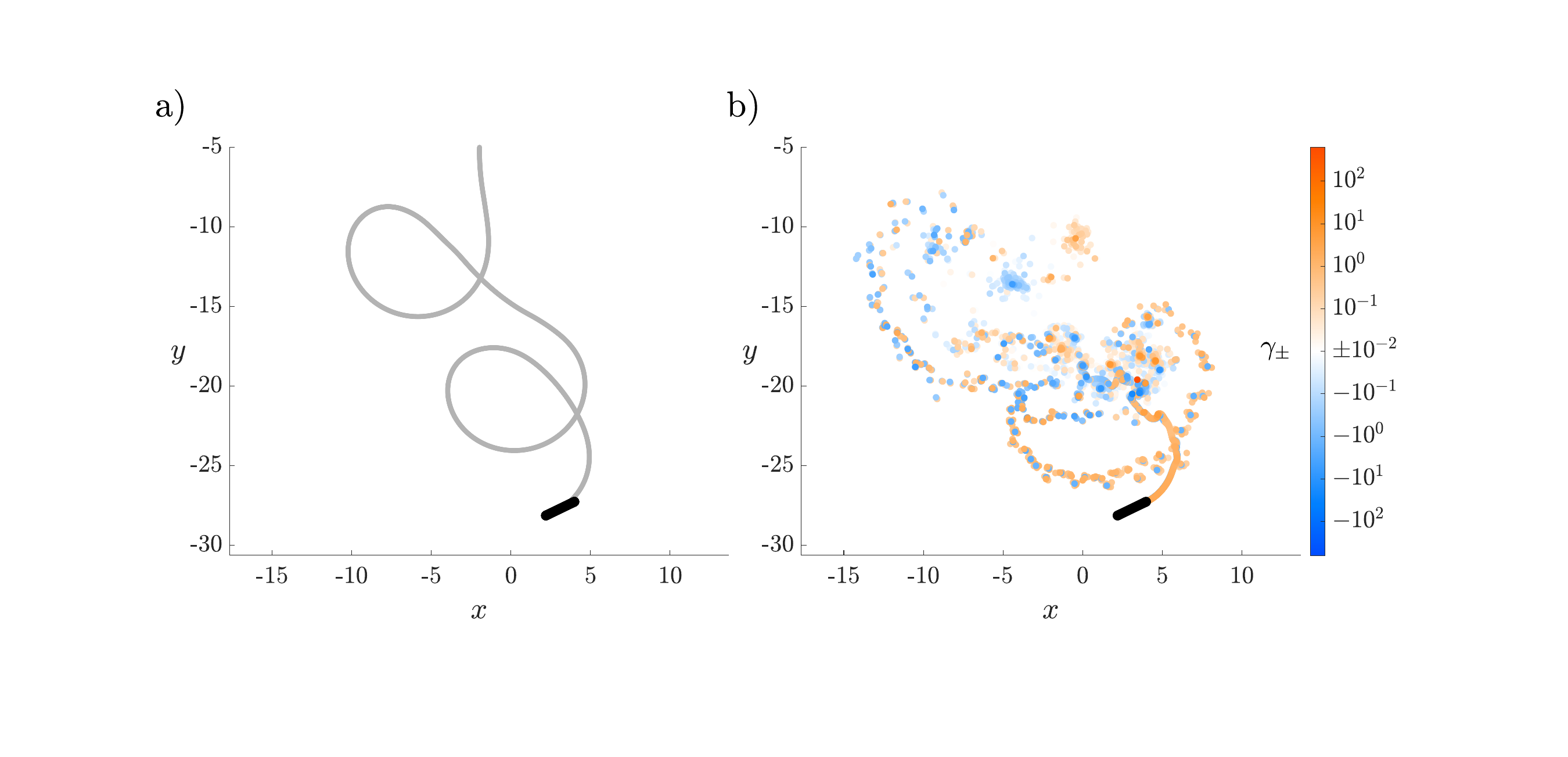}
    \caption{The trajectory of the center of mass (a) and the vortex wake (b) formed by a falling plate with $R_1 = 2.7$ undergoing a looping motion. The plate was released with initial angle $\beta(0) = 25^\circ$.} 
    \label{wake behind 2.7 figure}
\end{figure}
\begin{figure}[H]
    \centering
     \includegraphics[width=0.95\textwidth, trim=1.9cm 11cm 1.9cm 12cm, clip]{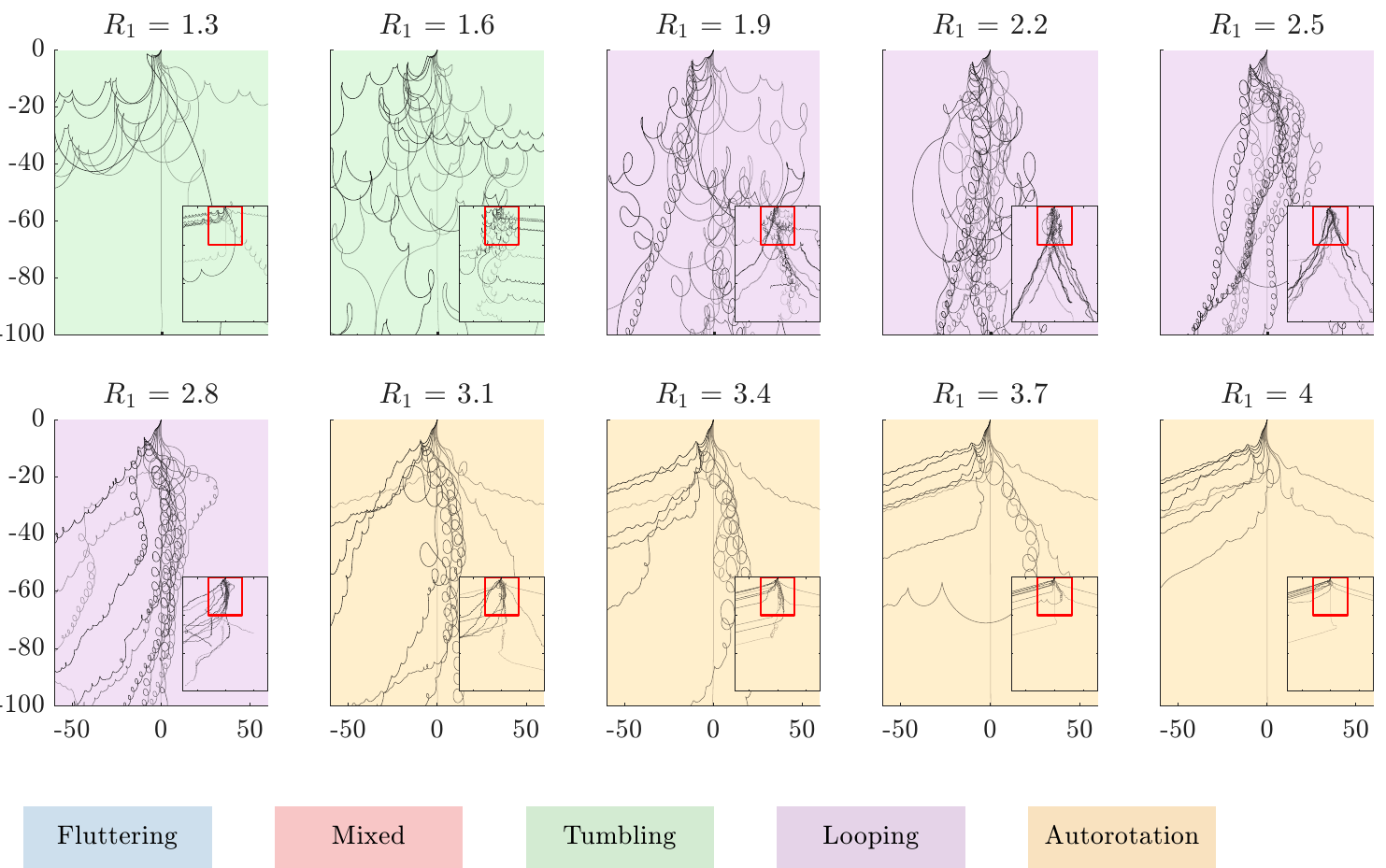}
    \caption{For $R_1 = 1.3, 1.6, \ldots , 4$, snippets of the center-of-mass trajectories of falling plates for eleven initial angles within $[0,\pi/4]$ during $0\leq t \lessapprox 500$. The darker trajectories correspond to larger initial angles. The plots are shaded according to the qualitative steady-state dynamics. In each panel an inset shows the large-scale features of the trajectories, with a red rectangle indicating the region shown in the main panel.}
    \label{large_R1_figure}
\end{figure}

As $R_1$ approaches 1.6, the stable periodic tumbling trajectories give way to irregular, erratic looping motions. Unlike tumbling motions, which do not change horizontal direction, these looping motions do, as in figure \ref{wake behind 2.7 figure}(a) where the trajectory shows loops. In this study, a plate is said to be undergoing a looping motion if its angular velocity never changes sign, but its horizontal velocity does. The fact that a looping body's horizontal velocity changes sign distinguishes it from tumbling. This motion resembles the ``circling state" observed for falling flexible sheets \cite{albenFall}. There, the flexible body bends into a circular arc with curvature approximating that of its looping trajectory. By conforming its body curvature to its trajectory, the flexible body is able to stabilize this falling motion. Instead of looping chaotically, the flexible body instead loops in a quasi-periodic fashion for extended time periods. As $R_1$ increases in the range $[1.6,2.8]$, the loop radius shrinks (figure \ref{large_R1_figure}, $R_1 = 1.9,\ 2.2,\ 2.5,\ 2.8$) until the plate begins to autorotate. 
\newline

\begin{figure}[H]
    \centering
    \includegraphics[width=1\textwidth, trim = 3cm 1cm 3.5cm 1.5cm,clip]{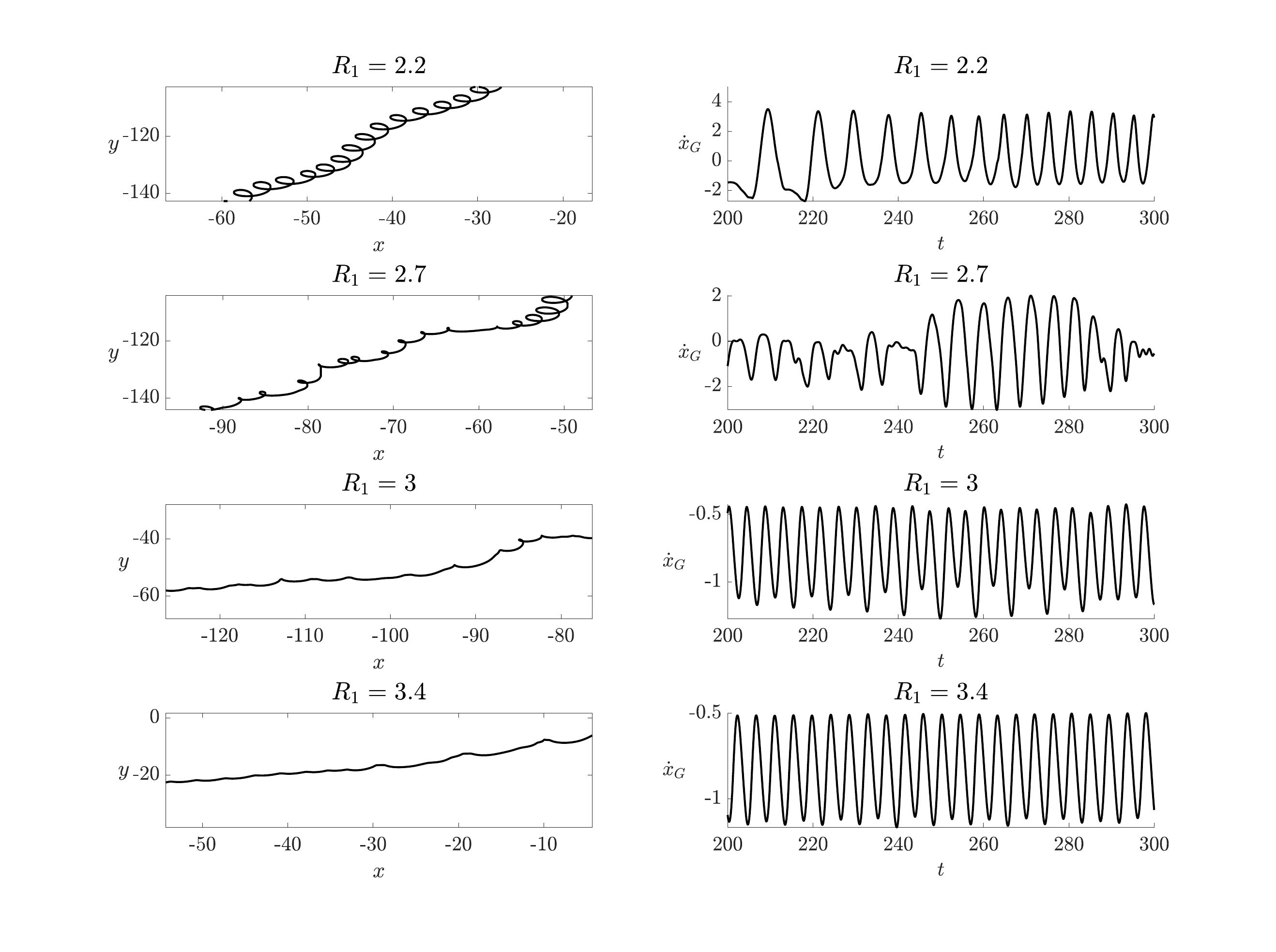}
    \caption{Motions at $R_1 = 2.2,\ 2.7,\ 3,\ 3.4$ showing the transition between looping and autorotation. Left column: snippets of center-of-mass trajectories. Right column: time series of horizontal velocity  $\dot x_G(t)$.}
    \label{transition between looping and autorotation figure}
\end{figure}

The transition between looping and autorotation is shown by motions at a sequence of $R_1$ values in figure \ref{transition between looping and autorotation figure}. 
Here we define autorotation as a plate motion in which both angular and horizontal velocities never change sign. The single-signed horizontal velocity distinguishes autorotation from tumbling and looping.
At $R_1 = 2.7$ (second row of figure \ref{transition between looping and autorotation figure}), near the transition between fluttering and tumbling, the horizontal velocity $\dot x_G(t)$ (right column) oscillates with a single sign for much of the first half of the time series (autorotation) but often reverses sign during the second half (looping). When $R_1 > 2.8$ (figure \ref{transition between looping and autorotation figure}, $R_1 = 3,\ 3.4$) the $\dot x_G(t)$ time series is single-signed, showing autorotation only.
As $R_1$ increases within this regime, the time-averaged horizontal velocity magnitude also increases.

\subsubsection{$2.8 < R_1$, The autorotation regime}
\begin{figure}[H]
    \centering
    \includegraphics[width=1\textwidth, trim=8cm 1.5cm 8cm 0.5cm, clip]{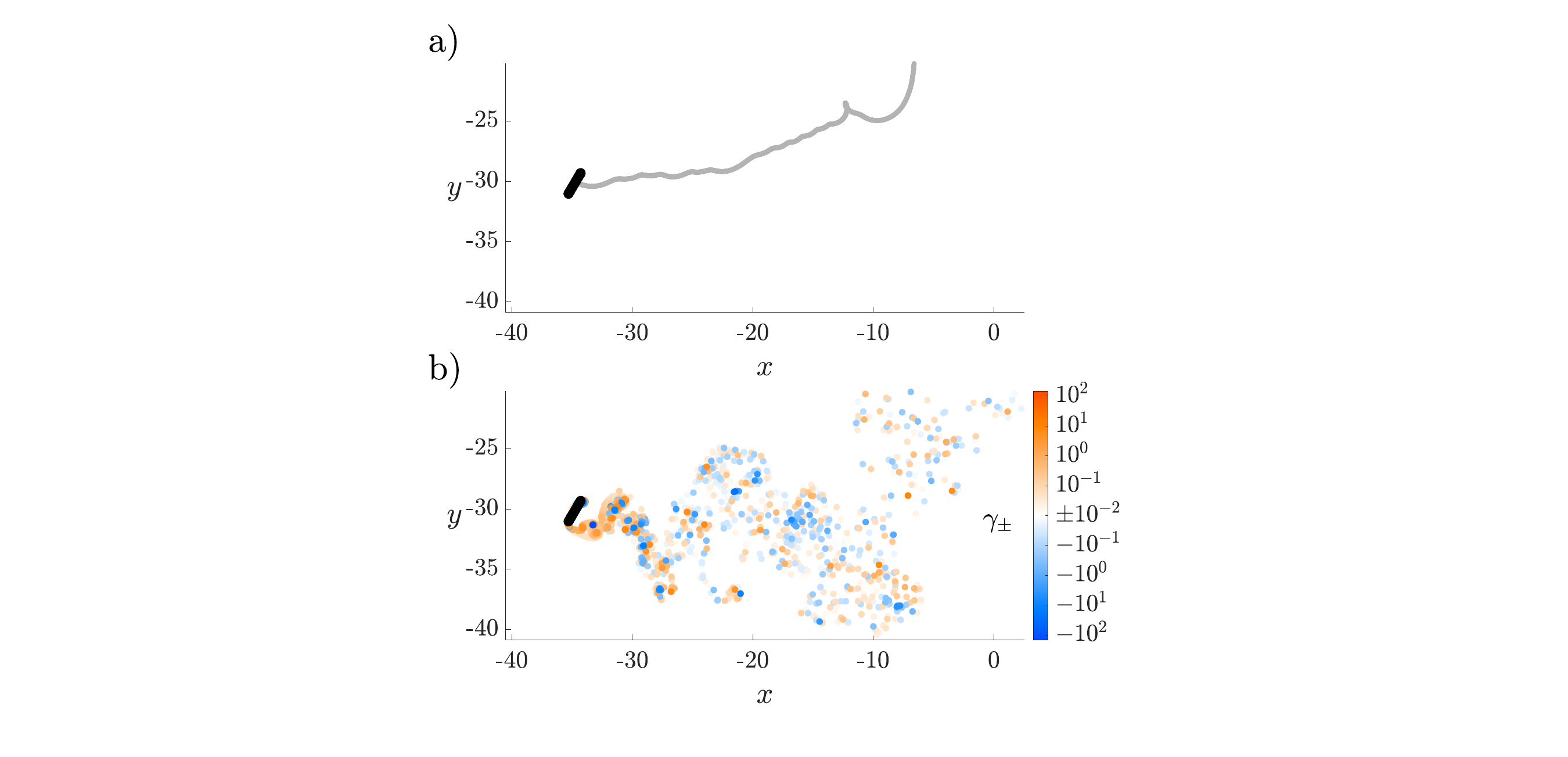}
    \caption{The center-of-mass trajectory (a) and vortex wake (b) of an autorotating plate with $R_1 = 4$. The plate was released with initial angle $\beta(0) = 25^\circ$.}
    \label{wake behind 4 figure}
\end{figure}

\begin{figure}[H]
    \centering 
\includegraphics[width=1\textwidth, trim=0cm 0cm 0cm 0.5cm, clip]{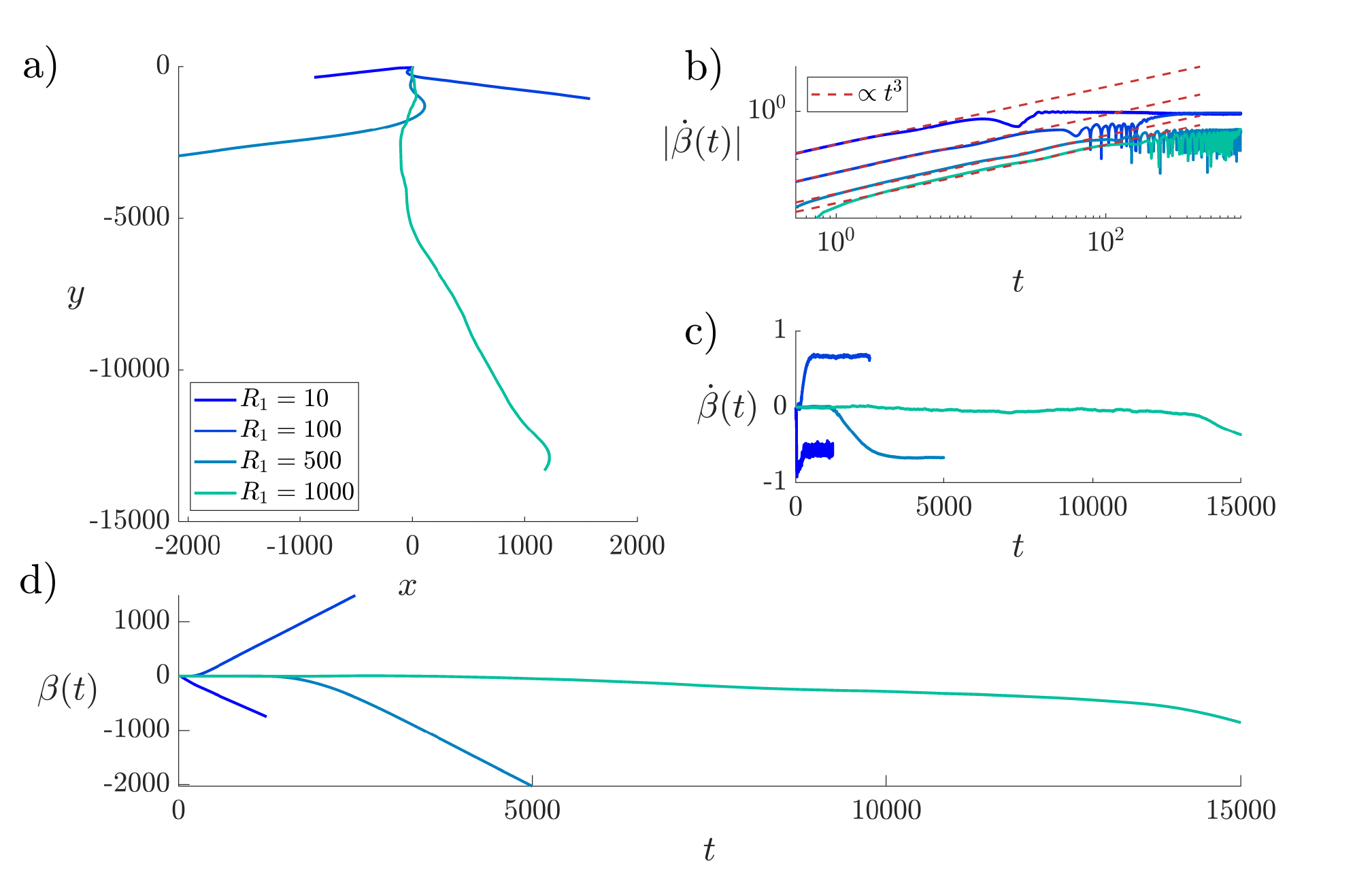}
    \caption{(a) The trajectory of the center of mass $\zeta_G(t)$, (b) angular velocity magnitude $|\dot\beta(t)|$ on logarithmic scale with red dotted lines showing curves proportional to $t^3$, (c) the angular velocity, and (d) the orientation angle $\beta(t)$ of falling plates released at a fixed initial angle $\beta(0) = 25^\circ$ with $R_1 = 10,\ 100,\ 500,\ 1000$. }
    \label{very large R1 behaviour figure}
\end{figure}

For $R_1 > 2.8$ the plate primarily autorotates quasi-periodically. The typical vortex wake behind an autorotating plate is shown in figure \ref{wake behind 4 figure}. In figure \ref{very large R1 behaviour figure} we show the effect of increasing $R_1$ from 10 to 1000, within the autorotation regime. In this wide range of $R_1$ the plates reach similar steady-state trajectories, shown in panel (a) (except for $R_1 = 1000$, not yet at steady state). Panel (b) shows that shortly after being released from rest ($t \approx$~0.1--1), the plates' angular velocity magnitudes increase as a power law $\propto t^3$ before saturating at $O(1)$ values across all tested $R_1$ values. 
The duration of the transient growth phase increases with increasing $R_1$.
In the saturated regime, the
angular velocity magnitude has oscillations whose amplitudes decrease with $R_1$ (as shown in panel (c)), but whose time averages are approximately $0.65$, independent of $R_1$ in this regime.

\subsubsection{Summary of flow characteristics.}
\begin{figure}[H]
    \centering
    \includegraphics[width=1\linewidth]{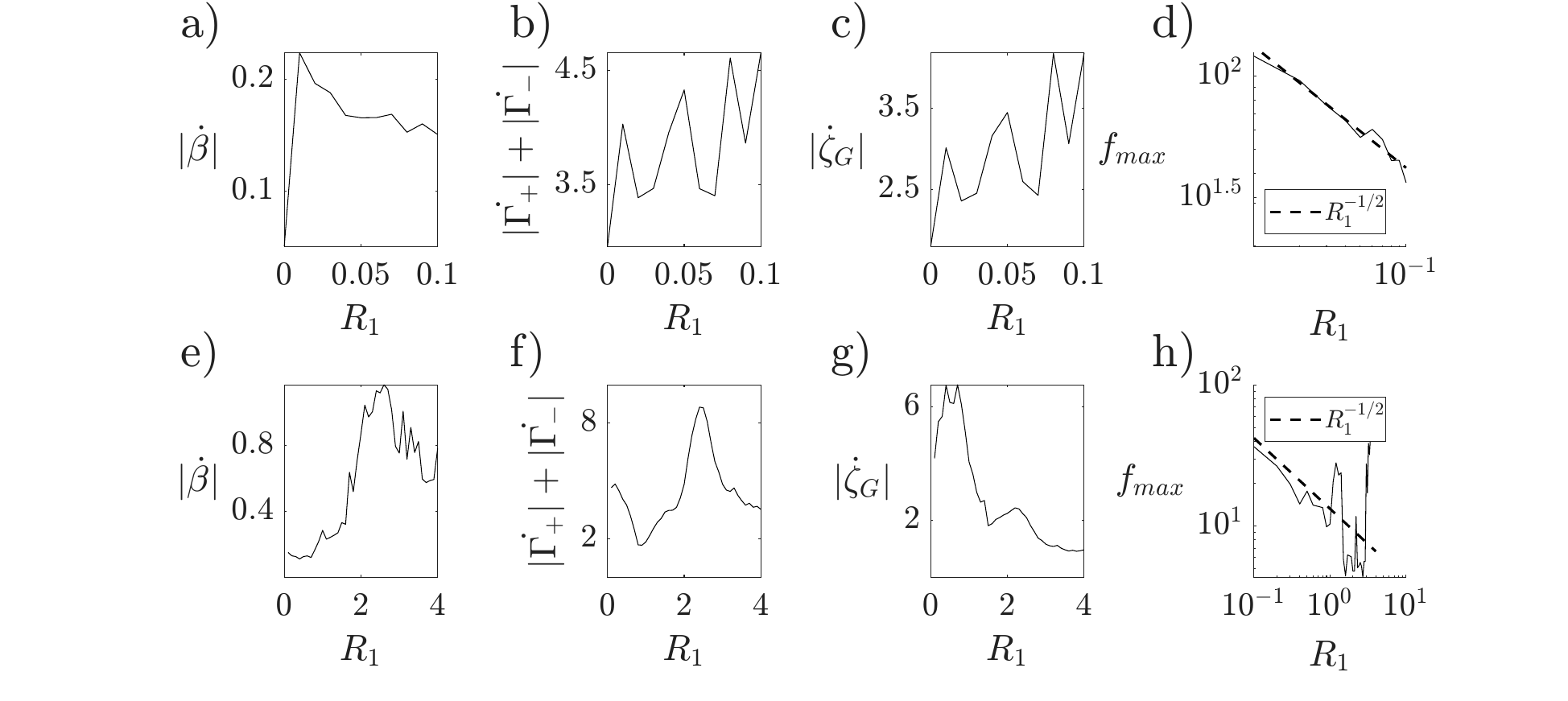}
    \caption{Dynamical quantities versus $R_1$ at small and large $R_1$ values (top and bottom rows respectively): (a,e) angular velocity $|\dot\beta|$, (b,f) circulation change $|\dot\Gamma_+| +|\dot\Gamma_-|$, (c,g) center-of-mass speed $|\dot\zeta_G|$, (d,h) angular velocity frequency $f_{max}$. Top row: $0 \leq R_1 \leq 0.1$. Bottom row:$0.1 \leq R_1 \leq4$}
    \label{key flow characteristics figure}
\end{figure}

Figure \ref{key flow characteristics figure} presents key flow characteristics across $R_1$ and the five regimes comprising fluttering ($0 \leq R_1 < 0.2$), mixed ($0.2 \leq R_1 <0.7$), tumbling ($0.7 \leq R_1 < 1.6$), looping ($1.6 \leq R_1 < 2.8$) and autorotating ($R_1 \geq 2.8$) falling motions.  We examine two parameter ranges: $0\leq R_1 \leq 0.1$ (panels (a–d), first row) and $0.1\leq R_1 \leq 4$ (panels (e–h), second row). Each column displays a different flow metric: the average magnitude of angular velocity $|\dot\beta|$, the average rate of circulation change $|\dot\Gamma_+| +|\dot\Gamma_-|$, the average center-of-mass speed $|\dot\zeta_G|$, and the spectral peak frequency of the angular velocity, $f_{max}$. To calculate $f_{max}$ we compute the power spectrum (Fourier transform) of the angular velocity time series at each initial angle, compute the frequency where the power spectrum is maximum, and average over the initial angles, obtaining one 
$f_{max}$ value at each $R_1$ value.
\newline
Panels (a) and (e) show that $|\dot\beta|$ is relatively small ($\approx 0.2$) in the fluttering and tumbling regime, but rapidly jumps to $\approx1$ in the looping regime before decreasing to $\approx 0.8$ in the autorotation regime. By comparing panel (b) to (c) and panel (f) to (g), we observe that when $R_1$ is small ($R_1\leq 0.1$), the average rate of circulation change and the speed of the plate are strongly correlated, but are much less so when $R_1$ is larger ($R_1 > 0.1$). This makes sense intuitively since $R_1 = \frac{\rho_b h}{\rho_f L}$ is the relative density of the plate. Hence when $R_1 \ll 1$ we expect fluid forces to dominate the plate motion (so its speed would be correlated with the circulation change in the fluid), and when $R_1 \gg 1$ we expect inertial effects to dominate instead. As shown in panels (d,h), the angular velocity oscillation frequency $\sim R_1^{-1/2}$ in the fluttering regime $0\leq R_1 <0.2$.  This scaling behavior is explored more broadly for V-shaped plates in section 4.2.3 as they are able to flutter for a larger range of $R_1$. This scaling law begins to break down at larger $R_1$ values (e.g. $R_1 = 0.3$) as the plate motions transition from fluttering to tumbling.

\subsection{V-shaped plates}
\label{vplate section}
\subsubsection{Physical and numerical parameters}
\vspace{-0.5cm}
\begin{figure}[H]
    \centering
\includegraphics[width=0.4\linewidth, trim = 0cm 4cm 0cm 2cm, clip]{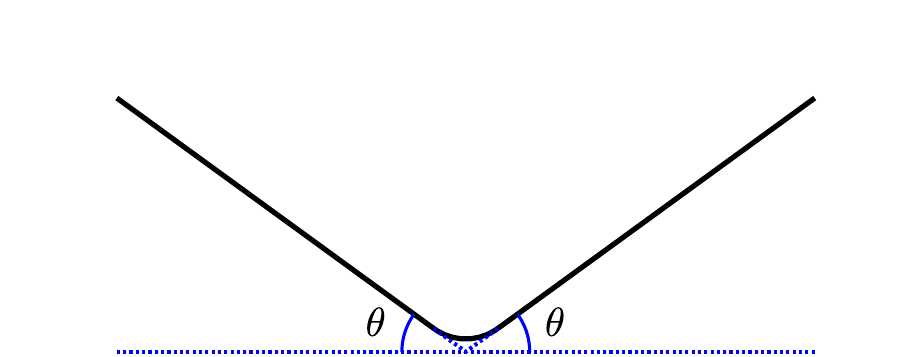}
    \caption{Illustration of the bending angle $\theta$.}
    \label{bending angle diagram}
\end{figure}

Having observed the effect of $R_1$ on the qualitative dynamics of falling flat plates, we extend our study to curved, V-shaped plates obtained by bending the flat plate symmetrically about its center. The main purpose of studying V-shaped plates is to show the effect of a simple change in plate shape on the motions. Such a transformation breaks the symmetry of the $\pm$ sides of the plate and introduces new falling dynamics while suppressing others. V-shaped plates can exhibit most of the falling behaviors observed in flat plates---specifically fluttering, looping, and autorotation---but notably cannot tumble. Unlike the flat plates, the V-shaped plates are also able to flutter periodically in a downward direction (i.e.~with zero average horizontal velocity). We consider a one-parameter family of bent, V-shaped plates indexed by $\theta$, the bending angle shown in figure \ref{bending angle diagram}. The sharp tip of the $V$ shape is smoothed by replacing it with a circular arc with radius $\frac{2}{5\pi}$. The bending angle $\theta$ is equal to half the exterior angle of the plate. This class of plates is unable to move in a purely tangential motion, so we now omit skin friction, which was introduced to suppress such motions for the flat plate. Hence, unlike that of the flat plate, the model for the V-shaped plates is purely inviscid. The results in what follows are drawn from the analysis of $\approx 1000$ simulations up to $t \approx 500$. In table \ref{V-plate parameters table} we summarize the numerical parameters considered in this portion of the study.

\begin{table}[H]
    \centering
    \renewcommand{\arraystretch}{1.2}
    \begin{tabular}{|c|c|}
        \hline
         Initial angles & $0^\circ, 4.5^\circ, 9^\circ, 13.5^\circ, \ldots, 45^\circ$  \\
         \hline
         $R_1$ & $0,\ 0.1,\ \ldots,\ 1,\ 2,\ \ldots,\ 5$\\ 
         \hline
         $\theta$ & $11.25^\circ,\ 15.47^\circ,\ 19.69^\circ,\ 23.91^\circ,\ 28.12^\circ,\ 32.34^\circ,\ 36.56^\circ,\ 40.78^\circ,\ 45^\circ$ \\
         \hline
    \end{tabular}
    \caption{Table showing the values of the physical parameters being varied for the falling V-plates. The bending angles are given by $9$ evenly spaced values between $\pi/16$ and $\pi/4$ radians.}
    \label{V-plate parameters table}
\end{table}

\subsubsection{The effect of bending angle}
\begin{figure}[H]
    \centering
    \includegraphics[width=1\textwidth, trim=0.5cm 0cm 2cm 0cm, clip]{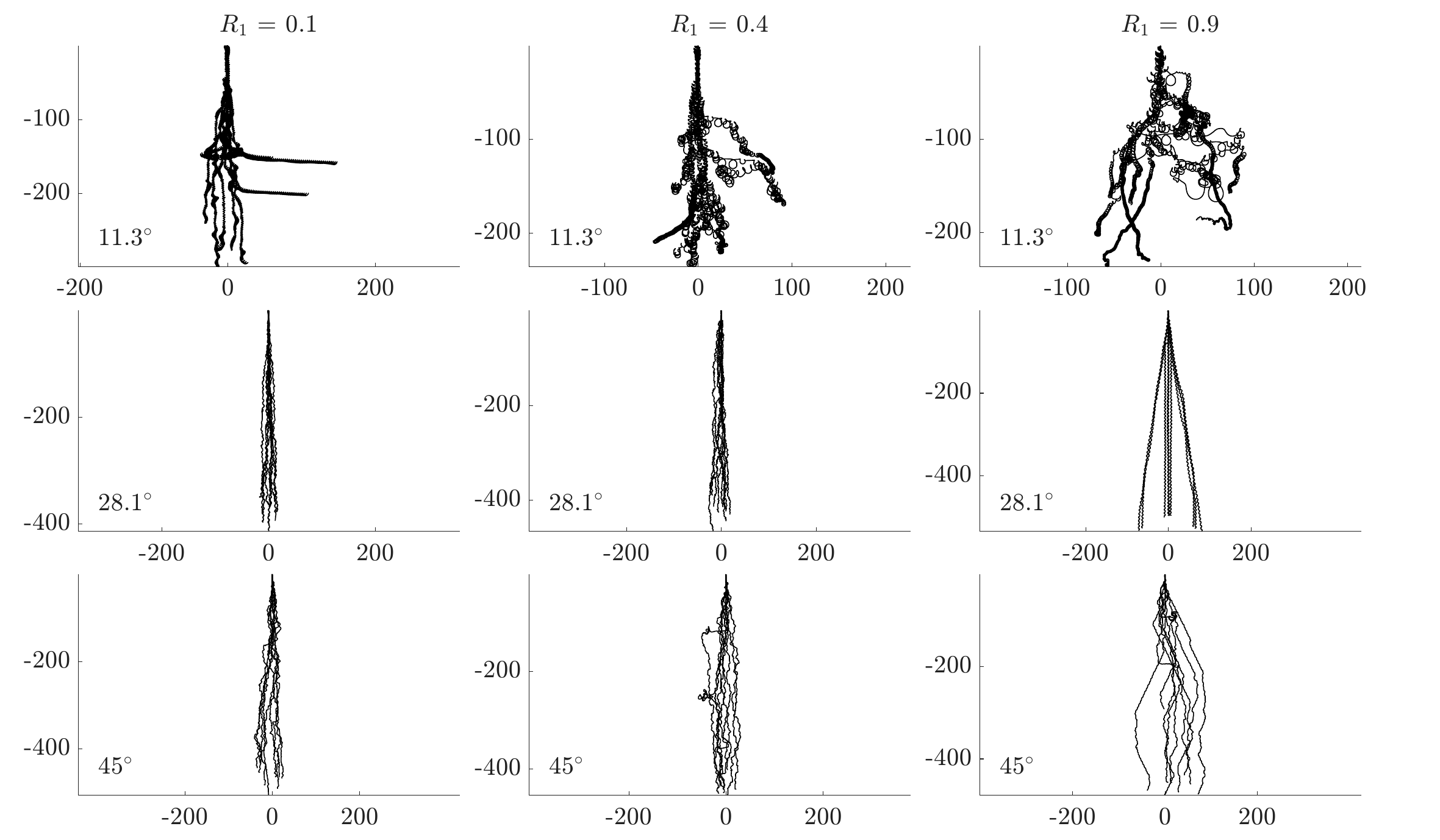}
    \caption{The center-of-mass trajectories of falling bent plates for $0\leq t \leq 500$, $R_1 = 0.1, 0.4, 0.9$, and bending angles $11.3^\circ, 28.1^\circ,$ and $45^\circ.$}
    \label{VPlate Small R1}
\end{figure}

First, we examine global trends in how V-shaped plates fall, with particular attention to the effect of bending angle. In figure \ref{VPlate Small R1}, we show the center-of-mass trajectories of various falling V-shaped plates for $0\leq t \leq 500$. Each column corresponds to a single value of $R_1 \in [0.1, 0.9]$, with the bending angle $\theta$ (listed at the bottom left of each panel) increasing downward. For these small values of $R_1$, the flattest plates ($\theta \leq 11.3^\circ$) behave chaotically, forming loops instead of fluttering. This is significantly different from what is observed for the flat plate. Without the presence of skin friction, the plate has to be either bent sufficiently, $\theta \geq 23.1^\circ$, or made sufficiently less dense, $R_1 \leq 0.1$, before the restoring force induced by the pressure alone is sufficient to induce flutter. Due to increased pressure drag, the more bent plates typically flutter instead, with quasi-periodic and sometimes periodic dynamics. The periodic dynamics correspond to approximately straight-line trajectories (e.g.~$R_1 =0.9$, $\theta = 28.1^\circ$). During each period, the body travels fixed distances $d_v$ and $d_h$ in the vertical and horizontal directions, yielding a straight-line path with slope $d_v/d_h.$
\newline

\begin{figure}[H]
    \centering
    \includegraphics[width=1\textwidth, trim=1cm 0cm 2cm 0cm]{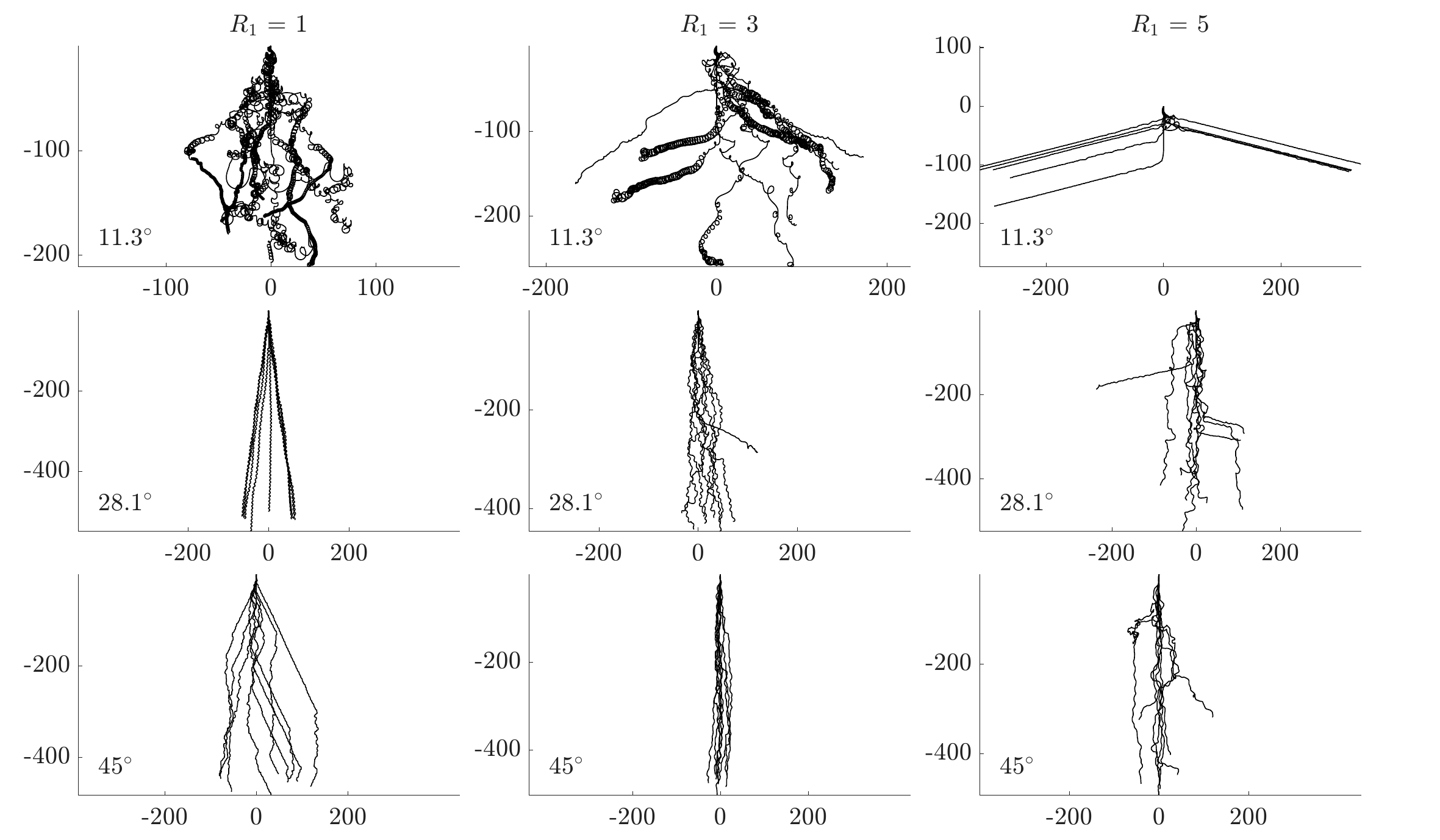}
    \caption{The trajectories of falling bent plates up to $t \approx 500$ for $R_1 = 1, 3, 5$ (one per column) and bending angles $\theta = 11.3^\circ,\ 28.1^\circ,\ 45^\circ$ (one per row, shown at the bottom left of each panel).}
    \label{VPlate Large R1}
\end{figure}

Figure \ref{VPlate Large R1} is analogous to figure \ref{VPlate Small R1} except with larger $R_1$, $\in [1, 5]$. 
In this $R_1$ range, the plates with small bending angle (top row) transition from chaotic motions at smaller $R_1$ to periodic autorotating motions at larger $R_1$. However, for larger bending angles (second and third rows), the plates transition from 
quasi-periodic fluttering at smaller $R_1$ to chaotic motions at larger $R_1$.
In the supplementary movie ``\href{https://drive.google.com/file/d/1hL_ran9ih_fGl7xXPKio2hVRLc6nAfKc/view?usp=drive_link}{movie\_5\_11.3\_vplate.avi}" the periodic autorotation of a V-shaped plate at large $R_1$ is shown.
Both figures \ref{VPlate Small R1} and \ref{VPlate Large R1} seem to suggest that increasing the bending angle stabilizes fluttering motions at small $R_1$ and decreasing the bending angle stabilizes autorotating motions at large $R_1$. This trend is more apparent in the expanded versions of figures \ref{VPlate Small R1} and \ref{VPlate Large R1} located in appendix \ref{com trajectories of v shaped plates appendix}.

\begin{figure}[H]
    \centering
    \includegraphics[width=1\textwidth, trim=0.5cm 0.3cm 2cm 0, clip]{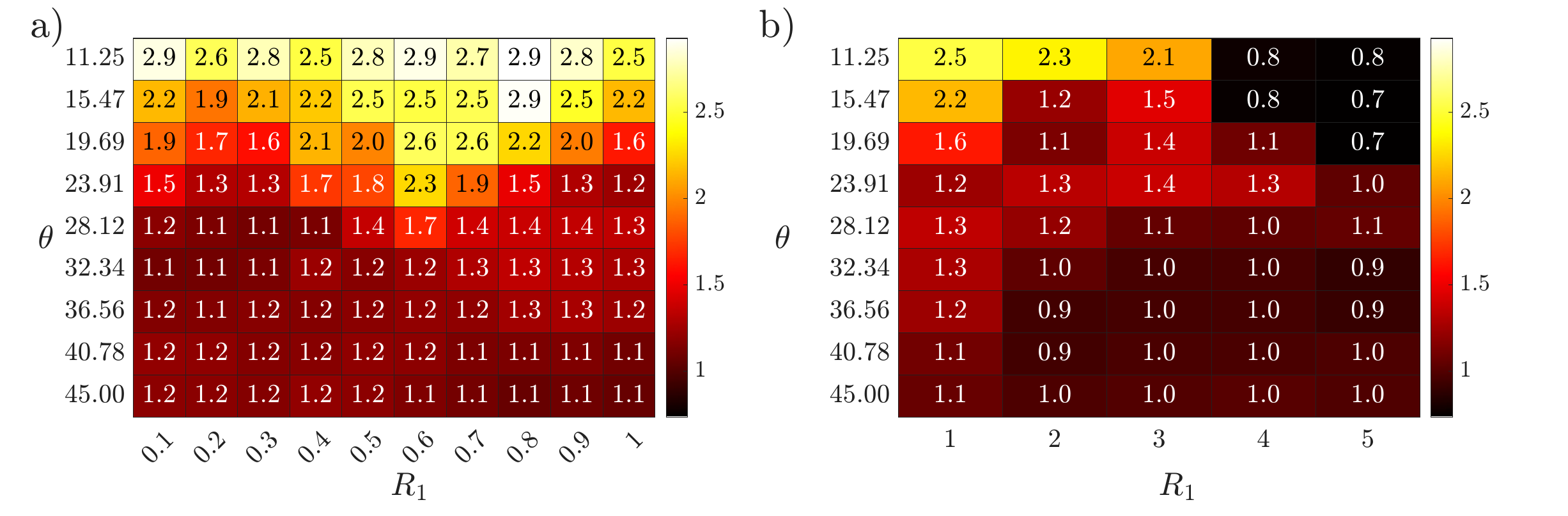 }
    \caption{Heat maps showing the average center-of-mass speed  versus $R_1$ and bending angles $\theta$ for the bent plates, with $0.1 \leq R_1 \leq 1$ (a) and $1 \leq R_1 \leq 5$ (b).}
    \label{Heat map average speed}
\end{figure}

An important quantity that varies strongly among the falling motions is the plate's average speed. This shown by the heat maps of  figure \ref{Heat map average speed}, in the space of $R_1$ and the bending angle $\theta$.
Generally, white, yellow, and orange correspond to (fast) chaotic looping motions, dark red corresponds to (slower) periodic fluttering, and (even slower) black  corresponds to periodic autorotation.
The regions of relatively constant speeds $\approx 1-1.2$ for larger $\theta$ and $R_1$ have similar slow fluttering motions. The heatmap in figure \ref{Heat map average speed} thus represents a phase diagram of the falling V-shaped plates. The proximity of the average speeds to $1$ is consistent with our nondimensionalization, which used an estimate of the terminal velocity as the characteristic speed.

\begin{figure}[H]
    \centering
\includegraphics[width=1.1\textwidth, trim = 2cm 0cm 0cm 0cm]{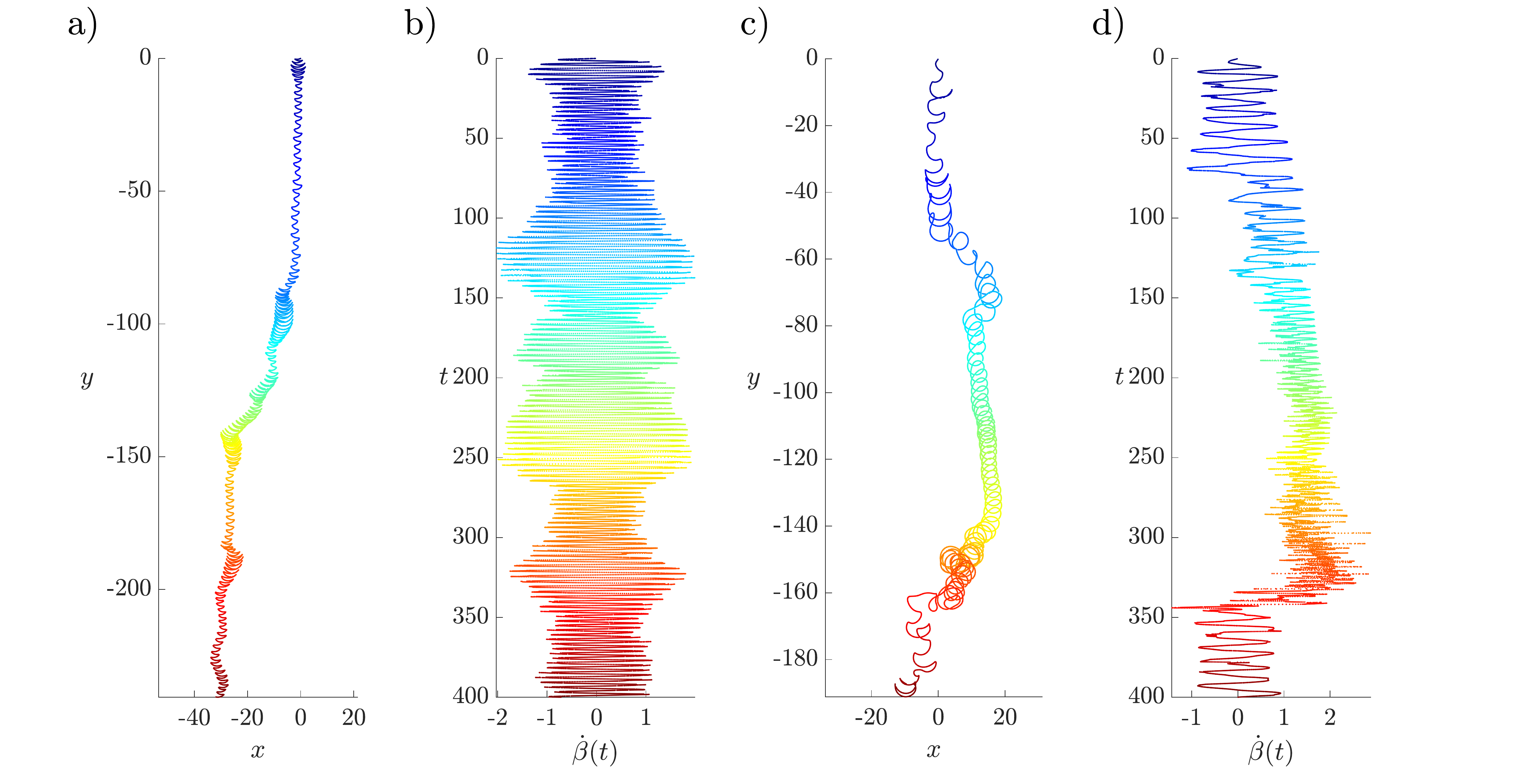}
    \caption{(a),(c) Center-of-mass trajectories and (b),(d) angular velocities of V-shaped plates with densities $R_1 = 0.1$ and $0.7$ respectively, and bending angle $\theta = 11.3^\circ$, for $0 \leq t \leq 500$. In each panel, the colors correspond to time, with the mapping given by the vertical axes of panels (b) and (d).}
    \label{v-plate looping}
\end{figure}

We now examine the model's behavior in greater detail with particular attention to when $R_1$ is small. For such $R_1$, the plate flutters for a range of $\theta \leq 11.3^\circ$,  as in figure \ref{VPlate Small R1} as well as figure \ref{v-plate looping}(a). In this figure, we compare the center-of-mass trajectories (panels (a) and (c)) and the angular velocities (panels (b) and (d)) for plates with $R_1 = 0.1$ and 0.7, and bending angle $11.3^\circ$. The plots are colored by time $t$. Interestingly, the bent plates flutter with an angular velocity whose amplitude varies in time, as shown in panel (b). By contrast, the angular velocity amplitude is approximately constant for flat-plate fluttering. For the example in figure \ref{v-plate looping}(b), the angular velocity amplitude is small near $t = 50$ (i.e.~slow fluttering) and large near $t = 125$ (i.e.~fast fluttering). 

The gradual increase in angular velocity from $t$ = 50 to 150 corresponds to the accumulation of vorticity behind the plate. This can be seen more clearly in the supplementary movie ``\href{https://drive.google.com/file/d/1OGfW0hYRWXaRih7rFVNHoJieoN6cjzK2/view?usp=drive_link}{movie\_0.1\_11.3\_vplate.avi}". In response, the plate oscillates faster and faster until it escapes from its wake, and then flutters slowly again near $t = 160$ before repeating the process, alternating between slow and fast fluttering at irregular time intervals. As $R_1$ increases past $0.2$, the fluttering amplitude increases until the plate begins to overturn and then loop chaotically, as shown in figure \ref{v-plate looping}(c). The main point of figure \ref{v-plate looping} and the associated discussion is that without the restoring force of skin friction that stabilizes the fluttering motion, the V-shaped plates with small bending angle loop instead of flutter as the flat plates do. To summarize, for V-shaped plates with  $\theta \leq 11.3^\circ$, the fluttering state is inherently stable for $R_1 \leq 0.1$, but for larger values of $R_1$ fluttering is unstable without skin friction. Sometimes, the looping plate spontaneously returns to fluttering, as shown at $t \approx 400$ in figure \ref{v-plate looping}(c).

\begin{wrapfigure}[27]{l}{0.55\textwidth}
\includegraphics[width= 0.55\textwidth, trim = 2cm 0cm 0cm 0cm]{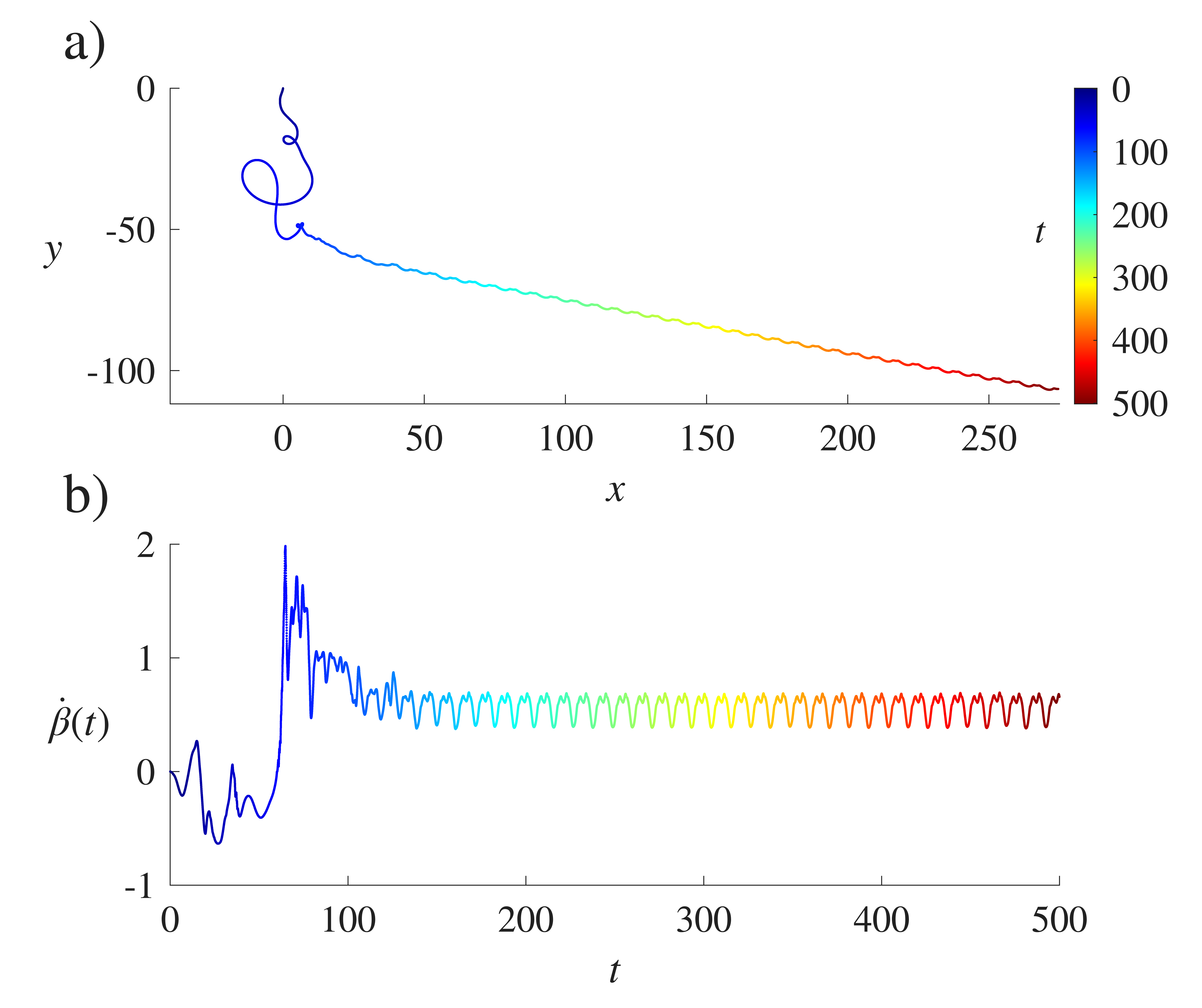}
    \caption{(a) Center-of-mass trajectory and (b) angular velocity of a V-shaped plate with density $R_1 = 4$ and bending angle $\theta = 11.3^\circ$, for $0 \leq t \leq 500$. In each panel, the colors correspond to time.}
    \label{v-plate autorotating}
\end{wrapfigure}
As mentioned at the beginning of this section, the tumbling motion (falling end-over-end), which occurred for the flat plate when $0.2 < R_1 < 1.7$, does not occur for the bent plates. This may be because the bent plate
is not symmetric after a $180^\circ$ rotation, so the fluid forces would be greatly altered on each half-cycle of a tumbling motion (which does not arise).
On the other hand, the bent plate has bilateral symmetry about the plate midpoint. A side-to-side fluttering motion has a similar symmetry in the fluid forces between times when the plate moves to one side and to the other, which may be one reason that symmetric fluttering can occur for V-shaped plates. Unlike the bent plates, The mirror and 180$^\circ$ symmetries of the flat plate may allow it to perform \textit{both} fluttering and tumbling motions. 
\newline 

Note, however, that the V-shaped plates are still able to autorotate when $R_1 \geq 4$. The center-of-mass trajectory and angular velocity of one such plate is shown in panels (a) and (b) of figure \ref{v-plate autorotating} respectively. After an initial period of looping, the plate converges to a periodic autorotating falling motion, with angular velocity oscillating about a fixed value. The analogous motion for $R_1 = 5$ is shown in the supplementary movie ``\href{https://drive.google.com/file/d/1hL_ran9ih_fGl7xXPKio2hVRLc6nAfKc/view?usp=drive_link}{movie\_5\_11.3\_vplate.avi}".

\subsubsection{$R_1^{-\frac{1}{2}}$-scaling of the dominant angular velocity frequency}
\label{vplate frequency scaling law section}
\begin{figure}[H]
    \centering
    \includegraphics[width=1\linewidth, trim = 1cm 0 1cm 0]{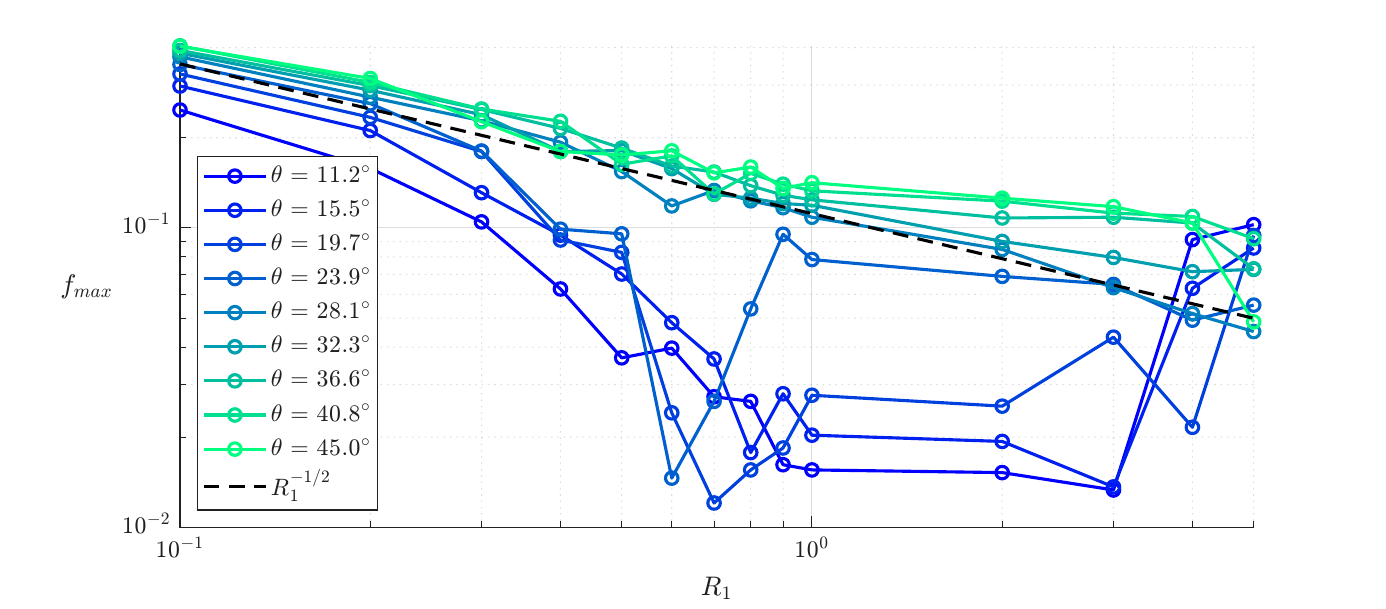}
    \caption{Log-log plot of the spectral peak frequency $f_{max}$ versus $R_1$. For $\theta \geq 28.1^\circ$ and $R_1 \leq 1$ the generic falling motion is fluttering with a frequency that scales as $R_1^{-\frac{1}{2}}$.}
    \label{frequency scaling}
\end{figure}

The generic behavior of the plates with large bending angles, $\theta > 28.12 ^\circ$, is quasi-periodic fluttering. We quantitatively study the effect of $R_1$ on the fluttering frequency as done previously for the flat plates, by computing the spectral peak frequency of the angular velocity, $f_{max}$. For all fluttering motions, a power law $f_{max} \sim R_1^{-1/2}$ shows good agreement with the data for $R_1 \leq 0.2$ (See figure \ref{frequency scaling}). For plates with bending angle $\theta \geq 28.12$ the validity of this scaling law extends to $R_1 \leq 1$, and remains a good approximation when $R_1 \leq 5$. Bending the plates extends the range of $R_1$ where fluttering occurs to larger values, extending the range of validity of this scaling law.  This scaling law was obtained for flat plates via dimensional analysis in \cite{belmonte1998flutter}. A similar scaling law is reported in \cite{Mavroyiakoumou_Alben_2020} for the oscillation frequency of a thin membrane in inviscid flow.
 
\begin{figure}[H]
    \centering
    \includegraphics[width=1\textwidth, trim=0.5cm 0.3cm 2cm 0, clip]{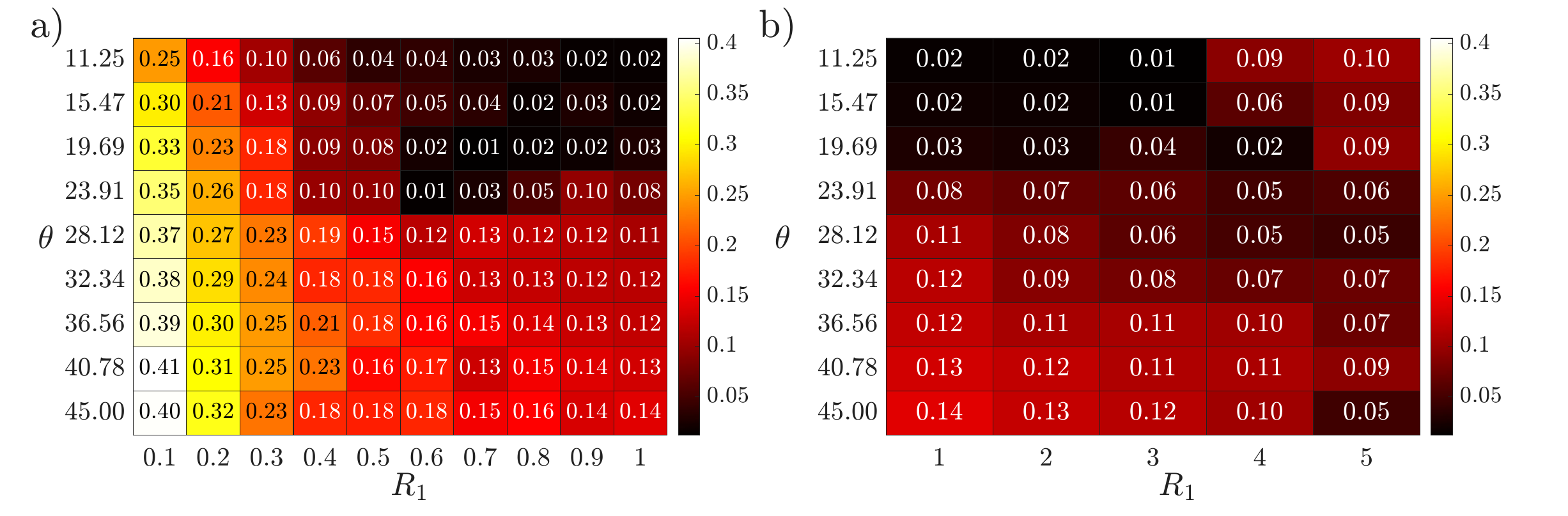}
    \caption{Heat maps of average peak frequency versus $R_1$ and $\theta$ for bent plates with (a) $0.1 \leq R_1 \leq 1$ and (b) $1 \leq R_1 \leq 5$.}
    \label{Heat map frequency}
\end{figure}

In figure \ref{Heat map frequency} we represent the data in figure \ref{frequency scaling} as a heat map analogous to figure \ref{Heat map average speed}. As before, the sharp changes in peak frequency (and color) represent sharp changes in the falling motion. The looping motions are predominantly aperiodic and have darker colors. On the other hand, low density, highly bent plates flutter rapidly, and have lighter colors. 
The sharp increase in $f_{max}$ for $\theta \leq 15.47^\circ$ when $R_1$ increases from 3 to 4 corresponds to the transition to autorotation, as can be seen in the first row of figure \ref{VPlate Large R1}. The heat map of frequency in figure \ref{Heat map frequency} sketches the phase diagram of falling motions of V-shaped plates and the sharp jumps in value are generally consistent with the similar sketch shown by the heat map of average speed in figure \ref{Heat map average speed}. However, some phase transitions are better captured by the heat map of average peak frequency. For example, when $\theta = 11.25^\circ$, as $R_1$ transitions from $0.1$ to $0.2$ the peak frequency drops from $0.25$ to $0.16$ suddenly. This corresponds to the transition from small amplitude fluttering to looping. By contrast, in figure \ref{Heat map average speed}, the average speed transitions smoothly from $2.9$ to $2.6$ over the same $R_1$ and $\theta$ values. Similarly, when $\theta = 45^\circ$, as $R_1$ transitions from $4$ to $5$ the spectral peak frequency $f_{max}$ jumps from $0.1$ to $0.05$ while the average center-of-mass speed remains constant at $1$. This transition can also be seen in the last row of columns $R_1 = 4,\ 5$ in figure \ref{VPlate Large R1}. From the center-of-mass trajectories shown in the figure, it can be seen that at $R_1= 5$, the trajectories are noticeably more erratic than at $R_1 = 4$. This corresponds to more overturning and longer phases of autorotation at $R_1 = 5$ than at $R_1 = 4$. The heat map of average center-of-mass speed thus sketches a phase diagram of lower resolution than that of the heat map of average peak frequency.

\clearpage

\subsubsection{Obtuse fluttering}
\begin{wrapfigure}[27]{l}{0.23\textwidth}  
    \includegraphics[width=0.17\textwidth, trim = 1.5cm 8cm 5cm 6cm, clip]{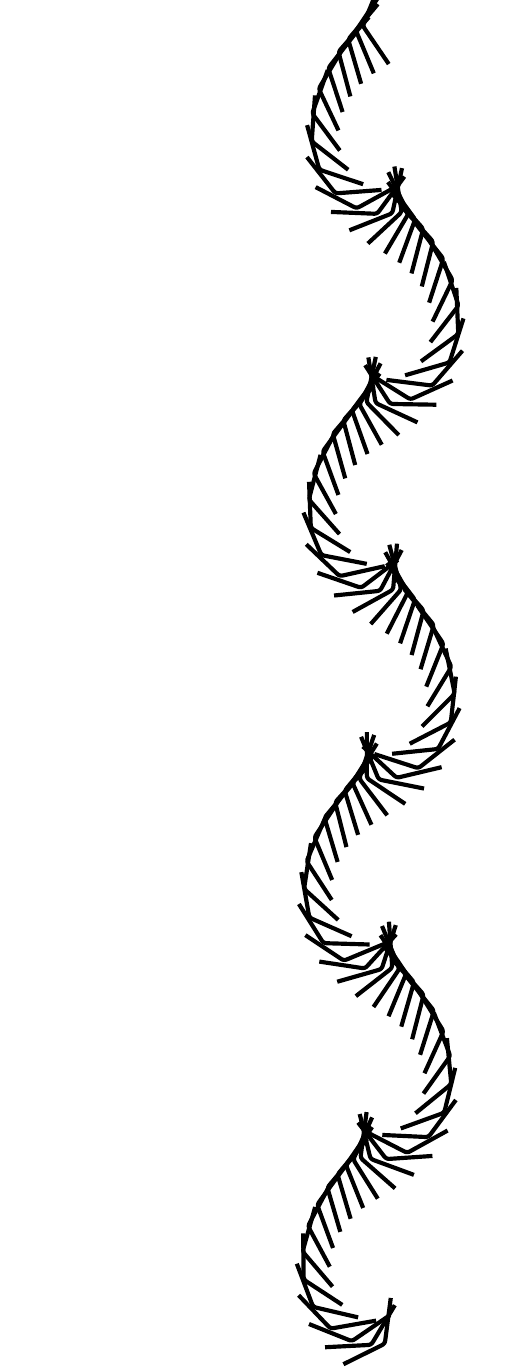}
    \caption{Snapshots of a V-shaped plate in obtuse fluttering with bending angle $\theta = 28.1^\circ$ and $R_1 = 0.9$.} 
    \label{obtuse fluttering}
\end{wrapfigure}

Most of the bent-plate dynamics were also seen for the flat plate. Here we show a distinct bent-plate motion that we call ``obtuse fluttering," shown in figure \ref{obtuse fluttering} at $R_1 = 0.9$ and $\theta = 28.1^\circ$.
This fluttering motion 
has cusps with obtuse rather than acute angles (as seen for the flat plate, e.g.~figures \ref{plate small R1}--\ref{wake behind 0.3 figure}), because the plate begins to overturn slightly at each cusp, with orientation angle magnitude $|\beta|$ peaking slightly above 90$^\circ$. With obtuse angles the trajectory appears smoother than those for flat-plate fluttering.
\newline

\begin{figure}[H]
    \centering
\includegraphics[width=1.1\linewidth, trim = 6cm 1cm 0cm 1cm]{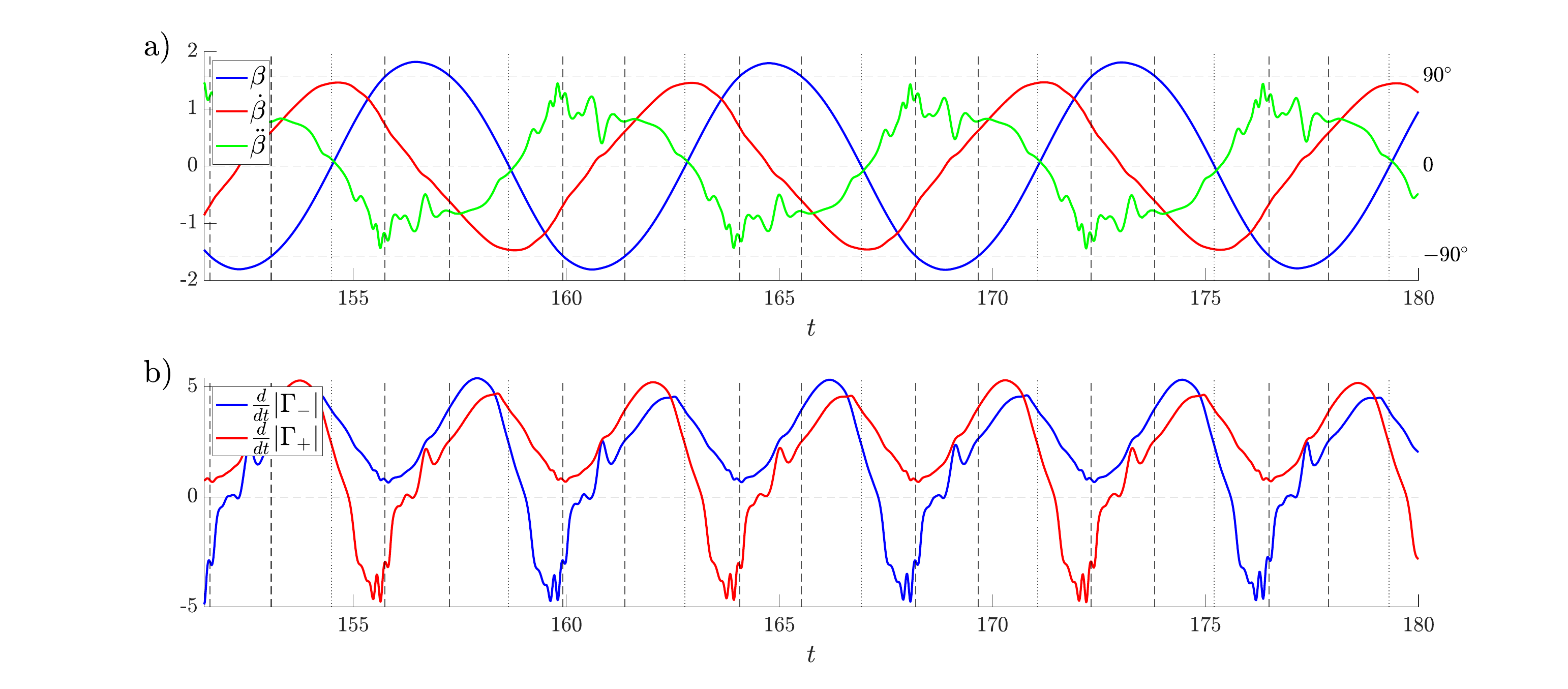}
    \caption{(a) The orientation angle $\beta$ (blue), angular velocity $\dot{\beta}$ (red) and angular acceleration $\ddot{\beta}$ (green) for a V-shaped plate in an obtuse fluttering motion with bending angle $\theta = 28.1^\circ$ and $R_1 = 0.9$. (b) The rate of change of the magnitude of the circulation of the $\pm$ sheet shown in red and blue respectively.}
    \label{obtuse fluttering torque}
\end{figure}

Unlike the fluttering motions of the flat plate, which are nonperiodic, this bent-plate fluttering motion is periodic. Without the rotational symmetry of the flat plate, the bent plate flutters through a substantially different mechanism. At $R_1 = 0.9$, the flat plates exhibit tumbling rather than fluttering because plate inertia is large relative to the restoring force of fluid pressure. However, bent plates experience very different fluid forces in an upside-down-V position ($\land$). Previous experiments \cite{hovering_childress, ristroph_falling_shapes} and figure \ref{obtuse fluttering torque} indicate that such a position is unstable and this may inhibit tumbling and stabilize the fluttering state. Supplementary movie ``\href{https://drive.google.com/file/d/1SjiTg3qTw6lRtvkmOn2cQvSTa2cG74qV/view?usp=drive_link}{movie\_0.9\_28.1\_vplate.avi}" shows an obtuse fluttering motion.
\newline

We illustrate the physical mechanism preventing overturning in figure \ref{obtuse fluttering torque} by plotting the orientation angle $\beta$, angular velocity $\dot\beta$ and angular acceleration $\ddot\beta$  (panel (a)) and the rates of change of the circulation magnitudes of the $\pm$ vortex sheets (panel (b)) for the obtuse fluttering plate in figure \ref{obtuse fluttering}. Note that the orientation angle $\beta$ is the angle formed by the line through the two edges of the plate and the horizontal axis.
\newline 

When $t \approx 164$ in figure \ref{obtuse fluttering torque}(a), $\beta$ has a maximum. The plate has rotated by a little more than $+\pi/2$ radians and the $+$ edge is trailing. At the same time, $t \approx 164$, in panel (b), the rate of change of circulation in the $+$ sheet is zero. In general, at the start of each half-period when $\beta$ has a maximum/minimum, no vorticity is released at the trailing edge. The restoring force inducing this obtuse fluttering motion thus appears to be primarily due to vorticity released at the leading edge. Panel (a) shows that, like a harmonic oscillator, the sign of $\ddot{\beta}$ (and therefore the torque) is opposite to the sign of the orientation angle. In other words, the fluid always applies a torque to the bent plate that restores it to the $\lor$ position with its tip pointed downward. When the plate overturns (i.e.~its angle magnitude exceeds $90^\circ$), there is a small local jump in torque corresponding to a sudden increase in restoring force. Note that when $\pm \beta < 0$, the $\pm$ edge is leading. At these times in panel (b), the rate of change of the magnitude of circulation in the $\pm$ sheet is positive, so circulation on the $\pm$ edge is increasing. During the sub-period where it is trailing, the circulation on the $\pm$ sheet decreases. Over each period, however, the magnitude of circulation on each sheet has a net increase. Intuitively, this corresponds to the transfer of gravitational potential energy to the kinetic energy of the fluid.
\newline

\section{Conclusion}
We demonstrated a new numerical model for falling bodies that includes the leading-edge shedding of vorticity. The inclusion of leading-edge vortex shedding allows for more realistic dynamics in the vortex sheet simulations of falling plates. It allows for long-time fluttering and tumbling dynamics previously only observed in direct Navier-Stokes simulations and experiments, while also predicting other falling motions that could potentially be observed in experiments. Some interesting special fluttering motions were observed, such as the progressive fluttering observed in \cite{pomerenk2024aerodynamicequilibriaflightstability} and the obtuse fluttering of the V-shaped plates. This obtuse fluttering is characterized by obtuse-angled cusps in the trajectory, and is driven primarily by leading-edge shedding effects, which provide a restoring force to the $\lor$ position with sign opposite to that of the plate orientation angle. For flat plates at low values of $R_1$ fluttering motions dominate. As $R_1$ increases, the dominant motion transitions to tumbling, looping, and then autorotation at key transition values. The transition between fluttering and tumbling dynamics observed at $R_1 \approx 0.7$ (equivalently, $I^* \approx 0.3$) 
is consistent with several prior experimental and viscous numerical investigations. While the amplitude of side-to-side fluttering motions increases with increasing $R_1$, the amplitude of tumbling motions decreases. For small $R_1$, bending the plate stabilizes its motion, yielding quasi-periodic instead of chaotic fluttering, while for large $R_1$, bending the plate destabilizes it from quasi-periodic autorotation to chaotic fluttering. When $\theta > 28.12 ^\circ$ only fluttering modes are observed, with frequencies that scale as $R_1^{-\frac{1}{2}}$ when $R_1 \leq 1$.

\section*{Acknowledgments}
This research was supported by the NSF-DMS Applied Mathematics program under
award number DMS-2204900.

\appendix
\section{The logarithmic singularity \label{log singularity appendix}}
Here we rigorously justify the presence of the logarithmic singularities at the endpoints of the Cauchy integral 

\begin{equation} \label{cauchy integral appendix}
\frac{1}{2\pi i}\int_{-1}^{1}{\frac{\gamma_b(s,t)}{z - \zeta(s,t)}\, ds}.
\end{equation}

The results found here are well known, and can be found in much greater generality in \cite{Muskhelishvili}. For completeness, however, we provide simpler arguments using only standard complex analysis. We require only a very weak form of the Sokhotski-Plemelj formula which can be stated as follows. Given any $C^1$ (continuously differentiable) curve $C$ and $C^1$ function $g(z)$ on the complex plane, then for any interior point $c \in C$, we have:
$$ \lim_{z \rightarrow c^\pm} \frac{1}{2\pi i}\int_C \frac{g(\zeta)}{\zeta - z}d\zeta = \frac{1}{2\pi i}\dashint_C \frac{g(\zeta)}{\zeta - c}d\zeta  \pm \frac{1}{2}g(c).$$ Here the $\pm$ superscript indicates the side from which $c \in C$ is approached.
Under these stronger assumptions, this formula is immediate. Indeed, when $g(z) \equiv 1$ we may integrate exactly, and obtain the formula via contour integration. Fix $c$, an interior point of $C$, and let $C_\epsilon$ be the curve $C$ after removing a symmetric segment of length $\epsilon$ centered at $c$. Observe that:
\begin{align*}
\lim_{z \rightarrow c^\pm} \frac{1}{2\pi i}\int_C \frac{g(\zeta)}{\zeta - z}d\zeta 
&= \lim_{z \rightarrow c^\pm} \frac{1}{2\pi i}\int_C \frac{g(\zeta) - g(z)}{\zeta - z}d\zeta + \lim_{z \rightarrow c^\pm}g(z)\frac{1}{2\pi i}\int_C \frac{1}{\zeta - z}d\zeta \\
&= \frac{1}{2\pi i}\int_C \frac{g(\zeta) - g(c)}{\zeta - c}d\zeta + g(c)\frac{1}{2\pi i}\dashint_C \frac{1}{\zeta - c}d\zeta  \pm \frac{1}{2}g(c)\\
&= \lim_{\epsilon \rightarrow 0}\frac{1}{2\pi i}\int_{C_\epsilon} \frac{g(\zeta) - g(c)}{\zeta - c}d\zeta + g(c)\frac{1}{2\pi i}\dashint_C \frac{1}{\zeta - c}d\zeta  \pm \frac{1}{2}g(c)\\
&= \frac{1}{2\pi i}\dashint_C \frac{g(\zeta)}{\zeta - c}d\zeta - g(c)\frac{1}{2\pi i}\dashint_C \frac{1}{\zeta - c}d\zeta + g(c)\frac{1}{2\pi i}\dashint_C \frac{1}{\zeta - c}d\zeta  \pm \frac{1}{2}g(c)\\
&= \frac{1}{2\pi i}\dashint_C \frac{g(\zeta)}{\zeta - c}d\zeta  \pm \frac{1}{2}g(c).
\end{align*}
The second equality follows from the fact that $|\frac{g(\zeta) - g(z)}{\zeta - z}| \leq \max_{\zeta \in C}(\|\nabla g(\zeta)\|)$ (which exists as $g\in C^1$), and dominated convergence. The third equality is a consequence of dominated convergence as well, while the fourth follows from the definition of the principal valued integral.
\newline

We now turn our attention to \eqref{cauchy integral appendix} and begin the argument. Recall that $\zeta(s,t)$ parameterizes the curve representing the thin plate in the complex plane. Let $C_b$ denote this curve, and let $c_\pm = \zeta(\pm1,t)$ denote its endpoints. Let us focus on the positive edge $c_+$ and define the function $$g(\zeta(s,t)) = \gamma_b(s,t)\partial_s\overline{\zeta(s,t).}$$ Notice that $$|g(c\pm)| = |\gamma_b(\pm1,t)|.$$ Hence $g$ vanishes at the endpoints if and only if $\gamma_b$ does. Let $w = \zeta(s,t)$. Then we may rewrite the contribution of the body to the conjugate fluid velocity as 
\begin{equation*}
    \begin{split}
        \int_{-1}^{1} \frac{\gamma_b(s,t)}{z - \zeta(s,t)}\, ds &= \int_{C_b}\frac{g(w)}{z - w}dw =
            \int_{C_b}\frac{g(w) - g(c_+) - \nabla g(c_+) \cdot(w - c_+)}{z - w}dw \\
            \ \ &\ \ \ \ \ \ \ \ \ - g(c_+)\int_{C_b}\frac{dw}{w - z}  - \nabla g(c_+) \cdot \int_{C_b}\frac{w - c_+}{w -z}dw \\
            &= \int_{C_b}\frac{g(w) - g(c_+) - \nabla g(c_+) \cdot(w - c_+)}{z - w}dw \\
            \ \ &\ \ \ \ \ \ \ \ \ - g(c_+)\big(\log(c_+ - z) - \log(c_- - z)\big)  \\ 
           \ \ &\ \ \ \ \ \ \ \ \ \ \ \ - \nabla g(c_+) \cdot \bigg( c_+ - c_- + (z-c_+)\big(\log(c_+ - z) - \log(c_- - z)\big)\bigg) \\ &=: I_1 + I_2 + I_3.
    \end{split}
\end{equation*}

Clearly, $I_3 = \Theta(1)$ as $z \rightarrow c_+$. If $\gamma_b(1,t) = 0$, then $g(c_+) = 0$, in which case $I_2 = 0$. When $\gamma_b(1,t) \not = 0$, however, $I_2 = \Theta\big(\log(c_+ - z)\big)$ as $z \rightarrow c_+$. The first term $I_1$, however, requires a more careful analysis.
First, we observe that we may extend the body $C_b$ along the positive edge linearly so that it remains a $C^1$ curve which we call $C_b^{ext}$. Let $h(w) = g(w) - g(c_+) - \nabla g(c_+) \cdot(w - c_+)$. Then $h(c_+) = 0$ and $\nabla h(c_+) = 0$ automatically. Therefore, by letting $h$ vanish on the extended part of the body, $h$ admits a $C^1$ extension to the new body $C_b^{\text{ext}}$ such that we may write
\begin{eqnarray}
    I_1 = \int_{C_b}\frac{h(w)}{z - w}dw = \int_{C_b^{\text{ext}}}\frac{h(w)}{z - w}dw.
\end{eqnarray}
The point of this is that now $c_+$ is an interior point of the curve $C_b^{\text{ext}}$, so that the Sokhotski-Plemelj formulae apply. Hence,

\begin{equation}
    \lim_{z \rightarrow c_+}\int_{C_b^{\text{ext}}}\frac{h(w)}{z - w}dw = \int_{C_b^{\text{ext}}}\frac{h(w)}{c_+ - w}dw \pm  \pi i h(c_+) = \int_{C_b^{\text{ext}}}\frac{h(w)}{c_+ - w}dw.
\end{equation}
Here the last equality follows from the fact $h(c_+) = 0$. Therefore, $I_1 = \Theta(1)$ as $z \rightarrow c_+$ as well. 

Hence, we see that 
\begin{equation}
    \int_{-1}^{1} \frac{\gamma_b(s,t)}{z - \zeta(s,t)}\, ds = 
    \begin{cases}
        \Theta\big(\log(c_+ - z)\big) \text{ as } z \rightarrow c_ + \text{, if }\gamma_b(1,t) \not= 0. \\
        \Theta(1) \text{ as } z \rightarrow c_ + \text{, if }\gamma_b(1,t) = 0.
    \end{cases} \label{Singularity}
\end{equation}

In fact, this shows that near the $+$ edge, we may find bounded function $\Phi_\pm$ holomorphic away from the body such that
\begin{equation}
    \int_{-1}^{1} \frac{\gamma_b(s,t)}{z - \zeta(s,t)}\, ds = \Phi_+(z) - g(c_+)\log(c_+ - z) - \nabla g(c_\pm)\cdot\big((z - c_\pm)(\log(c_\pm - z)\big) .
\end{equation}

A similar equation holds at the $-$ edge with signs reversed.
\clearpage

\section{Log quadrature with blob regularization \label{log quadrature with regularization}}

Here, we describe briefly how the log quadrature rule \eqref{log quadrature for singular body} can be modified to obtain \eqref{log quadrature for smooth sheets}. This quadrature rule is used in both the velocity smoothed Birkhoff-Rott \eqref{BirkhoffRottSmoothed} and pressure-jump equations \eqref{smooth bernoulli pressure jump}. Like in section \ref{LogDesingularizationOfBody}, we proceed by assuming that the position of the free vortex sheet is given as a piecewise linear function of circulation. This assumption allows us to compute the integral analytically on each subinterval before summing the contributions. More explicitly, suppose that we have $k_\pm$ points on the $\pm$ sheet. Denote their positions $\zeta_{\pm,j}$ and their corresponding circulations by $\Gamma_{\pm,j}$ then we have that 
\begin{align*}
    \int_0^{\Gamma_\pm(t)}\frac{\overline{z - \zeta_\pm(\Gamma',t)}}{|z - \zeta_\pm(\Gamma',t)|^2 + \delta^2}d\Gamma' &= \sum_{j=0}^{k_\pm}\int_{\Gamma_{\pm,j}}^{\Gamma_{\pm,j+1}}\frac{\overline{z - \zeta_\pm(\Gamma',t)}}{|z - \zeta_\pm(\Gamma',t)|^2 + \delta^2}d\Gamma' \\
    &\approx -\sum_{j=0}^{k_\pm}\int_{\Gamma_{\pm,j}}^{\Gamma_{\pm,j+1}}\frac{\overline{a_j\Gamma' + b_j -z}}{|a_j\Gamma' + b_j - z|^2 + \delta^2}d\Gamma' \\
    &= \sum_{j=0}^{k_\pm}\frac{-1}{a_j}\int_{\Gamma_{\pm,j}}^{\Gamma_{\pm,j+1}}\frac{\overline{\Gamma' - c_j}}{|\Gamma' - c_j|^2 + \delta^2}d\Gamma'.
\end{align*}
Here, $c_j = \frac{z -b_j}{a_j}$, and $a_j$, $b_j$ are given as before. Let $c_j^R = \Re{c_j}$ and $c_j^I = \Im{c_j}$ so that $c_j = c_j^R + ic_j^I$. By splitting the integrand into real and imaginary parts, 
\begin{equation}
    \frac{\overline{\Gamma' - c_j}}{|\Gamma' - c_j|^2 + \delta^2}d\Gamma' = \frac{\Gamma' - c^R_j}{|\Gamma' - c_j|^2 + \delta^2}d\Gamma' + i\frac{c^I_j}{|\Gamma' - c_j|^2 + \delta^2}d\Gamma',
\end{equation}

it is easy to compute that
\begin{multline}
    \int_{\Gamma_{\pm,j}}^{\Gamma_{\pm,j+1}}\frac{\overline{\Gamma' - c_j}}{|\Gamma' - c_j|^2 + \delta^2}d\Gamma' = \frac{1}{2}\log{\frac{(\Gamma_{\pm,j+1} -c_j^R)^2 + (c_j^I)^2 + \delta^2}{(\Gamma_{\pm,j} -c_j^R)^2 + (c_j^I)^2 + \delta^2}} \\ + \frac{ic_j^I}{\sqrt{(c_j^I)^2 + \delta^2}}\bigg( \arctan\big(\frac{\Gamma_{\pm,j+1}-c_j^R}{\sqrt{(c_j^I)^2 +\delta^2}}\big) - \arctan\big(\frac{\Gamma_{\pm,j}-c_j^R}{\sqrt{(c_j^I)^2 +\delta^2}}\big) \bigg).
\end{multline}

Hence, the contributions to the conjugate fluid velocity given by the free sheets can be approximated by the formula

\begin{multline}
    \frac{1}{2\pi i}\mathlarger{\int_0^{\Gamma_\pm(t)}}\frac{\overline{z - \zeta_\pm(\Gamma',t)}}{|z - \zeta_\pm(\Gamma',t)|^2 + \delta^2}d\Gamma' \approx \mathlarger{\sum_{j=0}^{k_\pm}}\frac{-1}{a_j 2\pi i}\mathlarger{\Bigg(} \frac{1}{2}\log{\bigg(\frac{(\Gamma_{\pm,j+1} -c_j^R)^2 + (c_j^I)^2 + \delta^2}{(\Gamma_{\pm,j} -c_j^R)^2 + (c_j^I)^2 + \delta^2}\bigg)} \\ + \frac{ic_j^I}{(c_j^I)^2 + \delta^2}\bigg( \arctan\big(\frac{\Gamma_{\pm,j+1}-c_j^R}{\sqrt{(c_j^I)^2 +\delta^2}}\big) - \arctan\big(\frac{\Gamma_{\pm,j}-c_j^R}{\sqrt{(c_j^I)^2 +\delta^2}}\big) \bigg)\mathlarger{\Bigg).} \label{FreeSheetContribution}
\end{multline}
\clearpage

\section{Numerical parameters and grid refinement studies}
\label{numerical parameters and grid refinement studies}
Here we summarize the numerical parameters used to obtain the results in the main body of the paper.
\begin{table}[H]
    \centering
    \begin{tabular}{|c|c|}
        \hline
         Time step, $dt$ & 0.012 \\
         \hline
         Number of grid points on the body, $n$ & 100 \\
         \hline
         Number of grid points on the free sheets in the far field, $k_\pm$ & 1000 \\
         \hline
         Vortex sheet regularization parameter, $\delta$ & 0.2 \\
         \hline
    \end{tabular}
    \caption{Summary of the numerical parameters used in the main results of the paper.}
    \label{tab:my_label}
\end{table}
Here, the far field refers to the portions of the $\pm$ sheets that have arc-length distances from the edges \textit{greater} than $20$  (i.e.~$10$ times the plate length). Similarly, the near field is defined as the portions of the $\pm$ sheets that have arc-length distances from the edges \textit{less} than $20$. To ensure the far-field portions of the free sheets contain at most $k_\pm$ points, we perform the following sheet pruning procedure. At the end of each time step, for each point in the far field, we compute the change in velocity on the body incurred by removing said grid point using equation \eqref{ConjFluidVel}. The grid point that incurs the smallest change in velocity when removed is then deleted. This procedure is iterated until the total number of points in the far field is equal to $k_\pm$. A similar criterion for point deletion was used in \cite{alben2013efficient}. The log quadrature rule described in section \ref{LogDesingularizationOfBody} and appendix \ref{log quadrature with regularization} is used only to compute the velocity induced by the near field, while the usual trapezoidal rule in used to compute the velocity induced by the far field. Hence, while the near field is treated as a vortex sheet, the far field is treated more as a collection of point vortices. Typically, the near field is represented with between $250$ and $500$ points with $dt = 0.012$, but this number increases as the time step size decreases.
\newline

\begin{figure}[H]
    \centering
    \includegraphics[width=1\linewidth, trim = 1cm 0cm 1cm 0cm, clip]{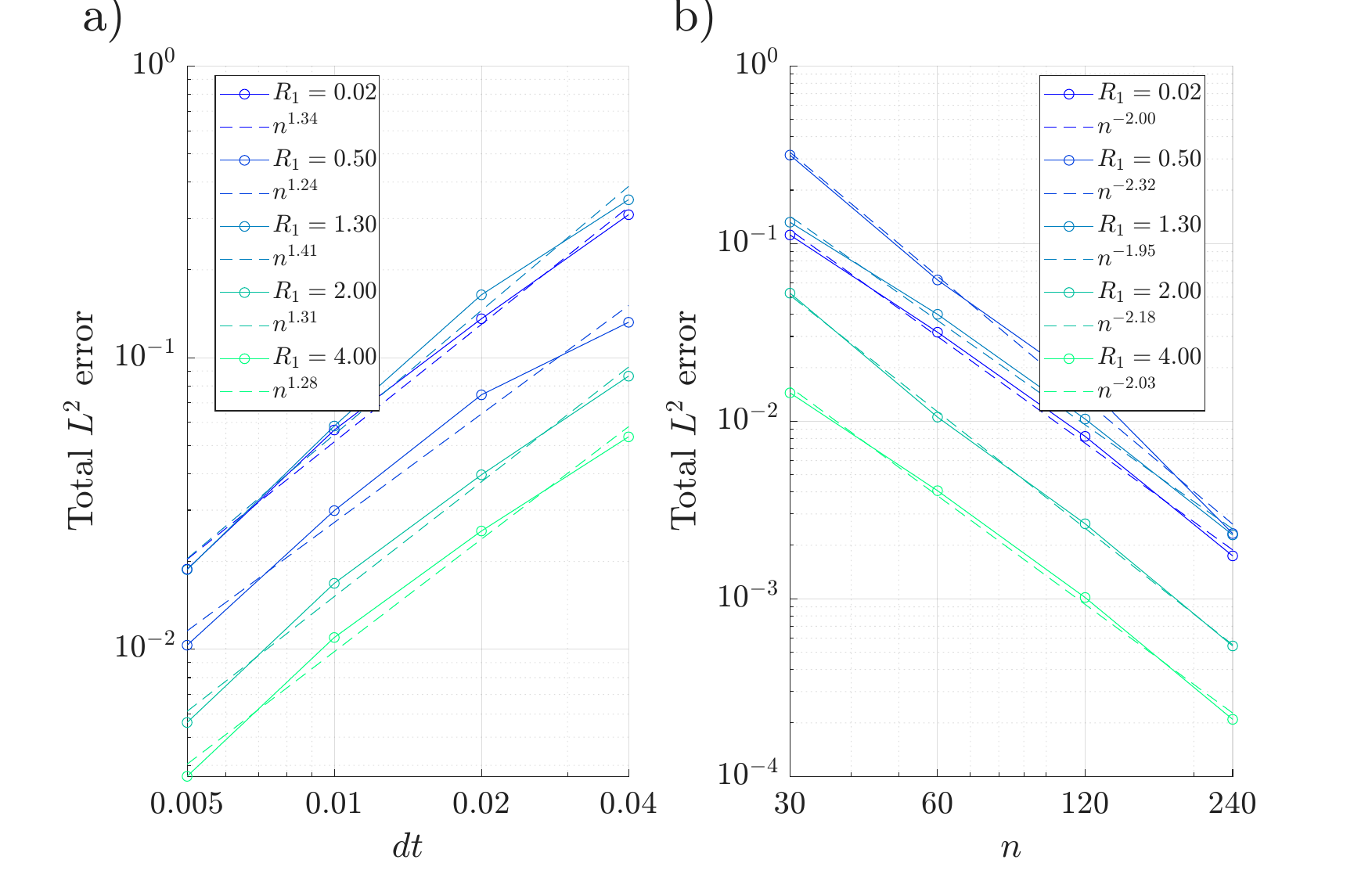}
    \caption{The total $L^2$ error versus $dt$ (panel a), and $n$ (panel b) for the flat plate at each $R_1$. The corresponding best-fit power law is plotted with dashed lines. The error is measured relative to the solution with $dt = 0.0025$ and $n = 480$ respectively.}
    \label{convergence_study}
\end{figure}

Formally, the algorithm is at least second order in both time and space.  However, as figure \ref{convergence_study} shows, the order of convergence as $dt \rightarrow 0$ is at worst $\approx1.24$ for the range of $R_1$ considered. The order of convergence in $\frac{1}{n}$ as the number of grid points $n \rightarrow \infty$ is $\approx 2$. These estimates were obtained via linear regression in logarithmic coordinates. The second-order convergence in the number of body grid points is expected, since when computing the strength $\gamma_b$, the trapezoidal rule is used on the Chebyshev grid which is second order in the number of body grid points. Indeed, since the Chebyshev grid is obtained by applying the function $-\cos(x)$ to the uniform grid with $n$ points on the interval $[0,\pi]$, the error incurred from the trapezoidal rule is $\sum^n_{k = 0} O(|\cos(k \pi/n) - \cos((k + 1)\pi/n)|^3) = 1/n^3 \sum^n_{k = 0} O(|\sin(k \pi/n)|^3) = O(1/n^2)$. This is because although the Chebyshev grid mesh size is of size $O(1/n^2)$ near both the edges of the plate,  the error contributed by the segments near the center of the plate with mesh size $O(1/n)$ dominates. 

\begin{figure}[H]
    \centering
    \includegraphics[width=1\linewidth, trim = 1cm 0cm 1cm 0cm, clip]{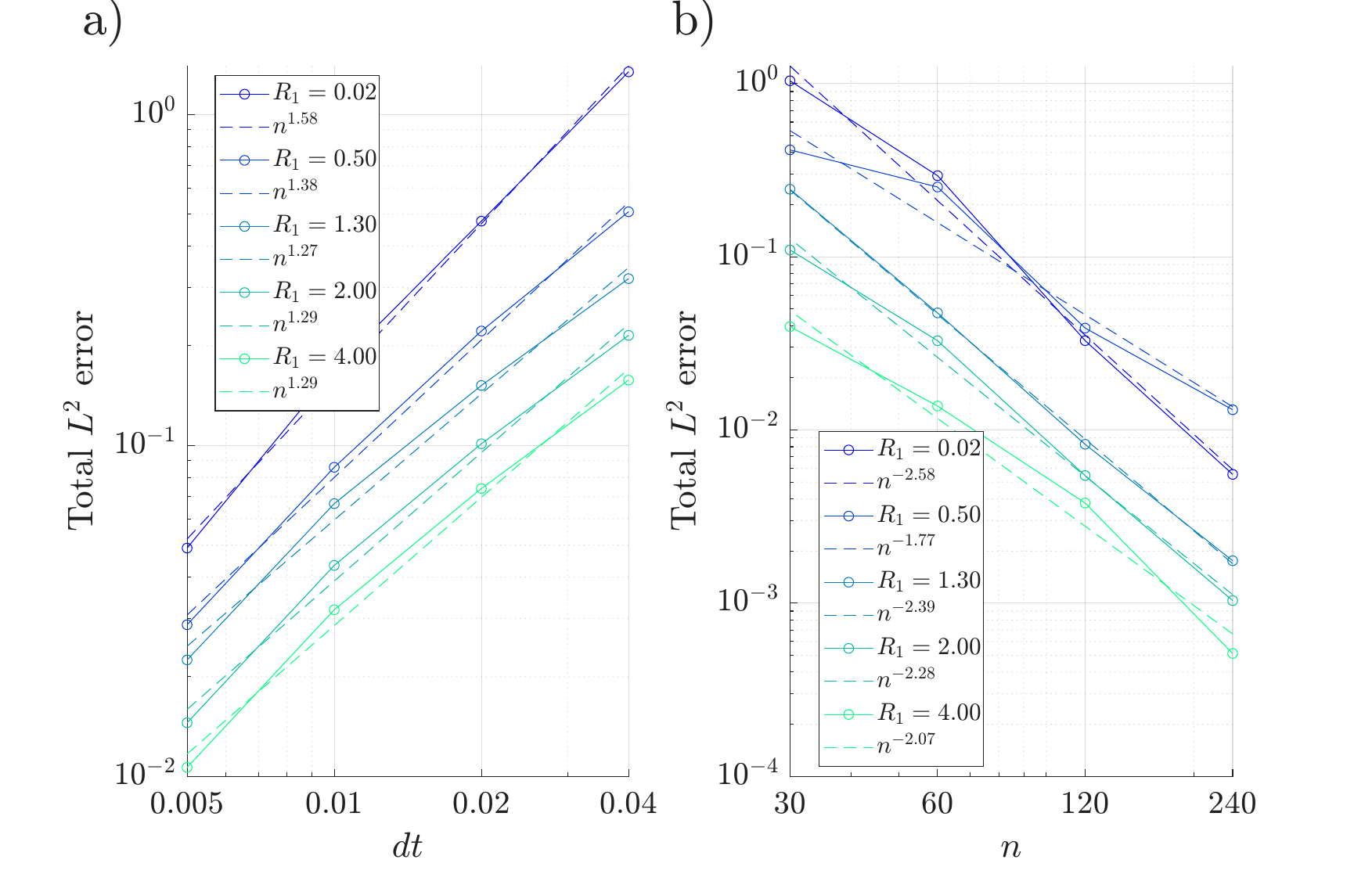}
    \caption{The total $L^2$ error versus $dt$ (panel a), and $n$ (panel b) for the V-shaped plates with bending angles $11.25^\circ$, $28.13^\circ$, and $45^\circ$. The corresponding best-fit power law is plotted with dashed lines. The error is measured relative to the solution with $dt = 0.0025$ and $n = 480$ respectively.}
    \label{v_convergence_study}
\end{figure}

The sum total error for V-shaped plates ($11.25^\circ,\  28.13^\circ,\  45^\circ$ bending angles combined) is shown in figure \ref{v_convergence_study}. The error follows scaling laws similar to the case of the flat plate (figure \ref{convergence_study}). Indeed, both models differ only in the inclusion of skin friction effects.

\begin{figure}[H]
    \centering
    \includegraphics[width=0.9\linewidth, trim = 2cm 1cm 2cm 1cm, clip]{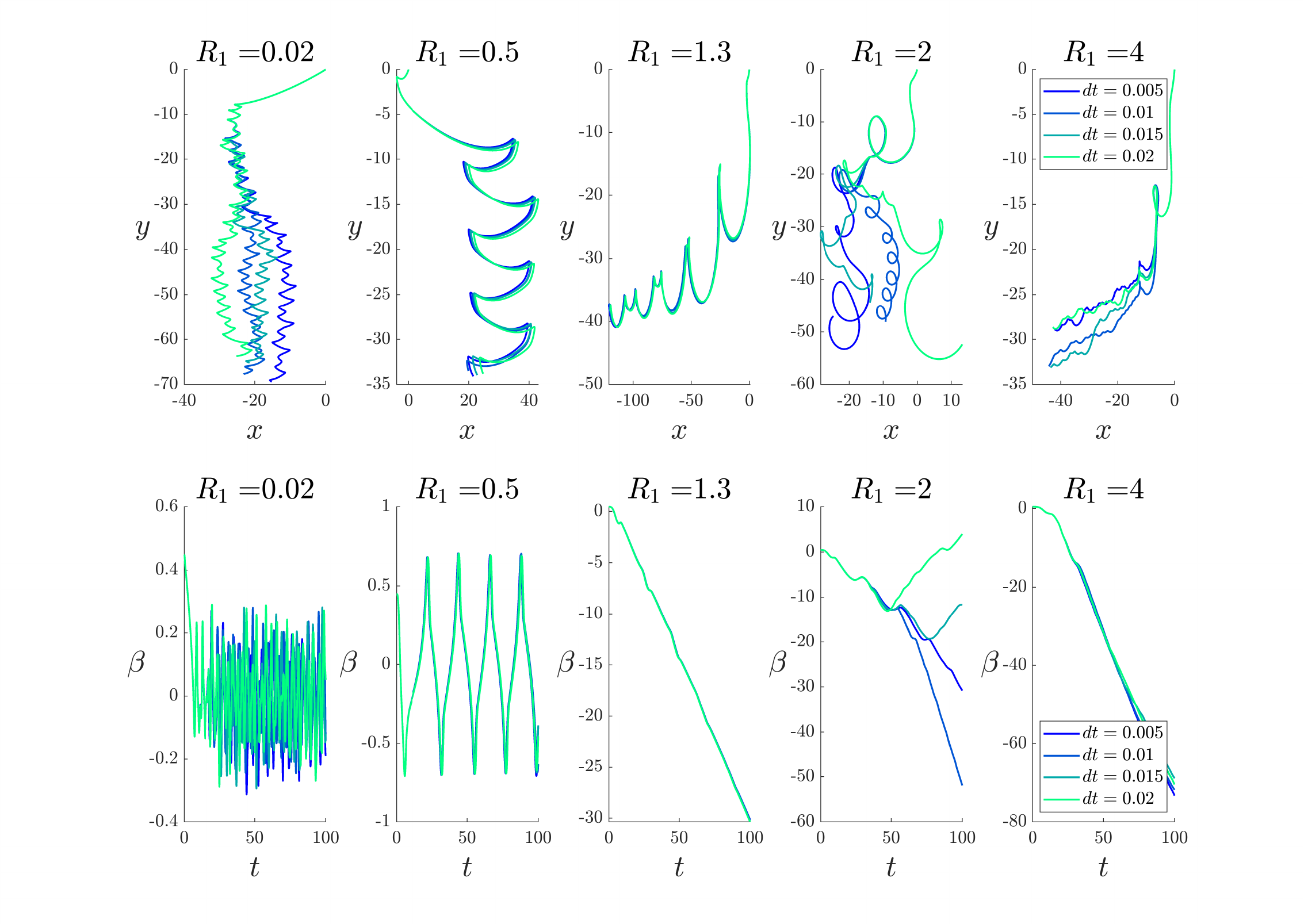}
    \caption{ The center-of-mass trajectories $\zeta_G(t)$ (top row) and the orientation angle $\beta(t)$ (bottom row) for $dt = 0.005,\ 0.01,\ 0.015,\ 0.02$ for the flat plate with $R_1 = 0.02,\ 0.5, \ 1.3,\ 2, \ 4$ up to $t = 100$. Here, $n = 100, k_\pm = 1000$. For all $dt \leq 0.02$ the qualitative dynamics exhibited are nearly identical.}
    \label{refinement in time figure}
\end{figure}

\begin{figure}[H]
    \centering
    \includegraphics[width=0.9\linewidth, trim = 2cm 1cm 2cm 1cm, clip]{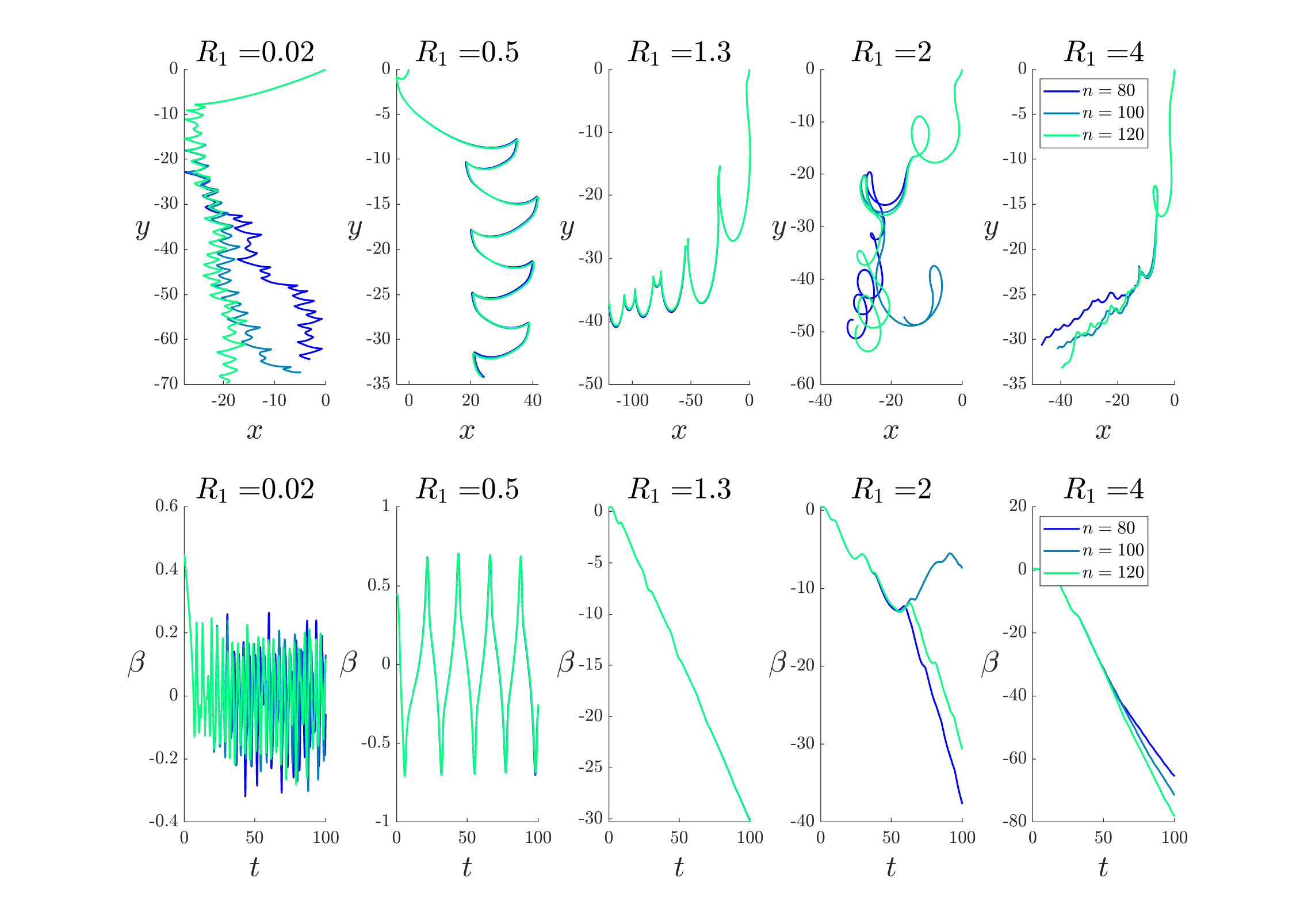}
    \caption{The center-of-mass trajectories $\zeta_G(t)$ (top row) and the orientation angle $\beta(t)$ (bottom row) for $n = 80,\ 100,\ 120$ for the flat plate with $R_1 = 0.02,\ 0.5, \ 1.3,\ 2, \ 4$ up to $t = 100$. Here, $dt = 0.01,\ k_\pm = 1000$. For all $n \geq 80$ the qualitative dynamics exhibited are nearly identical.}
    \label{refinement in body grid points figure}
\end{figure}

The convergence study in figure \ref{convergence_study} was limited only to short times. Indeed, as noted in \cite{aref1983integrable}, the motion of two-dimensional vortex structures is inherently chaotic, in the sense that it is highly sensitive to initial conditions. For any chaotic dynamical system, any change to numerical parameters, including grid refinement, can dramatically affect the computed motions of the free vortex sheets over long times. The motion of the plate, being coupled to the chaotic motion of the free vortex sheets, is chaotic as well. 
Due to the chaotic nature of the system, the error grows exponentially over long times. In figures \ref{refinement in time figure}, \ref{refinement in body grid points figure}, \ref{refinement in resolution of free sheets.} and \ref{refinement in resolution of delta figure} below, we instead consider how the large-scale qualitative features of simulation change under grid refinement. For illustration, we once again focus only on the flat plate. In all cases observed, only when tumbling occurs the system is not chaotic $(R_1 = 1.3)$. This is likely because the tumbling motion exhibits the most regular vortex wake, as can be seen from figure \ref{wake behind 1 figure}.

\label{refinement study}
\subsection{Refinement of the time step $dt$}
\subsection{Refinement in the number of body grid points, $n$}
\subsection{Refinement in the number of vortex sheet grid points in the far field, $k_\pm$}
\begin{figure}[H]
    \centering
    \includegraphics[width=0.9\linewidth, trim = 2cm 1cm 2cm 1cm, clip]{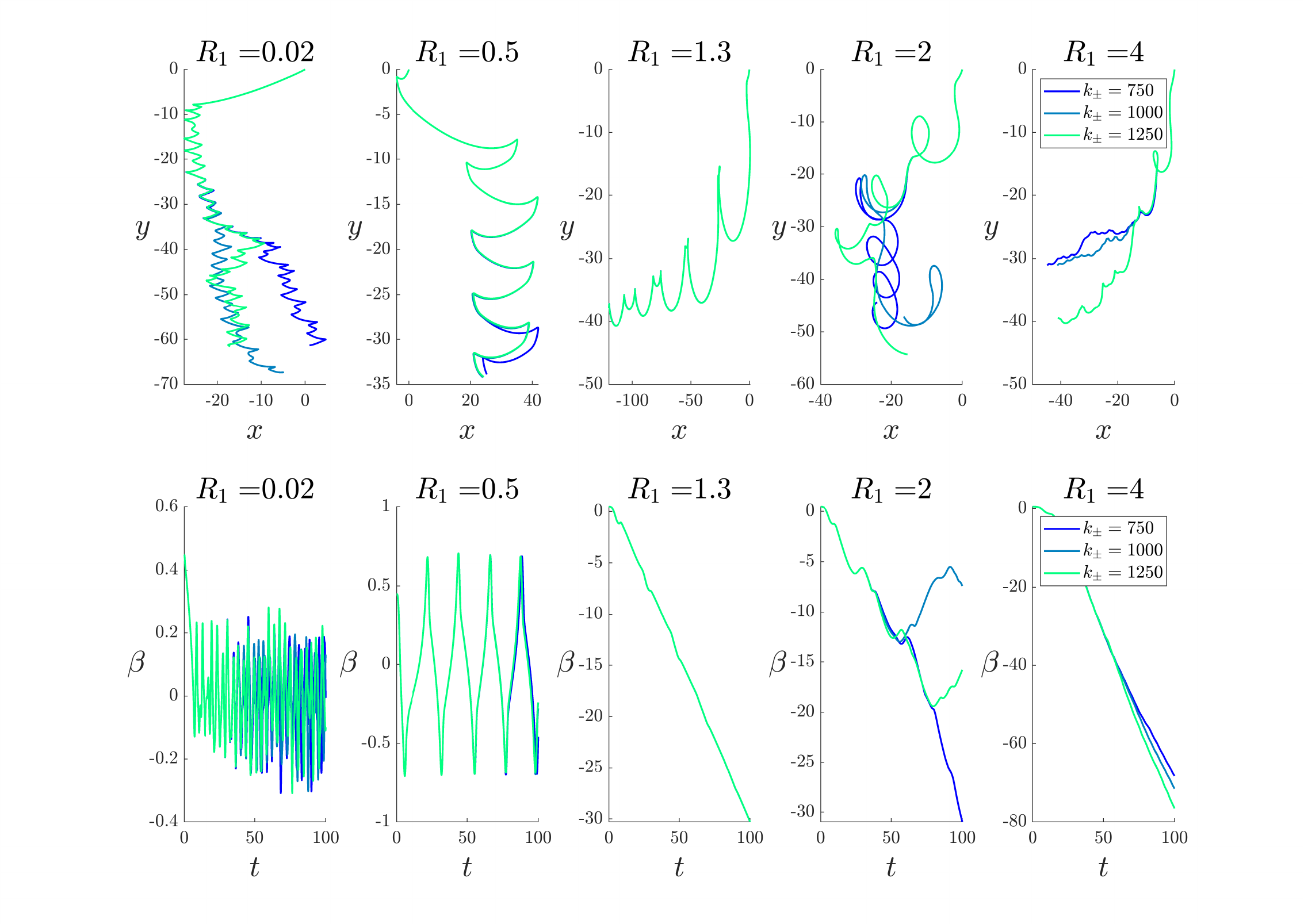}
    \caption{The center-of-mass trajectories $\zeta_G(t)$ (top row) and the orientation angle $\beta(t)$ (bottom row) for $k\pm = 750,\ 1000,\ 1250$ for the flat plate with $R_1 = 0.02,\ 0.5, \ 1.3,\ 2, \ 4$ up to $t = 100$. Here, $dt = 0.01,\ n = 100$. For all $n \geq 80$ the qualitative dynamics exhibited are nearly identical.}
    \label{refinement in resolution of free sheets.}
\end{figure}

\subsection{Refinement in $\delta$}
\label{refinement in delta section}
\begin{figure}[H]
    \centering
    \includegraphics[width=0.9\linewidth, trim = 2cm 1cm 2cm 1cm, clip]{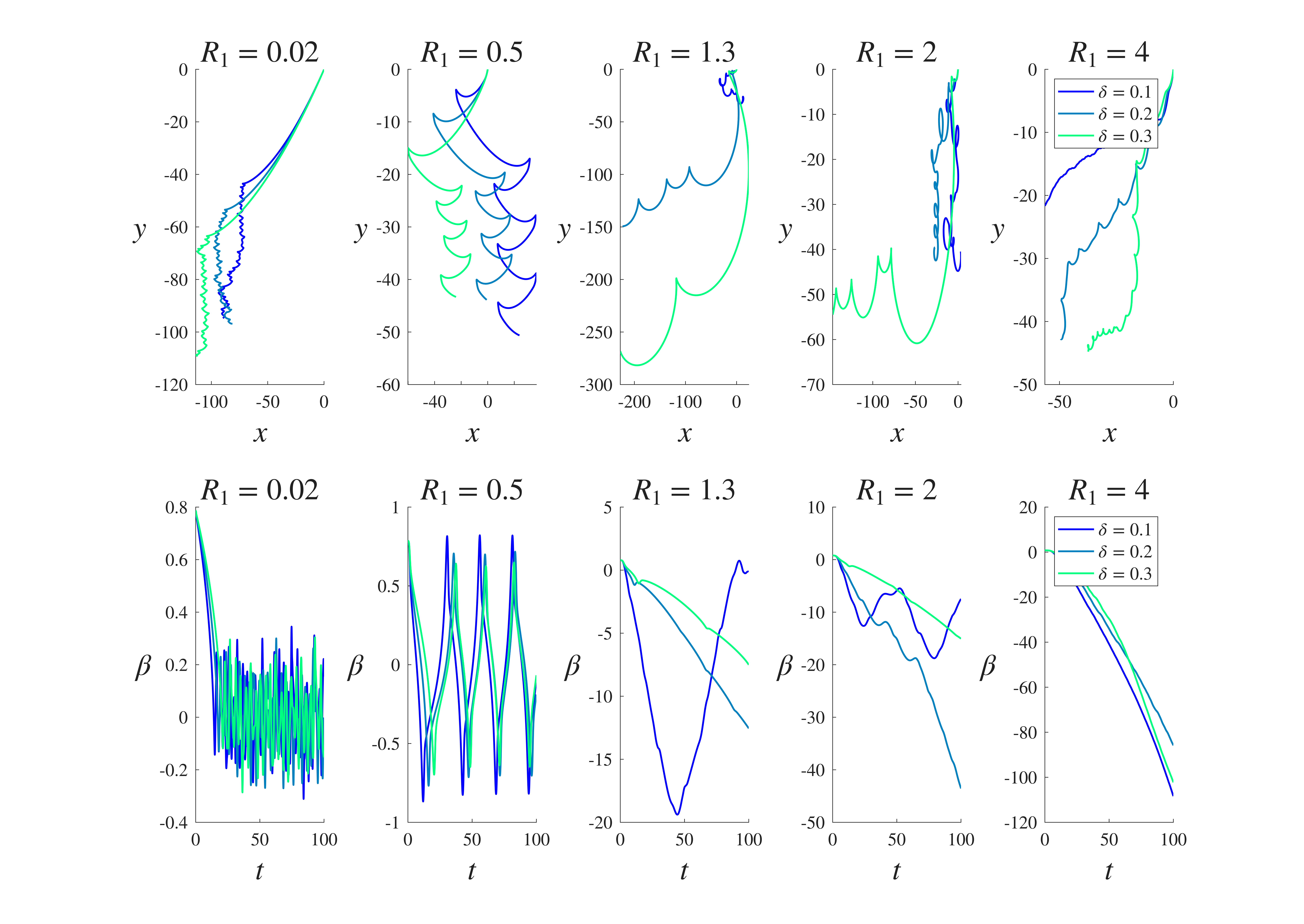}
    \caption{The center-of-mass trajectories $\zeta_G(t)$ (top row) and the orientation angle $\beta(t)$ (bottom row) for $\delta$ = 0.1, 0.2, and 0.3, up to $t = 100$. Here we take $dt = 0.01$, $n = 100$, and $k_\pm = 1000$. The qualitative dynamics are often similar when $\delta \leq 0.2$.} 
    \label{refinement in resolution of delta figure}
\end{figure}
\clearpage

\section{Center-of-mass trajectories of $V$-shaped plates}
\label{com trajectories of v shaped plates appendix}
\begin{figure}[H]
    \centering
    \includegraphics[width=1\textwidth, trim=0.5cm 0cm 2cm 0cm, clip]{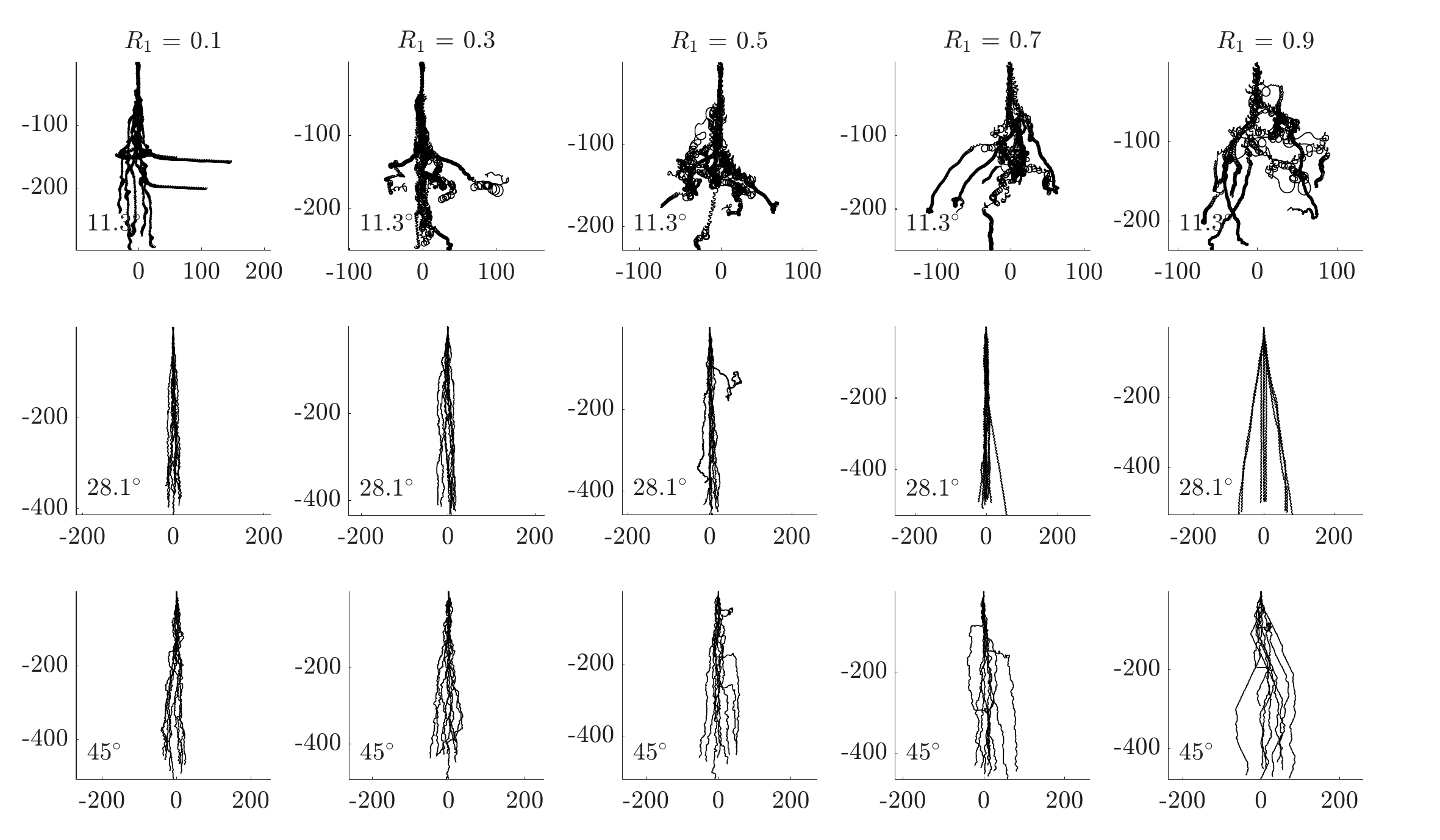}
    \caption{The center-of-mass trajectories of falling bent plates for $0\leq t \leq 500$, $R_1 = 0.1,\ 0.3,\ 0.5,\ 0.7,\ 0.9$, and bending angles $11.3^\circ, 28.1^\circ,$ and $45^\circ.$}
    \label{VPlate Small R1 expanded}
\end{figure}

\begin{figure}[H]
    \centering
    \includegraphics[width=1\textwidth, trim=1cm 0cm 2cm 0cm]{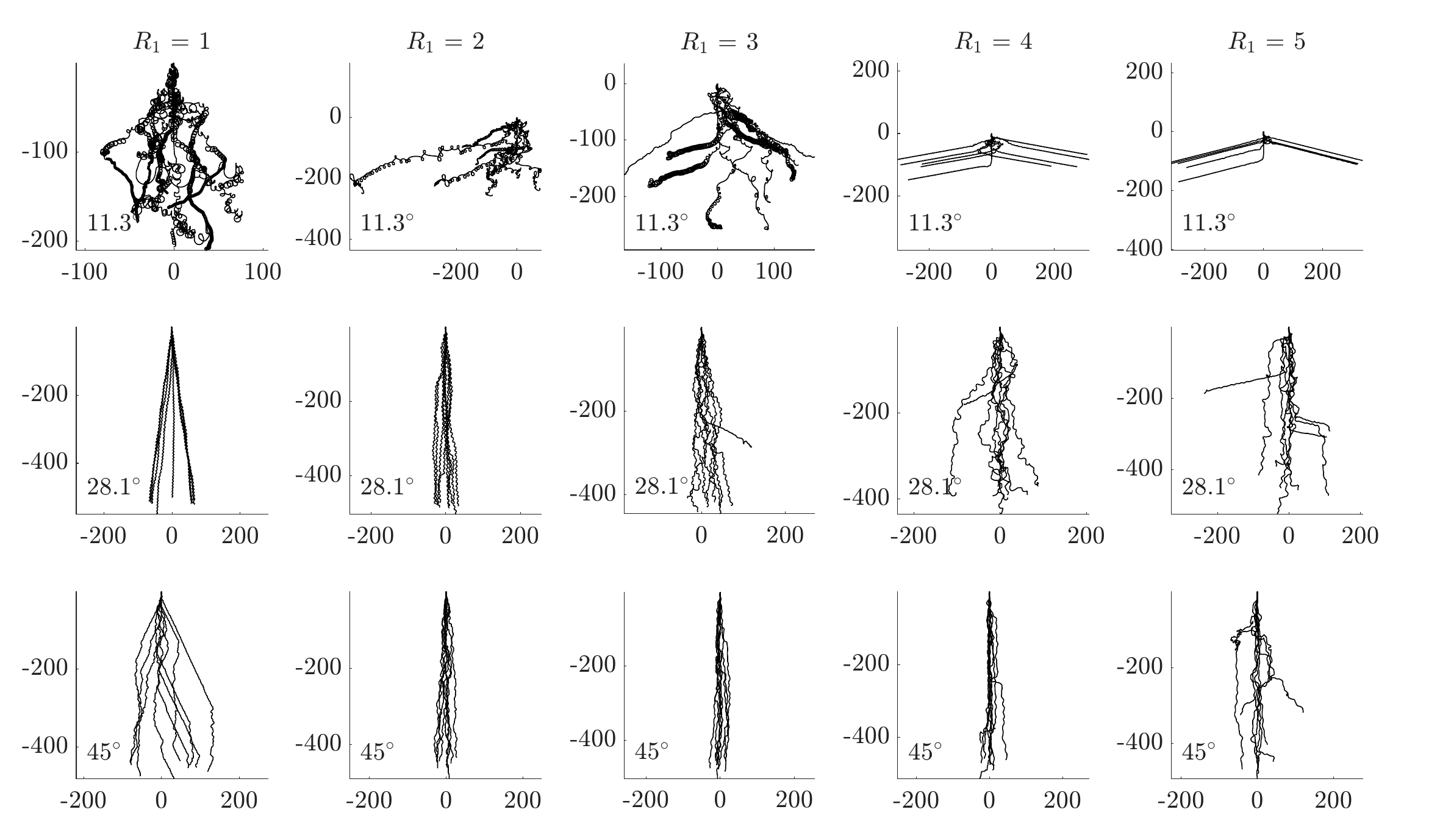}
    \caption{The trajectories of falling bent plates up to $t \approx 500$ for $R_1 = 1,\ 2,\ 3,\ 4,\ 5$ (one per column) and bending angles $\theta = 11.3^\circ,\ 28.1^\circ,\ 45^\circ$ (one per row, shown at the bottom left of each panel).}
    \label{VPlate Large R1 expanded}
\end{figure}

The center-of-mass trajectories of $V$-shaped plates are shown in figures \ref{VPlate Small R1 expanded} and \ref{VPlate Large R1 expanded}, which provide expanded views of figures \ref{VPlate Small R1} and \ref{VPlate Large R1} covering more extensive $R_1$ ranges: $R_1 = 0.1,\ 0.3,\ 0.5,\ 0.7,\ 0.9$, and $R_1 = 1,\ 2,\ 3,\ 4,\ 5$, respectively.

\clearpage
\printbibliography[
heading=bibintoc,
title={References}
]
\end{document}